\documentclass[a4paper,12pt,final]{article}
\def\ftype{png}
\pdfoutput=1
\usepackage{geometry}
\usepackage[T1]{fontenc}
\usepackage{ifthen}
\usepackage{amsmath}
\usepackage{amsfonts}
\usepackage{amsthm}
\usepackage{amssymb}
\usepackage{enumerate}
\usepackage{dsfont}
\usepackage[all]{xy}
\usepackage{epsfig}
\usepackage{color}
\usepackage{hyperref}
\hypersetup{colorlinks=true,linkcolor=black,citecolor=black}
\graphicspath{{Graphics/}}
\newtheorem{Lem}{Lemma}[section]
\newtheorem{Pro}[Lem]{Proposition}

\newtheorem{Cor}[Lem]{Corollary}
\newtheorem{Thm}[Lem]{Theorem}
\newtheorem*{Thm*}{Theorem}
\theoremstyle{definition}

\newtheorem{Def}[Lem]{Definition}
\newtheorem{Exa}[Lem]{Example}
\newtheorem*{Exa*}{Example}
\newtheorem{Rem}[Lem]{Remark}
\newtheorem*{Rem*}{Remark}

\newtheorem*{Sch*}{Schedule}
\newcommand{\bA}{{\mathbb A}}

\newcommand{\bC}{{\mathbb C}}
\newcommand{\bE}{{\mathbb E}}
\newcommand{\bF}{{\mathbb F}}

\newcommand{\bM}{{\mathbb M}}
\newcommand{\bN}{{\mathbb N}}
\newcommand{\bP}{{\mathbb P}}
\newcommand{\bR}{{\mathbb R}}

\newcommand{\cA}{{\mathcal A}}
\newcommand{\cB}{{\mathcal B}}

\newcommand{\cE}{{\mathcal E}}

\newcommand{\cF}{{\mathcal F}}

\newcommand{\cL}{{\mathcal L}}
\newcommand{\cM}{{\mathcal M}}

\newcommand{\cP}{{\mathcal P}}

\newcommand{\cS}{{\mathcal S}}
\newcommand{\cT}{{\mathcal T}}

\newcommand{\cl}{{\rm cl}}
\newcommand{\sa}{\cA_{\rm sa}}
\newcommand{\id}{\mathds{1}}
\newcommand{\tr}{\operatorname{tr}}
\newcommand{\az}{{\cA_0}}
\newcommand{\ao}{{\cA_1}}
\newcommand{\rk}{\operatorname{rk}}
\newcommand{\ri}{\operatorname{ri}}
\newcommand{\rb}{\operatorname{rb}}

\newcommand{\aff}{\operatorname{aff}}
\newcommand{\lin}{\operatorname{lin}}
\begin{document}
\thispagestyle{empty}
\begin{center} 
\textbf{\Large{Information topologies on non-commutative state spaces}}\\
\vspace{.3cm}
Stephan Weis\footnote{\texttt{sweis@mis.mpg.de}}\\
Max Planck Institute for Mathematics in the Sciences\\
Leipzig, Germany\\
\vspace{.1cm}
January 18, 2013
\end{center}
\noindent
{\small\textbf{\emph{Abstract --}}
We define an {\it information topology} (I-topology) and a
{\it reverse information topology} (rI-topology) on the state space of a 
C*-subalgebra of ${\rm Mat}(n,\bC)$. These topologies arise from sequential 
convergence with respect to the relative entropy. We prove that open disks, 
with respect to the relative entropy, define a base for them, while 
Csisz\'ar has shown in 1967 that the analogue is wrong for probability 
measures on a countably infinite set. The I-topology is finer than the norm 
topology, it disconnects the convex state space into its faces. 
The rI-topology is intermediate between these topologies.
We complete two fundamental theorems of information geometry to the full 
state space, by taking the closure in the rI-topology. The norm topology 
is too coarse for this aim only for a non-commutative algebra, so its 
discrepancy to the rI-topology belongs to the quantum domain. We apply 
our results to the maximization of the von Neumann entropy under linear 
constraints and to the maximization of quantum correlations.\\[3mm]
}
%
%
%
%
\tableofcontents
%
%
%
%
\section{Introduction}
\label{sec:intro}
\par
Pythagorean and projection theorems in information geometry make statements 
about the distance of a probability measure from a family of probability 
measures, see e.g.\ Amari and Nagaoka \cite{Amari_Nagaoka}~\S3 and Csisz\'ar 
and Mat\'u\v s \cite{Csiszar03}~\S{}I.C. The theorems provide a geometric 
frame for applications in large deviation theory or maximum-likelihood 
estimation. While information geometry is often confined to families of 
mutually absolutely continuous probability measures, some theorems have been 
extended \cite{Barndorff,Cencov,Csiszar03,Csiszar05} using the 
I-/rI-convergence\footnote{%
Here and in the sequel ``\,I\,'' stands for ``\,information\,'' and
``\,rI\,'' for ``\,reverse information\,''.}
with respect to the relative entropy, also known as Kullback-Leibler
divergence. In quantum information theory, see e.g.\ 
\cite{Amari_Nagaoka,Benatti,Bengtsson,Hayashi,Holevo,Ingarden,Nielsen,Petz08},
there is also a relative entropy, the Umegaki relative entropy, and one can 
ask the analogue questions as in classical probability theory.
\par
An $n$-level quantum system is described by an algebra of complex 
$n\times n$-matrices which includes the setting of probability measures on the
sample space $\{1,\ldots,n\}$ in form of the commutative algebra of diagonal 
matrices. It was discovered by Weis and Knauf \cite{Weis_Knauf} for a
$3$-level quantum system that the above theorems of information geometry can 
not be extended using the norm topology, because this topology is too coarse 
and its closures are too large. We believe that the convex geometry of a 
quantum state space already makes the norm topology unsuitable, which can 
not distinguish between the state space, a unit ball, a simplex or any
other convex body. 
In fact, in the commutative setting, the space of probability measures on a 
finite measurable space is a simplex, and the norm topology does have suitable 
closures in order to extend e.g.\ maximum-likelihood estimation for exponential 
families, see \cite{Barndorff}~p.~155. On the other hand, a quantum state 
space is a convex body but neither a ball nor a simplex \cite{BWZ}. It has a 
Lie group symmetry \cite{Bengtsson} and is studied under the name of 
{\it free spectrahedron} \cite{Sanyal} in the field of convex algebraic 
geometry, using techniques of algebraic geometry.
\par 
This article has two expository sections,  \S\ref{sec:exp1} and 
\S\ref{sec:news_rI_geometry}, including the main ideas and results. 
The preparatory section \S\ref{sec:matrix_analysis} follows and
provides techniques for subsequent analysis. The sections 
\S\ref{sec:topo} and \S\ref{sec:exp} collect the main proofs. 
The dependence on the matrix representation is investigated in 
\S\ref{sec:conc}.
\par
As we shall see in section \S\ref{sec:exp1}, the rI-topology is always first 
countable because the open disks of the relative entropy are its base. This 
is also true for the I-topology which will be discussed already in 
\S\ref{intro:hist}. In a sense, the I-/rI-topology combines simple properties 
of a metric topology with a special compatibility for geometric structure 
(decreasing topologies with respect to inclusion): 
\begin{enumerate}[1.]
\item
The I-topology recognizes the facial structure of the convex quantum
state space, which is split into the connected components of 
(relative interiors of) its faces. 
\item
The rI-topology is adjusted to information geometry, it extends the
Pythagorean and the projection theorem in topological closures. 
\item
The norm topology sees the state space as an arbitrary convex body. 
\end{enumerate}
In the setting of a non-commutative algebra of $n\times n$-matrices
the I-topology is too big and the norm topology is too small 
to extend the Pythagorean or the projection theorem. In the commutative 
setting the rI-topology equals the norm topology.
\par
In \S\ref{sec:news_rI_geometry} we show that the rI-topology has the perfect 
closures to extend the Pythagorean and projection theorem to the full state 
space of an $n$-level quantum system. However, the extensions are proved using
a combination of convex geometry and calculus of matrices, without taking the 
rI-topology into account. 
\par
The Pythagorean theorem implies the first solution 
in the literature to the maximization of the von Neumann entropy under linear 
constraints. This completes partial results from 1963 in Wichmann's article 
\cite{Wichmann}.
The projection theorem has applications to quantum correlations. 
In \S\ref{sec:max} we generalize several ideas from Ay's article \cite{Ay} about 
local maximizers of correlation into the quantum setting. An essential part
of our proof of the projection theorem is to study non-exposed faces of 
linear images of state spaces using Gr\"unbaum's notion of {\it poonem} 
\cite{Gruenbaum}. We argue in \S\ref{sec:why-poonems} why poonems are needed.
An analogous approach to exponential families of probability measures was 
taken in \cite{Csiszar05} with the concept of {\it access sequence}. 
\par
A broader usefulness of the 
I-/rI-topology in quantum statistics and quantum hypothesis testing is not yet 
clarified. In contrast to the commutative setting there is no canonical choice 
of a computational basis and the consequences for measurement and observation will 
have to be taken into consideration. One advance is that the infimum of the 
relative entropy does not decrease under information closures 
(\ref{eq:cl-under-inf}). This is useful e.g.\ in the Sanov theorem in quantum 
hypothesis testing \cite{Bjelakovic}.
%
%
%
%
\section{Information convergence, information topology}
\label{sec:exp1}
\par
This is an expository section. We recall literature on information 
convergence and topology in \S\ref{intro:hist} where we also define 
quantum state spaces and have a first discussion about topologies on 
the state space. After recalling some properties of the relative entropy 
in \S\ref{sec:entropy_properties} we give an overview of our purely 
topological results for finite-level quantum systems in 
\S\ref{sec:news_top}. 
\par
An $n$-level quantum system is described by the algebra ${\rm Mat}(n,\bC)$. 
We consider a C*-subalgebra $\mathcal A$ of ${\rm Mat}(n,\bC)$, i.e.\ a 
complex subalgebra $\cA$ of ${\rm Mat}(n,\bC)$ closed under the adjoint map 
$a\mapsto a^*$. The definition of a C*-subalgebra includes completeness with 
respect to a norm, but this is clear in finite dimensions. We prefer the term 
C*-subalgebra because it reminds us of the complex field $\bC$ and of the 
closure under the adjoint map. 
Unless otherwise stated, $\cA$ is a C*-subalgebra  of ${\rm Mat}(n,\bC)$.
%
%
%
\subsection{Spaces of probability measures and quantum states}
\label{intro:hist}
\par
We discuss convergence with respect to the relative entropy, called 
{\it information convergence}. For finite measurable spaces we discuss the 
associated topology in detail. Then we generalize to finite-level quantum 
systems and we finish with a short discussion of infinite-dimensional 
commutative von Neumann algebras. All proofs and further issues follow in 
\S\ref{sec:topo}. A review of information convergence and its topology in 
probability theory is given in \S{}I.C in \cite{Csiszar03}. 
\par
Let $\mathcal M$ be a set of 
probability measures on a measurable space $(X,\mathcal X)$. If 
$P,Q\in\mathcal M$ are absolutely continuous with respect to a 
$\sigma$-finite measure $\lambda$ and $p(x)$ resp.\ $q(x)$ is the 
Radon-Nikodym derivative of $P$ resp.\ $Q$, then the 
{\it relative entropy} is
\begin{equation}
\label{def:kl}\textstyle
D(P||Q)
\;:=\;
\int_Xp(x)\log\tfrac{p(x)}{q(x)}d\lambda\,.
\end{equation}
This equals zero if and only if $P=Q$ and otherwise $D(P||Q)$ is strictly 
positive or $+\infty$, see \cite{Kullback}. The {\it total variation} is
\begin{equation}
\label{def:total}\textstyle
\|P-Q\|_1
\;:=\;
\int_X|p(x)-q(x)|d\lambda\,.
\end{equation}
It is well-known that total variation defines a norm on the space of 
signed measures having a density with respect to $\lambda$. The 
{\it Pinsker-Csisz\'ar inequality} \cite{Gilardoni} shows
\begin{equation}
\label{def:Pinsker-Csiszar-Original}\textstyle
\|P-Q\|_1^2
\;\leq\;
2 D(P||Q)\,.
\end{equation}
Given a sequence
$(P_k)_{k\in\mathbb N}\subset\mathcal M$ and a probability measure
$P\in\mathcal M$ we have, according to \cite{Csiszar03},
{\it I-convergence} resp.\ {\it rI-convergence} of $(P_k)_{k\in\mathbb N}$
to $P$ if
\begin{equation}
\label{intro:info_conv}\textstyle
\lim_{k\to\infty}D(P_k||P)\;=\;0
\qquad
\text{resp.}
\qquad
\lim_{k\to\infty}D(P||P_k)\;=\;0\,.
\end{equation}
Csisz\'ar has studied these convergences in the context of the 
$f$-divergence, generalizing the relative entropy. He has proved in 
Theorem 3 in \cite{Csiszar67} that the {\it information neighborhoods},
defined for $P\in\mathcal M$ and $\epsilon>0$ by
\begin{equation}
\label{intro:neigh}\textstyle
\{Q\in\mathcal M\mid D(Q||P)<\epsilon\}
\qquad
\text{resp.}
\qquad
\{Q\in\mathcal M\mid D(P||Q)<\epsilon\}\,,
\end{equation}
are not a base of a topology if $(X,\mathcal X)=(\bN,2^\bN)$
where $2^\bN$ is the power set of $\bN$.
\par
In spite of Csisz\'ar's negative result it is possible to define 
an {\it I-topology} resp.\ {\it rI-topology} on $\cM$ in terms of the 
convergence of (countable) sequences (\ref{intro:info_conv}), see Dudley 
and Harremo\"es \cite{Dudley98,Harremoes}. Here a subset $U\subset\cM$ is 
open if for each probability measure $P\in U$ and each sequence 
$(P_k)_{k\in\mathbb N}\subset\mathcal M$ that I- resp.\ rI-converges 
to $P$, there exists $N\in\bN$ such that for all $k\geq N$ we have 
$P_k\in U$. It follows from (\ref{def:Pinsker-Csiszar-Original}) that
the I-/rI-topology is finer than the norm topology of the total variation.
\par
We will take the approach by Dudley and Harremo\"es to study information 
topologies for an $n$-level quantum system. The common ground between 
finite-level quantum systems and spaces of probability measures are spaces 
of probability measures on a finite measurable space. The 
{\it probability simplex} of a non-empty (at most) countable set $X$ is
\begin{equation}
\label{def:prob_simplex}\textstyle
\bP(X)
\;:=\;
\{p=(p_x)_{x\in X}\in[0,1]^X
\mid \sum_{x\in X}p_x=1\}\,.
\end{equation}
Elements $p$ of $\bP(X)$ are called {\it probability vectors} on $X$ and 
can be identified with probability measures $P$ on $(X,2^X)$ using 
$P(A):=\sum_{x\in A}p_x$ for $A\subset X$. For a finite measurable space 
$X=\{1,\ldots,n\}$ the probability simplex $\bP(X)$ is a simplex of 
dimension $n-1$ and the information 
neighborhoods (\ref{intro:neigh}) are a base of a topology. The rI-topology 
on $\bP(X)$ equals the total variation topology, which is the restriction 
of the standard Euclidean topology on $\bR^X$. The I-topology splits $\bP(X)$ 
into connected components $C(X')$ of constant support $X'\subset X$,
\[\textstyle
C(X')\;:=\;\{p\in\bP(X)\mid p_x>0\iff x\in X'\}\,.
\]
On each connected component the I-topology equals the norm topology. This 
decomposition into connected components is the stratification 
(\ref{eq:stratification}) of the probability simplex into relative interiors 
of its faces. 
\par
To describe the I-/rI-topology for a C*-subalgebra $\mathcal A$ of 
${\rm Mat}(n,\bC)$ let us introduce some notation. We denote the identity 
in ${\rm Mat}(n,\bC)$ by $\id_n$ (the zero by $0_n$ or $0$) and the identity 
in $\cA$ by $\id$. A {\it state} on $\cA$ is a complex linear functional 
$f:\cA\to\bC$, such that $f(a^*a)\geq 0$ for all $a\in\cA$ and $f(\id)=1$. 
The standard trace $\tr$ turns ${\rm Mat}(n,\bC)$ into a complex Hilbert 
space with the {\it Hilbert-Schmidt} inner product
$\langle a,b\rangle\,:=\,\tr(ab^*)$ for $a,b\in{\rm Mat}(n,\bC)$ and we use 
the {\it two-norm} $\|a\|_2:=\sqrt{\langle a,a\rangle}$. By $\sa$ we denote 
the real vector space of self-adjoint matrices in $\cA$ and 
$(\sa,\langle\cdot,\cdot\rangle)$ is a Euclidean vector space. We 
call its norm topology on any subset simply {\it norm topology}
as all norms are equivalent in finite dimensions.
\par
There is a one-to-one correspondence between states $f$ on $\cA$ and matrices 
in $\cA$ which are positive semi-definite ($\rho\succeq 0$) and have trace one 
($\tr(\rho)=1$), see e.g.\ Theorem 2.4.21 in \cite{Bratteli}. The functional 
$f$ and the matrix $\rho$ are related by 
\begin{equation}
\label{eq:state-iso}\textstyle
f(a)\;=\;\langle a,\rho\rangle \qquad (a\in\cA)\,.
\end{equation}
The matrix representation $\rho$ of $f$ is called {\it density matrix} in 
quantum mechanics. We will use the terms of state and density matrix 
synonymously. The {\it state space} is
\begin{equation}
\label{def:state_space}\textstyle
\cS
\;=\;
\cS_\cA
\;:=\;
\{\rho\in\cA\mid\rho\succeq 0,\tr(\rho)=1\,\}\,.
\end{equation}
For the commutative subalgebra $\cA$ of complex diagonal matrices of size 
$n\times n$ the state space is the probability simplex (\ref{def:prob_simplex}),
\begin{equation}
\label{def:commutative_inclusion}\textstyle
\bP(\{1,\ldots,n\})
\;=\;
\cS_\cA
\;\subset\;
{\rm Mat}(n,\bC)\,.
\end{equation}
\par
We will show for a possibly non-commutative C*-subalgebra $\mathcal A$ of 
${\rm Mat}(n,\bC)$ that the analogues of information neighborhoods 
(\ref{intro:neigh}) are bases of two topologies. For a non-commutative 
algebra $\cA$ the analogue of the rI-topology is strictly finer than the 
norm topology (Corollary~\ref{cor:top}) and it defines---unlike the norm 
topology---useful closures in information theory, as we shall outline 
in \S\ref{sec:news_rI_geometry}.
\par
We show in Theorem~\ref{thm:info_top}.3 that the analogue of the I-topology 
splits $\cS$ into connected components of states $\rho\in\mathcal S$ of 
constant support $s(\rho)\in\cA$. Here we use the partial ordering $\preceq$ 
on $\cA$ defined by $a\preceq b$ for $a,b\in\cA$ if and only if $b-a$ is 
positive semi-definite. The {\it projection lattice} $\cP$ of the algebra $\cA$
is
\begin{equation}
\label{def:proj}\textstyle
\cP\;=\;\cP_\cA\;:=\;\{p\in\cA\mid p^2=p^*=p\}\,,
\end{equation}
its elements are {\it projections}. The partial ordering restricts to $\cP$, 
more details about lattices are discussed in 
\S\ref{ingredients:convex} and infinite-dimensional algebras are treated e.g.\ 
in \cite{Alfsen}.
The {\it support projection} of a self-adjoint matrix $a\in\cA$ is the 
infimum
\[\textstyle
s(a)
\;:=\;
\bigwedge\{p\in\cP\mid pa=a\}\,.
\]
Again, like for the probability simplex $\bP(X)$, the decomposition of $\cS$ 
into connected components is the stratification (\ref{eq:stratification}) of 
the state space $\cS$ into relative interiors of its faces. Of course, $\cS$ 
is homeomorphic in the norm topology to the closed Euclidean unit ball. This 
is a property of any convex body, i.e.\ compact and convex subset of Euclidean 
space, known as the Theorem of Sz.\ Nagy, see e.g.\ \S{}VIII.1 in \cite{Berge}.
\par
In the algebraic formalism, the measurable space $(\bN,2^\bN)$
corresponds to the von Neumann algebra of bounded sequences
\[\textstyle
l^\infty
\;:=\;
\{x=(x_i)_{i\in\bN}\in\bC^\bN\mid\sup_{i\in\bN}|x_i|<\infty\}
\]
acting by multiplication on the Hilbert space 
$l^2:=\{x\in l^\infty\mid\sum_{i\in\bN}|x_i|^2<\infty\}$ of square
summable sequences. The space
$l^1:=\{x\in l^\infty\mid\sum_{i\in\bN}|x_i|<\infty\}$ of absolutely
summable sequences contains the probability simplex\footnote{%
The probability simplex $\bP(\bN)$ corresponds to the {\it normal states} 
on $l^\infty$ (see e.g.\ Theorem 2.4.21 in \cite{Bratteli}). The space of 
positive linear maps $f:l^\infty\to\bC$ with $f(\id)=1$ is strictly larger 
than $\bP(\bN)$ and can be represented by bounded additive measures which 
are not necessarily $\sigma$-additive (see e.g.\ p.\ 89 in \cite{Werner} 
and p.\ 296 in \cite{Dunford_Schwartz}).}
$\bP(\bN)$. This has, for the algebra $\cA=l^\infty$, the form 
(\ref{def:state_space}) of a state space if we denote for $x\in l^1$ the set 
of inequalities $x_i\geq 0$ for all $i\in\bN$ simultaneously by $x\succeq 0$ 
and if we use the {\it trace} ${\rm tr}:l^1\to\bC$, 
$x\mapsto\sum_{i\in\bN}x_i$,
\[\textstyle
\bP(\bN)
\;=\;
\cS_\cA
\;=\;
\{x\in l^1\mid x\succeq 0, {\rm tr}(x)=1\}\,.
\]
The discussion above shows that the information neighborhoods 
(\ref{intro:neigh}) do not define a topology on $\bP(\bN)$ but two 
topologies are defined in terms of the convergences (\ref{intro:info_conv}). 
%
%
%
%
\subsection{The relative entropy}
\label{sec:entropy_properties}
\par
The relative entropy is a measure of distance between states. It has an 
operational meaning e.g.\ in hypothesis testing \cite{Petz08}.
We recall well-known convexity and continuity properties. Although 
the relative entropy is not continuous in the norm topology, we point
out that it is continuous in the I-topology in its first argument and 
continuous in the rI-topology in its second argument (\ref{eq:omegaconts}).
\begin{Def}
\label{def:entropy}
The {\it relative entropy} of a density matrix $\rho\in\cS$ from 
$\sigma\in\cS$ is 
\begin{equation}\textstyle
\label{eq:relative_entropy}
S(\rho,\sigma)\;:=\;\tr \rho(\log(\rho)-\log(\sigma))
\end{equation}
if ${\rm Im}(\rho)\subset{\rm Im}(\sigma)$. Otherwise 
$S(\rho,\sigma):=+\infty$. The logarithm can be defined by functional 
calculus, see Remark~\ref{rem:func_calc}.3.
\end{Def}
\par
The relative entropy (\ref{eq:relative_entropy}) satisfies 
$S(\rho,\sigma)\geq 0$ for all $\rho,\sigma\in\cS$ with equality if and 
only if $\rho=\sigma$, see e.g.\ \S11.3 in \cite{Petz08} or \S11.3 of 
\cite{Nielsen}. It is discontinuous (in the norm topology) in the first
argument already for the algebra $\cA=\bC^2$ of a bit and in the second
argument for the algebra $\cA={\rm Mat}(2,\bC)$ of a qubit.
\begin{Exa}
\label{ex:norm_coarseness}
If $\cA=\bC^2$  then 
$S\left((\tfrac{n-1}{n},\tfrac 1{n}),(1,0)\right)=\infty$ for all 
$n\in\bN$ while $S\left((1,0),(1,0)\right)=0$.
If $\cA={\rm Mat}(2,\bC)$, then for real $\alpha$ we have
\[\textstyle
S\left(\tfrac 1{2}(\id_2+\sigma_1),
\tfrac 1{2}(\id_2+\cos(\alpha)\sigma_1+\sin(\alpha)\sigma_2)
\right)
\;=\;
\left\{\begin{array}{rl}0 & \text{if } \alpha=0\mod 2\pi,\\
\infty & \text{else.}\end{array}\right.
\]
Example 11 in \cite{Weis_Knauf} is less trivial: A smooth curve 
$t\mapsto\sigma_t$ 
converging in norm to $\rho$ on the boundary of the Bloch ball
$\cS_{{\rm Mat}(2,\bC)}$ can have any non-negative limit of 
$S(\rho,\sigma_t)$.
\end{Exa}
\par
Let us now turn to some well-known properties of the relative entropy.
\begin{Def}
\label{def:conti}~
\begin{enumerate}[1.]
\item
A function $f:X\to(-\infty,\infty]$, defined on a convex subset $X$ of a 
finite-dimensional Euclidean vector space $\bE$, is {\it convex} if for
$x_1,x_2\in X$ and $\lambda\in[0,1]$
\[\textstyle
f((1-\lambda)x_1+\lambda x_2)\;\leq\;(1-\lambda)f(x_1)+\lambda f(x_2)\,.
\]
In the special case that $Y$ is another convex subset of $\bE$ and 
$f:X\times Y\to(-\infty,\infty]$ is defined, such that
for $x_1,x_2\in X$, $y_1,y_2\in Y$ and $\lambda\in[0,1]$ we have
\[\textstyle
f((1-\lambda)x_1+\lambda x_2,(1-\lambda)y_1+\lambda y_2)
\;\leq\;(1-\lambda)f(x_1,y_1)+\lambda f(x_2,y_2)
\]
then $f$ is called {\it jointly convex}.
A function $f:X\to\bR$ is {\it strictly convex} if for
$x,y\in X$, $x\neq y$ and $\lambda\in(0,1)$
\[\textstyle
f((1-\lambda)x+\lambda y)\;<\;(1-\lambda)f(x)+\lambda f(y)\,.
\]
If $f$ is (strictly) convex, we say that $-f$ is {\it (strictly) concave}.
\item
If $(X,d)$ is a metric space and $f:X\to(-\infty,\infty]$ then
$f$ is {\it lower semi-continuous} if for all $x\in X$ and every
sequence $(x_i)_{i\in\mathbb N}\subset X$ converging to $x$ we have
\[\textstyle
\liminf_{i\to\infty}f(x_i)
\;\geq\;
f(x)\,.
\]
\end{enumerate}
\end{Def}
\begin{Rem}
\label{rem:conv_entropy}~
\begin{enumerate}[1.]
\item
The lower semi-continuity of the relative entropy (in the norm 
topology) is proved e.g.\ by 
Wehrl in \S{}III.B in \cite{Wehrl}, using Lindblad's representation 
of the relative entropy \cite{Lindblad}. Ohya and Petz give another
proof in \S5 in \cite{Ohya}. They use Kosaki's formula and write the 
relative entropy as a supremum of affine functionals.
\item
The joint convexity of the relative entropy follows from Lieb's theorem 
\cite{Lieb}, see e.g.\ \S11.4 in \cite{Nielsen} or \S{}III in \cite{Wehrl} 
for proofs and the historic context. Convexity of the relative entropy
is a special case of the joint convexity of quasi-entropies \cite{Petz86}.
Also, it follows easily from the monotonicity of the relative entropy 
under quantum operations, see \S3.4 in \cite{Petz08}.
\item
A convex lower semi-continuous function is 
continuous along straight lines. More precisely, let 
$f:X\to(-\infty,\infty]$ be a convex and lower semi-continuous function 
defined on a closed convex subset $X$ of a finite-dimensional Euclidean 
vector space $\bE$. We extend $f$ to $\bE$ by setting its value to
$+\infty$ outside of $X$. Since $X$ is convex, the extension $\widetilde{f}$
is convex. Since $X$ is closed, $\widetilde{f}$ is lower semi-continuous. 
Thus, Corollary 7.5.1 in \cite{Rockafellar} shows for $x,y\in\bE$, 
subject to $\widetilde{f}(x)<+\infty$, that
\[\textstyle
\widetilde{f}(y)\;=\;
\lim_{\lambda\nearrow 1}\widetilde{f}((1-\lambda)x + \lambda y)\,.
\]
Here the values of $\widetilde{f}$ converge in the Alexandroff
compactification $(-\infty,\infty]$ of $\bR$, see
Example~\ref{exa:alexandroff}. For example, if $\tau\in\cS$ is any 
invertible density matrix, then for arbitrary $\rho,\sigma\in\cS$ we 
have $S(\rho,\tau)<\infty$ so
\begin{equation}\textstyle
\label{eq:continuous_straight}
S(\rho,\sigma)\;=\;
\lim_{\lambda\nearrow 1}S(\rho,(1-\lambda)\tau + \lambda\sigma)\,.
\end{equation}
We use (\ref{eq:continuous_straight}) in Theorem~\ref{thm:info_top}.5
to prove that the state space $\cS$ is connected in the rI-topology%
\footnote{%
The analogue of (\ref{eq:continuous_straight}) with flipped
arguments is wrong: If $\sigma$ is not invertible, then 
$S((1-\lambda)\tau + \lambda\rho,\sigma)=\infty$ for $\lambda<1$ while 
the limit $S(\rho,\sigma)$ can be arbitrary.}.
\end{enumerate}
\end{Rem}
%
%
%
%
%
%
\subsection{New results about the I- and the rI-topology}
\label{sec:news_top}
\par
This section summarizes properties of the I-/rI-topology of a C*-subalgebra 
$\cA$ of ${\rm Mat}(n,\bC)$ with focus on similarities to a metric topology. 
Reasoning is done within the theory of sequential convergence, recalled in 
\S\ref{sec:topo}, exceptions are the Pinsker-Csisz\'ar inequality 
(\ref{eq:pinsker}) and the continuity result of (\ref{eq:omegaconts}) which 
are from matrix theory. Since the I-topology and the rI-topology share many 
properties, we use a pre\-fix variable $\omega\in\{{\rm I},{\rm rI}\}$ to 
denote
\[\textstyle
\text{$\omega$-topology, $\omega$-closure, etc.}  
\]
Unless otherwise specified we always use the norm topology. We end the 
section with an application in information theory. 
\begin{Def}[Information topology]~
\begin{enumerate}[1.]
\item
We use short-hand notation for the two possible variable orderings
of the relative entropy (\ref{eq:relative_entropy}),
\[\textstyle
S^{\rm I}(\rho,\sigma)\;:=\;S(\sigma,\rho)
\qquad\text{and}\qquad
S^{\rm rI}(\rho,\sigma)\;:=\;S(\rho,\sigma)\,.
\]
If $\{A(i)\}_{i\in\bN}$ is a sequence of statements, then we shall say that
$A(i)$ is true {\it for large $i$} if there is $N\in\bN$ such that $A(i)$
holds for all $i\geq N$. We define a family of subsets of the state space 
$\cS$ by
\[\textstyle
\cT^\omega
\;:=\;
\left\{U\subset\cS\;\mid
\begin{array}{l}
\text{if } \rho\in U, (\rho_i)_{i\in\bN}\subset\cS
\text{ and }\lim_{i\to\infty}S^\omega(\rho,\rho_i)=0\,,\\
\text{then } \rho_i\in U\text{ for large }i
\end{array}
\right\}\,.
\]
The {\it open $\omega$-disk} about $\rho\in\cS$ with radius 
$\epsilon\in(0,\infty]$ is
\begin{equation}\textstyle
\label{eq:open_disks}
V^\omega(\rho,\epsilon)
\;:=\;\{\sigma\in\cS\mid S^\omega(\rho,\sigma)<\epsilon\}
\end{equation}
and the {\it closed $\omega$-disk} about $\rho\in\cS$ with radius 
$\epsilon\in(0,\infty]$ is
\begin{equation}\textstyle
\label{eq:closed_disks}
W^\omega(\rho,\epsilon)
\;:=\;\{\sigma\in\cS\mid S^\omega(\rho,\sigma)\leq\epsilon\}\,.
\end{equation}
We denote the (unique) norm topology on $\cS$ by $\cT^{\|\cdot\|}$. 
For $a\in\cA$ and $(a_i)_{i\in\bN}\subset\cA$ we denote by
$\lim_{i\to\infty}a_i=a$ the convergence of $(a_i)_{i\in\bN}$ to $a$ 
in norm.
\item
Let $(X,\mathcal T)$ be a topological space. The topology $\mathcal T$ 
is a {\it Hausdorff topology} if each two distinct points of $X$ belong 
to two disjoint open sets. A family $\mathcal B\subset\mathcal T$ is a 
{\it base} for $(X,\mathcal T)$ if any non-empty open subset of $X$ is a
union of a subfamily of $\mathcal B$. A family $\mathcal B(x)$ of open 
sets containing $x\in X$ is called a {\it base} for $(X,\mathcal T)$ 
at $x$ if for any open set $V$ containing $x$ there exists 
$U\in\mathcal B(x)$ such that $U\subset V$. The topological space 
$(X,\mathcal T)$ is {\it first-countable} if there exists a countable 
base at every point $x\in X$, it is {\it second-countable} if it has a 
countable base. 
\end{enumerate}
\end{Def}
\par
The family $\cT^\omega$ is easily seen to be a topology on $\cS$, 
which we call the $\omega$-{\it topology}. The inclusion 
$\cT^{\|\cdot\|}\subset\cT^\omega$ follows directly from the 
{\it Pinsker-Csisz\'ar inequality}, which confirms for 
$\rho,\sigma\in\cS$ that 
\begin{equation}\textstyle
\label{eq:pinsker}
\|\rho-\sigma\|_1^2
\;\leq\;
2S(\rho,\sigma)\,.
\end{equation}
Here the trace norm from Definition~\ref{def:represent}.2 is used,
see e.g.\ \S3.4 in \cite{Petz08} for a proof. The inclusion 
$\cT^{\|\cdot\|}\subset\cT^\omega$ implies that the $\omega$-topology
is a Hausdorff topology.
The convergence of sequences {\it a priori}, in terms of the relative 
entropy, is equivalent to the convergence {\it a posteriori}, in terms 
of the $\omega$-topology. This is formalized as the equivalence a) 
below, which in Theorem~\ref{thm:info_top} takes the form of 
$C(\cT^\omega)=C^\omega$. For sequences $(\rho_i)_{i\in\bN}\subset\cS$ 
and states $\rho\in\cS$ we have
\begin{align}
\label{eq:metric-cond}\textstyle
\lim_{i\to\infty}S^\omega(\rho,\rho_i)=0
\;\stackrel{\textup{a)}}{\iff}\;&\textstyle
\forall U\in\cT^\omega\text{ with }\rho\in U\text{ we have }
\rho_i\in U\text{ for large }i\\\nonumber
\textstyle\;\stackrel{\textup{b)}}{\iff}\;&\textstyle
\forall\epsilon\in(0,\infty]\text{ we have }
\rho_i\in V^\omega(\rho,\epsilon)\text{ for large }i\,.
\end{align}
Equivalence a) holds more generally for any divergence function in the 
sense of \S\ref{sec:topo_seq}. In particular, a) holds for 
infinite-dimensional algebras. 
\par
The equivalence b) is more restrictive. It follows from a continuity
property. In Proposition~\ref{pro:continuous_1st_2nd}
we use (\ref{eq:pinsker}), and for the case $\omega={\rm rI}$ some
perturbation theory, to show for all $\rho,\sigma\in\cS$ and 
$(\sigma_i)_{i\in\bN}\subset\cS$
\begin{equation}
\label{eq:omegaconts}\textstyle
\lim_{i\to\infty}S^\omega(\sigma,\sigma_i)\;=\;0
\quad\implies\quad
\lim_{i\to\infty}S^\omega(\rho,\sigma_i)\;=\;S^\omega(\rho,\sigma)\,.
\end{equation} 
In Theorem~\ref{thm:info_top}.2 we show that (\ref{eq:omegaconts}) means that 
the relative entropy is continuous in the first argument for the I-topology 
and in the second argument for the rI-topology. Therefore the open $\omega$-disks 
are a base of $\cT^\omega$, which is equivalent to b) in (\ref{eq:metric-cond}). 
Hence $\cT^\omega$ is first-countable while Corollary~\ref{cor:top} shows
that $\cT^\omega$ is second-countable if and only if $\mathcal A$ is 
commutative. The continuity (\ref{eq:omegaconts}) is wrong for the 
infinite-dimensional algebra $l^\infty$, where the open $\omega$-disks are 
not a base of a topology, see \S\ref{intro:hist}.
\par
In addition to proving distance-like properties (\ref{eq:metric-cond}), 
we use (\ref{eq:omegaconts}) in Theorem~\ref{thm:info_top}.4  to show
\begin{equation}
\label{eq:top-inclusion}\textstyle
\cT^{\|\cdot\|}
\;\subset\;
\cT^{\rm rI}
\;\subset\;
\cT^{\rm I}\,.
\end{equation}
Equivalently we have for all sequences $(\rho_i)_{i\in\bN}\subset\cS$ and 
states $\rho\in\cS$ the ordering of convergences
\[\textstyle
\lim_{i\to\infty}S(\rho_i,\rho)\;=\;0
\quad\implies\quad
\lim_{i\to\infty}S(\rho,\rho_i)\;=\;0
\quad\implies\quad
\lim_{i\to\infty}\rho_i\;=\;\rho\,.
\]
Later in Corollary~\ref{cor:top} we prove the proper inclusion
$\cT^{\|\cdot\|}\subsetneq\cT^{\rm rI}$ for non-commutative algebras, whereas 
$\cT^{\rm rI}\subsetneq\cT^{\rm I}$ holds already for the algebra $\bC^2$ of 
a bit. Other conditions for commutativity will be mentioned in 
\S\ref{sec:non-commutative}.
\par
The infimum of the relative entropy between a state and a set of states 
is useful in information theory and in quantum information theory, e.g.\ 
to compute optimal error rates in hypothesis testing \cite{Bjelakovic}.
\begin{Def}[$\omega$-closure]
For $\rho\in\cS$ and $X\subset\cS$ we write
\begin{equation}
\label{def:rI-distance}\textstyle
S^{\omega}(\rho,X)\;:=\;\inf_{\tau\in X}S^{\omega}(\rho,\tau)
\end{equation}
and we define the {\it $\omega$-closure} of $X\subset\cS$ by
\begin{equation}
\label{def:rI-closure}\textstyle
{\rm cl}^\omega(X)
\;:=\;
\{\rho\in\cS\mid S^{\omega}(\rho,X)=0\}\,.
\end{equation}
\end{Def}
\par
In a C*-subalgebra of ${\rm Mat}(n,\bC)$ we can show in 
Theorem~\ref{thm:info_top}.2 for an arbitrary subset $X\subset\cS$
of states that the $\omega$-closure ${\rm cl}^\omega(X)$ is 
the topological closure of $X$ with respect to $\cT^\omega$.
Differently frased, we have for all $\rho\in\cS$ 
\begin{equation}
\label{eq:cl-under-inf}\textstyle
S^\omega(\rho,{\rm cl}^\omega(X))
\;=\;
S^\omega(\rho,X)\,.
\end{equation}
This is also proved in Corollary~\ref{cor:closures} directly from 
(\ref{eq:omegaconts}). The analogue statement for spaces of probability 
measures on infinite $\sigma$-algebras is wrong by Example~\ref{exa:clcl}.
%
%
%
%
%
%
%
%
%
%
%
%
\section{New results about exponential families}
\label{sec:news_rI_geometry}
\par
This is an expository section about exponential families in a C*-subalgebra 
$\cA$ of ${\rm Mat}(n,\bC)$. With the exception of some corollaries, all 
proofs are done in \S\ref{sec:exp}. We begin in \S\ref{intro:pyth} by
explaining and proving the Pythagorean theorem and the projection theorem 
in a restriction where this is easy. In \S\ref{sec:poonem-algorithm} 
and \S\ref{sec:extension} we define an extension of every exponential family, 
by lifting faces of a convex parameter space. In \S\ref{sec:informal-pythagorean}
we explain a new Pythagorean theorem, valid for this extension. A 
corollary solves the problem of maximizing the von Neumann entropy under 
linear constraints. This is the first complete solution in the literature. 
In \S\ref{sec:select-com-pro} we explain a new projection theorem, valid 
for the extension. A corollary shows that the extension is the rI-closure 
of the exponential family. 
\par
The {\it Staffelberg family} \cite{Weis_Knauf} in \S\ref{sec:select-com-pro}
shows that the norm topology is too coarse to extend an exponential family 
appropriately, its closures are too large. The {\it Swallow family} 
\cite{Weis_Knauf} in \S\ref{sec:why-poonems} demonstrates why poonems
\cite{Gruenbaum} are essential in our proof of the projection theorem.
\par
Issues proved in \S\ref{sec:exp} but not covered in the present section
include applications to quantum correlations in \S\ref{sec:max} and
equality conditions for closures of exponential families in
\S\ref{sec:non-commutative}.
%
%
%
%
%
%
\subsection{A recap of elementary information geometry}
\label{intro:pyth}
\par
We recall the Pythagorean theorem and the projection theorem for the 
relative entropy in a C*-subalgebra $\cA$ of ${\rm Mat}(n,\bC)$. The aim of 
this article is to extended them from the invertible states to the whole 
state space. These theorems are exemplary for many elegant ideas in 
information geometry \cite{Amari_LNS}. The following Pythagorean 
theorem has first appeared in articles \cite{Petz94,Nagaoka} by 
Petz and Nagaoka. It is well-known \cite{Nagaoka,Amari_Nagaoka,Jencova} 
that it fits into the differential-geometric context of dually flat 
spaces, initiated by Amari \cite{Amari_LNS}. 
\par
The {\it Pythagorean theorem of relative entropy} applies to states 
$\rho,\sigma,\tau\in\cS$ where $\sigma,\tau$ are invertible and 
$\rho-\sigma$ is orthogonal to $\log(\tau)-\log(\sigma)$ with respect
to the Hilbert-Schmidt inner product. An elementary calculation shows
for the relative entropy $S$
\begin{equation}\textstyle
\label{eq:pythagoras}
S(\rho,\sigma)\,+\,S(\sigma,\tau)\;=\;S(\rho,\tau)\,.
\end{equation}
With relative entropy (\ref{eq:relative_entropy}) replaced by squared 
Euclidean distance, this equation reminds 
us of the Pythagorean theorem in Euclidean geometry. 
See also \S3.4 in \cite{Petz08} and \S3.4 in \cite{Amari_Nagaoka} for
further information, as well as \S7 in \cite{Amari_Nagaoka} for an
overview of applications in estimation theory.
\begin{Def}
We use the real analytic function $R_\cA:\sa\to\sa$, 
\begin{equation}
\label{def:R}\textstyle
R(\theta)
\;=\;
R_\cA(\theta)
\;:=\;
\exp_\cA(\theta)/\tr(\exp_\cA(\theta))\,.
\end{equation}
The exponential $\exp_\cA$ is defined by functional calculus\footnote{%
We have $\exp_\cA(a)=\id+\sum_{i=1}^\infty a^i/{i!}$ where the identity
$\id$ in $\cA$ can differ from the identity $\id_n$ of ${\rm Mat}(n,\bC)$.}
in the algebra $\cA$, see Definition~\ref{def:functional_calc}.3 and 
Remark~\ref{rem:func_calc}.2. For a non-empty real affine subspace 
$\Theta\subset\sa$ we define an {\it exponential family} in $\cA$ by
\begin{equation}
\label{def:E}\textstyle
\cE
\;:=\;
R_\cA(\Theta)
\;=\;
\{R_\cA(\theta)\mid\theta\in\Theta\,\}\,.
\end{equation}
The parametrization $R_\cA$ is the {\it canonical parametrization} of 
$\cE$. We call a one-dimensional exponential family (with/-out
parametrization) {\it e-geodesic}. We use the translation vector 
space 
$U:=\lin(\Theta)=\Theta-\Theta=\{\theta_1-\theta_2\mid\theta_1,\theta_2\in\Theta\}$.
\end{Def}
\par
We mention vocabulary in the literature. The analogue of the 
parametrization $R_\cA$ of an exponential family is called 
{\it canonical parametrization} in probability theory, see \S20 in \cite{Cencov}. 
For an affine map $a:\bR\to\Theta$ the curve 
$\gamma:t\mapsto R_\cA\circ a(t)$ is called {\it e-geodesic} in \S3.4 in 
\cite{Petz08}. This curve is called {\it $(+1)$-geodesic} in Section~7.2 in 
\cite{Amari_Nagaoka} while an {\it e-geodesic} is a more general concept 
there, see also Remark~4 in \cite{Weis_Knauf}.
\begin{figure}
\begin{center}
\begin{picture}(4.4,3.5)
\ifthenelse{\equal{\ftype}{eps}}{%
\put(0,0){\includegraphics[height=3.5cm, bb=35 35 450 340, clip=]%
{Pythagoras.eps}}}{%
\put(0,0){\includegraphics[height=3.5cm, bb=35 35 450 340, clip=]%
{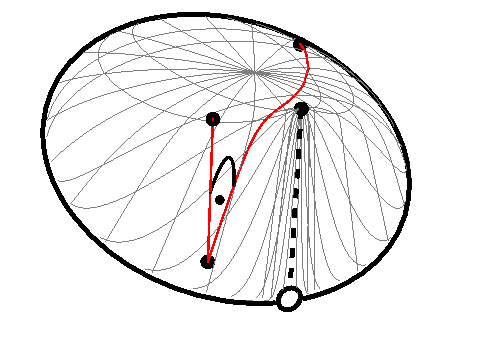}}}
\put(1.7,2.2){$\rho$}
\put(1.6,0.5){$\sigma$}
\put(3.0,3.3){$\tau$}
\put(3.2,2.1){$c$}
\put(2.8,-0.3){$\rho(0)$}
\end{picture}
\end{center}
\caption{\label{fig:Pythagoras}The {\it Staffelberg family} is sketched
by e-geodesics (thin curves). The Euclidean geodesic from $\sigma$
to $\rho$ meets the e-geodesic from $\sigma$ to $\tau$ (red) 
orthogonally with respect to the BKM-metric: The Pythagorean theorem 
$S(\rho,\sigma)+S(\sigma,\tau)=S(\rho,\tau)$ holds.
Closure components in different topologies are indicated (bold).}
\end{figure}
\begin{Exa}[The Staffelberg family]
\label{ex:staffel}
We shall use Pauli $\sigma$-matrices
$\sigma_1:=\left(\begin{smallmatrix}0&1\\1&0\end{smallmatrix}\right)$,
$\sigma_2:=\left(\begin{smallmatrix}0&-i\\i&0\end{smallmatrix}\right)$ and
$\sigma_3:=\left(\begin{smallmatrix}1&0\\0&-1\end{smallmatrix}\right)$. The
{\it Staffelberg family}, studied in \cite{Weis_Knauf}, is the
exponential family 
\[\textstyle
R({\rm span}_\bR(\sigma_1\oplus 0,\,\sigma_2\oplus 1\,))
\;\subset\;
{\rm Mat}(2,\bC)\oplus\bC
\;\cong\;
\left(\begin{smallmatrix}*&*&0\\{}*&*&0\\0&0&*\end{smallmatrix}\right)
\;\subset\;
{\rm Mat}(3,\bC)
\]
embedded into ${\rm Mat}(3,\bC)$ by block diagonal matrices. The Staffelberg 
family is depicted in Figure~\ref{fig:Pythagoras}. The pointed circle about
the family is an equator of the Bloch ball $\cS({\rm Mat}(2,\bC))$, parametrized
for real $\alpha$ by 
$\rho(\alpha):=\tfrac{1}{2}(\id_2+\sin(\alpha)\sigma_1
+\cos(\alpha)\sigma_2)\oplus 0$. The figure also shows
$c:=\tfrac{1}{2}(\rho(0)+0_2\oplus 1)$. Figure~\ref{fig:two_families}
shows the Staffelberg family inside the state space 
$\cS_\cB:=\{\rho\in\cB\mid\rho\succeq 0, {\rm tr}(\rho)=1\}$ of the real 
*-subalgebra $\cB$ spanned by $\sigma_1\oplus 0$, $\sigma_2\oplus 0$, 
${\rm i}\sigma_3\oplus 0$, ${\rm diag}(1,1,0)$ and $\id_3$. The algebra $\cB$
is closed under real scalar multiplication and under the adjoint map. Its
state space is a 3D cone with apex $0_2\oplus 1$ based on the circle
$\{\rho(\alpha)\mid\alpha\in\bR\}$.
\end{Exa}
\par
The Pythagorean theorem (\ref{eq:pythagoras}) applies to exponential 
families. For states $\rho\in\cS$ and $\sigma,\tau\in\cE$, such that 
$\rho-\sigma$ is perpendicular to the translation vector space $U$, 
with respect to the Hilbert-Schmidt scalar product, we have 
\begin{equation}
\label{thm:pythagoras_E}\textstyle
S(\rho,\sigma)+S(\sigma,\tau)=S(\rho,\tau)\,.
\end{equation}
The condition $\rho-\sigma\perp U$ means that the Euclidean straight
line from $\sigma$ to $\rho$ is perpendicular to the exponential 
family $\cE$ with respect to the BKM-Riemannian metric, see 
Remark~\ref{rem:bkm}. This is indicated by the right angle in 
Figure~\ref{fig:Pythagoras}.
\par
The {\it projection theorem}, is now an easy corollary. For every state 
$\rho\in\cE+U^\perp$ the intersection $(\rho+U^\perp)\cap\cE$ contains a 
unique state $\pi_\cE(\rho)$, which defines a projection to $\cE$
\begin{equation}
\label{eq:pro_thm}
\pi_\cE:\;(\cE+U^\perp)\cap\cS\;\to\;\cE\,,
\quad
\rho\;\mapsto\;\pi_\cE(\rho)\,.
\end{equation}
By the choice of $\rho$ the intersection is non-empty. If it contains 
two states $\sigma,\tau$, then $\rho-\sigma\perp U$ and $\rho-\tau\perp U$. 
Equality $\sigma=\tau$ follows if we add the two corresponding Pythagorean 
equations (\ref{thm:pythagoras_E}).
The minimal relative entropy of $\rho$ from $\mathcal E$, called
{\it entropy distance} in \cite{Weis_Knauf}, is
\begin{equation}
\label{def:entropy_dist}\textstyle
{\rm d}_\cE(\rho)
\;:=\;
\inf_{\tau\in\cE}S(\rho,\tau)\,.
\end{equation}
If $\rho\in\cE+U^\perp$ then (\ref{thm:pythagoras_E}) implies the 
projection theorem
\begin{equation}
\label{thm:projection_E}\textstyle
{\rm d}_\cE(\rho)\;=\;S(\rho,\pi_\cE(\rho))\,.
\end{equation}
A geometric optimization formula like (\ref{thm:projection_E}) is called
a projection theorem in \S3.4 in \cite{Amari_Nagaoka}. We will extend 
(\ref{thm:projection_E}) in this article to arbitrary states $\rho$.
%
%
%
%
%
%
\subsection{An algorithm for poonems of the mean value set}
\label{sec:poonem-algorithm}
\par
We use two equivalent descriptions of the convex set of mean values. 
Its boundary components are described algebraically in a lattice $\cP^U$ 
of projections.
\begin{Def}
The {\it mean value set} of a linear subspace $U\subset\sa$ of self-adjoint 
matrices is the orthogonal projection of the state space onto $U$
\begin{equation}
\label{def:m_set}\textstyle
\bM(U)
\;=\;
\bM_\cA(U)
\;:=\;
\pi_U(\cS_\cA)
\;\subset\;U\,.
\end{equation}
Here $\pi_U:\sa\to U$ denotes the orthogonal projection from $\sa$ onto 
$U$. This linear mapping is characterized for each $a\in\sa$ by the 
equation $a-\pi_U(a)\perp U$. For $u_1,\ldots,u_k\in\sa$ we abbreviate
${\bf u}:=(u_1,\ldots,u_k)$ and we define the {\it mean value mapping} by 
\begin{equation}
\label{def:mean_value_map}\textstyle
m_{\bf u}:\;\sa\;\to\;\bR^k\,,\quad a\;\mapsto\;
(\langle u_1,a\rangle,\ldots,\langle u_k,a\rangle)\,.
\end{equation}
The {\it convex support} of ${\bf u}$ is
\begin{equation}\textstyle
\label{def:cs}
{\rm cs}({\bf u})
\;=\;
{\rm cs}_\cA({\bf u})
\;:=\;
\{m_{\bf u}(\rho)\mid\rho\in\cS_\cA\}\;\subset\;\bR^k\,.
\end{equation}
\end{Def}
\par
The concept of convex support was first used by Barndorff-Nielsen 
\cite{Barndorff} in probability theory and later by \v{C}encov
\cite{Cencov}. It was refined by Csisz\'ar and Mat\'u\v s 
\cite{Csiszar03,Csiszar05} to investigate mean values of exponential 
families. Barndorff-Nielsen's definition for a finite measurable space is 
equivalent to (\ref{def:cs}) if probability measures are embedded into an 
algebra of diagonal matrices like in (\ref{def:commutative_inclusion}).
\par
The convex support introduces coordinates on the mean value set $\bM(U)$.
If $U:={\rm span}_\bR(u_1,\ldots,u_k)$, then the convex bodies 
$\bM(U)\cong{\rm cs}({\bf u})$ are ``affinely isomorphic'' 
(see Remark 1.1.1 in \cite{Weis_supp}): The mean value mapping restricts 
to the bijection
\begin{equation}
\label{eq:mean_iso}\textstyle
m_{\bf u}|_{\bM(U)}:\;\bM(U)\;\to\;{\rm cs}({\bf u})
\end{equation}
such that $m_{\bf u}\circ\pi_U=m_{\bf u}$. In the majority of all proofs 
we are going to use the Hilbert-Schmidt Euclidean geometry in $\sa$, using 
orthogonal projection $\pi_U$ rather than coordinates.
\begin{figure}
\centerline{\ifthenelse{\equal{\ftype}{eps}}{%
\includegraphics[height=3.5cm, bb=0 0 415 360, clip=]%
{faces.eps}}{%
\includegraphics[height=3.5cm, bb=0 0 415 360, clip=]%
{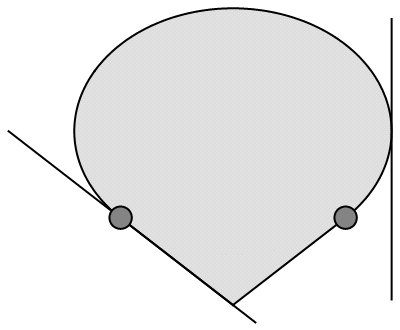}}}
\caption{\label{fig:faces}This clove shape is the mean value set of 
the {\it Swallow family}. It is the convex hull of an ellipse and a
point (Example~\ref{exa:swallow}). The supporting hyperplane to the left 
resp.\ right defines an exposed face which is a segment 
resp.\ a point. Two non-exposed faces (points) of the clove are 
indicated by small disks.}
\end{figure}
\par
The simplest boundary component of a convex set $C$ 
in Euclidean space is an {\it exposed face} of $C$, the set of 
maximizers in $C$ of a linear functional. The empty set is
an exposed face by definition.
Except for $\emptyset$ and $C$, every exposed face of $C$ is the
intersection of $C$ and a {\it supporting hyperplane} $H$, i.e.\ 
an affine subspace $H$ of codimension one which intersects $C$ 
such that $C\setminus H$ is convex. Figure~\ref{fig:faces} shows
examples.
\newsavebox{\poonem}
\sbox{\poonem}{\begin{picture}(0.45,0.45)\put(0.25,0.15){\circle*{0.28}}\end{picture}}
\begin{figure}
\begin{center}
\begin{picture}(10,2)
\ifthenelse{\equal{\ftype}{eps}}{%
\put(0,0){\includegraphics[height=2cm, bb=10 20 1000 180, clip=]%
{poonem.eps}}}{%
\put(0,0){\includegraphics[height=2cm, bb=10 20 1000 180, clip=]%
{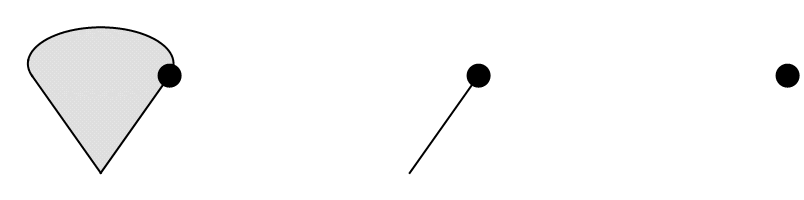}}}
\put(3.5,1){\huge$\supset$}
\put(7.5,1){\huge$\supset$}
\end{picture}
\end{center}
\caption{\label{fig:poonems}
Repeated inclusions of exposed faces define a poonem. Here a poonem of
point form is visualized by the disk \usebox{\poonem}.}
\end{figure}
\par
A {\it poonem} \cite{Gruenbaum} of $C$ is a member of a sequence 
$F_1\subset\cdots\subset F_k=C$, s.th.\ $F_i$ is an exposed face of 
$F_{i+1}$ for $i=1,\ldots,k-1$. A {\it non-exposed} face of $C$ is 
a poonem of $C$, which is not an exposed face of $C$. 
Figure~\ref{fig:poonems} shows an example of a poonem, which is
a non-exposed face. The notion of {\it poonem} is equivalent to the 
notion of {\it face} (see e.g.\ \S1.2.1 in \cite{Weis_touch} for a 
proof) which we will use in \S\ref{ingredients:convex}. The set of
exposed faces and the set of poonems are partially ordered by 
inclusion, they are lattices.
\par
The first step to the extension of $\cE$ is a lifting construction for
poonems of the mean value set $\bM(U)$. For every poonem $P$ of $\bM(U)$ 
the inverse image under projection $\{\rho\in\cS\mid \pi_U(\rho)\in P\}$ 
is an exposed face of the state space $\cS$. For every exposed face $F$ 
of the state space $\cS$ there exists a unique projection $p\in\cP_\cA$ 
in the projection lattice (\ref{def:proj}) of $\cA$, such that
\[\textstyle
F
\;=\;
\cS_{p\cA p}
\;=\;
\{\rho\in p\cA p\mid \rho\succeq 0, {\rm tr}(\rho)=1\}\,.
\]
The C*-subalgebra $p\cA p=\{pap\mid a\in\cA\}$ is called
{\it compressed algebra}. Let us collect in the {\it projection lattice} 
\begin{equation}
\label{eq:projection-lattice}\textstyle
\cP^U
\end{equation}
all the projections arising in this construction from poonems of $\bM(U)$. The 
projection lattice $\cP^U$ ordered by $\preceq$ is isomorphic to the lattice of 
poonems of $\bM(U)$ ordered by inclusion. An algorithm to compute $\cP^U$ is 
described in Remark~\ref{rem:pu}. This is the algebraic reformulation of the 
concept of poonem for the special case of a mean value set.
%
%
%
%
%
%
%
\subsection{The extension of an exponential family}
\label{sec:extension}
\par
We introduce an extension to the exponential family defined in (\ref{def:R}) 
in terms of a non-empty affine subspace $\Theta\subset\sa$ of self-adjoint 
matrices, 
\[\textstyle
\cE\;=\;R_\cA(\Theta)
\;=\;\{\exp_\cA(\theta)/{\rm tr}\exp_\cA(\theta)\mid\theta\in\Theta\}\,.
\]
The translation vector space of $\Theta$ is $U:=\Theta-\Theta$. 
To each poonem of the mean value set $\bM(U)$ corresponds a projection
$p\in\cP^U$, we associate an exponential family to $p$ and take the
union over all $p\in\cP^U$.
\begin{figure}
\begin{center}
\begin{picture}(9,3.5)
\ifthenelse{\equal{\ftype}{eps}}{%
\put(0,0){\includegraphics[height=3.5cm, bb=70 30 500 480, clip=]%
{Cone_and_cs_and_60.eps}}
\put(4.5,0){\includegraphics[height=3.5cm, bb=50 50 500 430, clip=]%
{Cone_and_cs_and_51.eps}}}{%
\put(0,0){\includegraphics[height=3.5cm, bb=70 30 500 480, clip=]%
{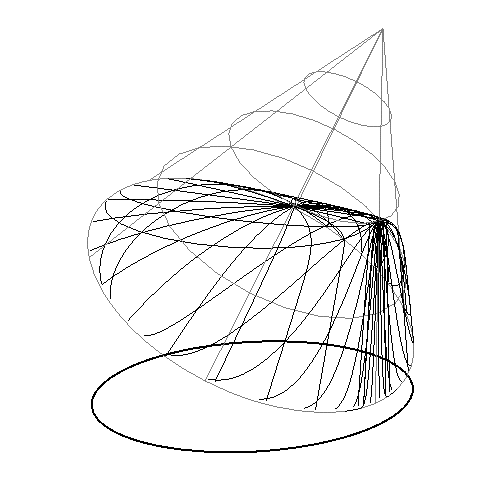}}
\put(4.5,0){\includegraphics[height=3.5cm, bb=50 50 500 430, clip=]%
{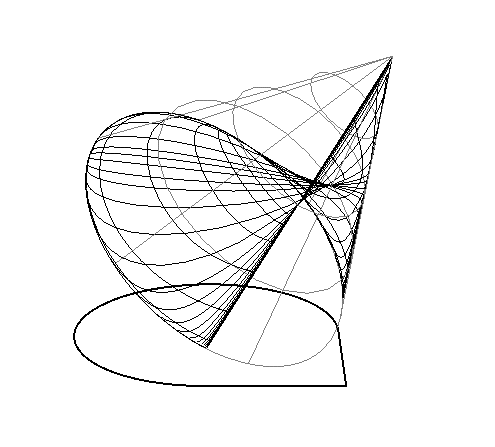}}}
\put(2.45,0.35){$\rho(0)$}
\put(2.65,1.75){$c$}
\put(2.50,3.2){$0_2\oplus 1$}
\put(5.9,0.25){$\rho(0)$}
\put(7.2,0.75){$\rho(\tfrac{\pi}{2})$}
\put(7.7,3.0){$0_2\oplus 1$}
\end{picture}
\caption{\label{fig:two_families}The {\it Staffelberg family} (left) 
and {\it Swallow family} (right) are sketched by e-geodesics inside the
conic state space (real algebra). The boundary of the mean 
value set (projection of the state space) is drawn underneath each family.}
\end{center}
\end{figure}
\begin{Exa}\label{exa:swallow}
The {\it Swallow family}, studied in \cite{Weis_Knauf}, is the
exponential family
\[\textstyle
R({\rm span}_\bR(\sigma_1\oplus 1,\,\sigma_2\oplus 1\,))\,.
\]
Figure~\ref{fig:two_families} shows two exponential families $\cE$
inside the conic state space $\cS_\cB$ of the real algebra $\cB$ from 
Example~\ref{ex:staffel}. The figure also shows the 
mean value sets $\bM(V)$, translated into the drawing frame, with 
respect to the {\it canonical tangent space} \cite{Weis_Knauf}
\[\textstyle
V
\;:=\;
\{\log(\rho)-{\rm tr}(\log(\rho))\id_3/3\mid\rho\in\cE\}\,.
\]
The projection of the 3D cone $\cS_\cB$ onto $V$ equals 
the mean value set $\bM_\cA(V)$ for algebras $\cA={\rm Mat}(3,\bC)$ 
or $\cA={\rm Mat}(2,\bC)\oplus\bC$. This follows from 
Theorem~\ref{thm:mean_value_chart} as	 $\mathcal E$ is included 
in the state space $\cS_\cB$. This also follows from simpler arguments
about state spaces, see Lemma 3.13 in \cite{Weis_supp}. 
So $\bM(V)=\pi_V(\cS_\cB)$ is the convex hull of an ellipse and a point.
\end{Exa}
\par
We now consider an affine space 
$p\Theta p=\{p\theta p\mid \theta\in\Theta\}$ for each $p\in\cP^U$ and we 
use it to define an exponential family in the compressed algebra $p\cA  p$ 
by
\begin{equation}
\label{def:ep}\textstyle
\cE_p
\;:=\;
R_{p\cA p}(p\Theta p)
\;=\;
\{p\exp(p\theta p)/{\rm tr}(p\exp(p\theta p))\mid\theta\in\Theta\}
\end{equation}
using the normalized exponential $R_{p\cA p}$ in (\ref{def:R}).
Noticing $\id\in\cP^U$ and $\cE_\id=\cE$, the disjoint union
\begin{equation}
\label{def:ext}\textstyle
{\rm ext}(\cE)
\;:=\;
\bigcup_{p\in\cP^U\setminus\{0\}}\cE_p
\end{equation}
contains $\cE$ and we will prove in Lemma~\ref{lem:ext-m}
a bijection between this extension and the mean value set
\begin{equation}
\label{def:bijection:extpi}\textstyle
\pi_U|_{{\rm ext}(\cE)}:\;{\rm ext}(\cE)\;\to\;\bM(U)\,.
\end{equation}
This allows us to define a projection
\begin{equation}
\label{def:projection:extpi}\textstyle
\pi_\cE:\;\cS\;\to\;{\rm ext}(\cE)\,,
\end{equation}
such that $\pi_U(\rho)=\pi_U\circ\pi_\cE(\rho)$ holds for all states $\rho$.
\par
The bijection (\ref{def:bijection:extpi}) is based on the lattice analysis 
outlined above and on the mean value chart of exponential families, proved 
in \cite{Wichmann} and generalized in \S\ref{sec:charts}.
\begin{Exa}[Extension]
\label{exa:extension}
The Staffelberg family, shown in Figure~\ref{fig:Pythagoras}, 
is extended by the pointed circle $\{\rho(\alpha)\mid \alpha\in(0,2\pi)\}$ of 
one-point exponential families, the states $\rho(\alpha)$ being defined in 
Example~\ref{ex:staffel}, and by
$c=\tfrac{1}{2}(0_2\oplus 1+\rho(0))$. Their union is in one-to-one 
correspondence with the elliptical boundary of the mean value set. 
Figure~\ref{fig:two_families} shows the Staffelberg family with its mean 
value set. It also shows the Swallow family, where the extension contains 
two one-dimensional exponential families, the open segments between 
$0_2\oplus 1$ and $\rho(0)$ resp.\ $\rho(\tfrac{\pi}{2})$.
The extension of the Swallow family is the norm closure. The extension of the 
Staffelberg family is strictly smaller than the norm closure. 
\par
The projection lattice $\cP^V$ of these exponential families $\cE$ is
computed in \S3.3 in \cite{Weis_supp}. The rI-closure
${\rm cl}^{\rm rI}(\cE)$ is computed in \S{}IV.B and \S{}IV.D in 
\cite{Weis_Knauf}. By Theorem~\ref{thm:ext_rI} the rI-closure
${\rm cl}^{\rm rI}(\cE)$ is the extension ${\rm ext}(\cE)$ which is
computed directly from the projection lattice $\cP^V$ in \cite{Weis_Knauf}.
\end{Exa}
%
%
%
%
%
%
%
\subsection{The Complete Pythagorean theorem}
\label{sec:informal-pythagorean}
\par
We will extend in Theorem~\ref{thm:pyth_with_proof} the Pythagorean theorem 
(\ref{thm:pythagoras_E}) to the full state space using the extension 
previously defined. A corollary is the maximization of the von Neumann entropy 
under linear constraints, a fundamental problem in quantum statistical mechanics, 
see e.g.\ \cite{Ingarden,Ruelle,Petz08,Amari_Nagaoka}. 
\begin{Thm*}[Complete Pythagorean theorem]
For any $\rho\in\cS$ and $\sigma,\tau\in{\rm ext}(\cE)$ such that
$\rho-\sigma\perp U$ we have
$S(\rho,\sigma)+S(\sigma,\tau)=S(\rho,\tau)$.
\end{Thm*} 
\begin{Def}
The {\it von Neumann entropy} of $\rho\in\cS$ is defined by
\begin{equation}
\label{def:von_Neumann}\textstyle
S(\rho)\;:=\;-\tr \rho\log(\rho)\,,
\end{equation}
using functional calculus, see Remark~\ref{rem:func_calc}.3. 
The {\it free energy} of $\theta\in\sa$ is 
\begin{equation}
\label{def:free_energy}\textstyle
F(\theta)
\;=\;
F_\cA(\theta)
\;:=\;
\log\tr\exp_\cA(\theta)\,,
\end{equation}
where $\exp_\cA$ is defined by
functional calculus in $\cA$, see (\ref{def:R}).
\end{Def}
\par
According to Jaynes \cite{Jaynes} the state which maximizes the von Neumann 
entropy under arbitrary constraints is the least biased choice of a state compatible
with the constraints. Exponential families with linear canonical parameter space 
$\Theta=U$ are called {\it Gibbsian families} \cite{Petz08}, they maximize 
the von Neumann entropy under linear constraints. We can show this for their
extension, too. Let $u_1,\ldots,u_k\in\sa$, put 
$\Theta:=U:={\rm span}_\bR(u_1,\ldots,u_k)$, $\cE:=R_\cA(U)$ and denote
${\bf u}=(u_1,\ldots,u_k)$. We consider the mean value map $m_{\bf u}$ and the 
convex support ${\rm cs}({\bf u})=m_{\bf u}(\cS)$, defined respectively in 
(\ref{def:mean_value_map}) and (\ref{def:cs}).
\begin{Cor}
\label{cor:maxent}
For all $x\in{\rm cs}({\bf u})$ there exists a unique state
$\sigma\in{\rm ext}(\cE)$ such that
$\sigma={\rm argmax}\{S(\rho)\mid\rho\in\cS, m_{\bf u}(\rho)=x\}$.
\end{Cor}
{\em Proof:\/}
Using the bijections (\ref{def:bijection:extpi}) and (\ref{eq:mean_iso})
there exists a unique state $\sigma\in{\rm ext}(\cE)$ such that
$m_{\bf u}(\sigma)=x$. Let $\rho\in\cS$ such that
$m_{\bf u}(\rho)=m_{\bf u}(\sigma)=x$. Then by (\ref{eq:mean_iso}) we have
$\rho-\sigma\perp U$. 
Using $S(\rho)=\log({\rm tr}\,\id)-S(\rho,\tfrac{1}{{\rm tr}\,\id}\id)$,
the Complete Pythagorean theorem, applied to the states $\rho,\sigma$ and
$\tau:=\tfrac{1}{{\rm tr}\,\id}\id\in\cE$, shows 
\[\textstyle
S(\sigma)-S(\rho)
\;=\;
S(\rho,\tau)-S(\sigma,\tau)
\;=\;
S(\rho,\sigma)\,.
\]
The claim follows from the distance-like properties of the relative entropy, 
see \S\ref{sec:entropy_properties}. 
\hspace*{\fill}$\Box$\\
\par
There exist coordinates $\beta_1,\ldots,\beta_k$ for a Gibbsian family $\cE$ 
analogous to {\it inverse temperatures} in statistical mechanics 
\cite{Ingarden}---this explains the sign convention of $-\beta$ (not $+\beta$)
below. The extension ${\rm ext}(\cE)$ has a second parameter specifying the 
support projection. We consider the projection lattice $\cP^U$ and the free
energy $F$, defined respectively in (\ref{eq:projection-lattice}) and 
(\ref{def:free_energy}).
\begin{Cor}
\label{cor:coordianates-on-maxent}
For every mean value tuple $x=(\xi_1,\dots,\xi_k)\in{\rm cs}_\cA({\bf u})$ 
there is a unique maximizer $\rho(x)$ of the von Neumann entropy among all 
states  $\rho\in\cS_\cA$ with mean values $m_{\bf u}(\rho)=x$. There exists 
a unique projection $p\in\cP^U\setminus\{0\}$ and there exist
(generally not unique) numbers $\beta_1,\ldots,\beta_k\in\bR$, such that
\[\textstyle
\tfrac{\partial}{\partial\beta_j}\,
F_{p\cA p}(-\sum_{i=1}^k\beta_ipu_ip)
\;=\;
-\xi_j\,,
\qquad j=1,\ldots,k\,.
\]
For each solution $p,\beta_1,\ldots,\beta_k$ we have 
\[\textstyle
\rho(x)
\;=\;
R_{p\cA p}(-\sum_{i=1}^k\beta_ipu_ip)
\]
and $\rho(x)$ has the von Neumann entropy
$S(\rho(x))=\sum_{i=1}^k\beta_i\xi_i+F_{p\cA p}(-\sum_{i=1}^k\beta_ipu_ip)$.
\end{Cor}
{\em Proof:\/}
Corollary~\ref{cor:maxent} shows that the extension ${\rm ext}(\cE)$ 
is the set of unique maximizers of the von Neumann entropy under the 
linear constraints. Corollary~\ref{cor:rI-coords} provides the described 
coordinates. It remains to compute the von Neumann entropy of 
$\rho:=\rho(x)=R_{p\cA p}(p(-\sum_{i=1}^k\beta_iu_i)p)$. By definition of 
$R_{p\cA p}$ in (\ref{def:R}) we have
\begin{align*}
S(\rho) & \; =\; -{\rm tr}\rho\log(\rho)\\
&\;=\;
{\textstyle -\tr\rho[-\sum_{i=1}^k\beta_ipu_ip
-F_{p\cA p}(-\sum_{i=1}^k\beta_ipu_ip)]}\\
&\;=\; 
{\textstyle \sum_{i=1}^k\beta_i\langle pu_ip,\rho\rangle
+
F_{p\cA p}(-\sum_{i=1}^k\beta_ipu_ip)}\,.
\end{align*}
We have $\langle pu_ip,\rho\rangle=\langle u_i,\rho\rangle=\xi_i$ for
$i=1,\ldots,k$ because $\rho$ has support $s(\rho)=p$.
\hspace*{\fill}$\Box$\\
\par
In the paragraph following Corollary~\ref{cor:rI-coords} we comment on
the uniqueness of parameters $\beta_1,\ldots,\beta_k$.
The special case of Corollary~\ref{cor:coordianates-on-maxent} for
states of full support $s(\rho)=\id$ in $\cA$ was known already in 1963. 
To explain it, let ${\rm ri}(C)$ denote the {\it relative interior} of a 
convex subset $C$ in Euclidean space, i.e.\ ${\rm ri}(C)$ is the interior 
of $C$ in the norm topology of the affine hull of $C$. Wichmann 
\cite{Wichmann} has proved the bijection
\begin{equation}
\label{eq:wichmanns-bijection}\textstyle
\pi_U|_\cE:\;\cE\;\to\;{\rm ri}\,\bM(U)\,.
\end{equation}
It implies the Pythagorean theorem (\ref{thm:pythagoras_E}) and the projection 
theorem (\ref{thm:projection_E}) for all states $\rho\in U+{\rm ri}\,\bM(U)$, 
in particular for all invertible states $\rho$. This holds since 
$\pi_U({\rm ri}\,\cS)={\rm ri}\,\bM(U)$ and because the relative interior 
${\rm ri}\,\cS$ consists of the invertible states, see e.g.\ 
Proposition~\ref{ss:st}. The map
\begin{equation}
\label{eq:inference}\textstyle
\Phi:\;{\rm cs}({\bf u})\;\to\;\cS\,,
\quad
x\;\mapsto\;{\rm argmax}\{S(\rho)\mid\rho\in\cS, m_{\bf u}(\rho)=x\}
\end{equation}
parametrizes ${\rm ext}(\cE)$ by mean values. Wichmann has shown the following:
\begin{enumerate}
\item[$\bullet$] $\Phi({\rm ri}\,{\rm cs}({\bf u}))=\cE$,
\item[$\bullet$] $\Phi|_{{\rm ri}\,{\rm cs}({\bf u})}:$
${\rm ri}\,{\rm cs}({\bf u})\to\cE$ is real analytic,
\item[$\bullet$] $\Phi({\rm cs}({\bf u}))\subset\overline{\cE}$, where
$\overline{\cE}$ denotes the closure of $\cE$ in the norm topology.
\end{enumerate} 
\par
Our result of $\Phi({\rm cs}({\bf u}))={\rm ext}(\cE)$ in 
Corollary~\ref{cor:coordianates-on-maxent} resolves 
$\Phi({\rm cs}({\bf u}))\subset\overline{\cE}$ to an equality. 
However we do not know how tight the upper bound 
$\Phi({\rm cs}({\bf u}))\subset\overline{\cE}$ is. This question is related 
to the continuity of $\Phi$. A discontinuous $\Phi$ is presented in 
Example~\ref{exa:staffelberg}, see
also Proposition~\ref{pro:continuity_with_proof}.
%
%
%
%
%
%
%
\subsection{The Complete projection theorem}
\label{sec:select-com-pro}
\par
The following theorem and its corollary hold for our standard assumptions
of an affine space $\Theta$ and $\cE=R_\cA(\Theta)$. We extend in 
Theorem~\ref{thm:ext_rI} the projection theorem (\ref{thm:projection_E}) to the 
full state space using the extension ${\rm ext}(\cE)$ previously defined. As 
a corollary we write ${\rm ext}(\cE)$ as the topological closure in the 
rI-topology. 
\par
For technical reasons we write $S_\rho(\sigma):=S(\rho,\sigma)$ for 
the relative entropy with fixed state $\rho$ and variable state $\sigma$. 
We use the entropy distance 
\[\textstyle
{\rm d}_\cE(\rho)
\;=\;
\inf_{\sigma\in\cE}S_\rho(\sigma)
\]
and the projection $\pi_\cE$ defined in (\ref{def:entropy_dist})
and (\ref{def:projection:extpi}) respectively.
\begin{Thm*}[Complete projection theorem]
For each $\rho\in\cS$ the relative entropy $S_\rho$ has 
on the extension ${\rm ext}(\cE)$ a unique local minimizer at 
$\pi_\cE(\rho)$. We have
\[\textstyle 
{\rm d}_\cE(\rho)
\;=\;
\min_{\sigma\in{\rm ext}(\cE)}S_\rho(\sigma)
\;=\;
S_\rho(\pi_\cE(\rho))\,.
\]
\end{Thm*}
\par
The Complete projection theorem shows 
${\rm ext}(\cE)=\{\rho\in\cS\mid{\rm d}_\cE(\rho)=0\}$, where the 
right-hand side is the rI-closure ${\rm cl}^{\rm rI}(\cE)$ of $\cE$ 
defined in (\ref{def:rI-closure}). Theorem~\ref{thm:info_top}.2 shows 
that ${\rm cl}^{\rm rI}(\cE)$ is the topological closure of $\cE$
with respect to the rI-topology. This topology can be defined by the
base $\{V^{\rm rI}(\rho,\epsilon)\mid\epsilon\in(0,\infty]\}$ of
open rI-disks $V^{\rm rI}(\rho,\epsilon)
=\{\sigma\in\cS\mid S(\rho,\sigma)<\epsilon\}$. To sum up:
\begin{Cor}
We have ${\rm ext}(\cE)={\rm cl}^{\rm rI}(\cE)$.
\end{Cor}
\par 
We have seen in (\ref{eq:top-inclusion}) that the rI-topology is finer 
than the norm topology. So ${\rm cl}^{\rm rI}(\cE)\subset\overline{\cE}$
is obvious. A proper inclusion is possible.
\begin{Exa}
\label{exa:staffelberg}
The {\it Staffelberg family} $\cE$ from Example~\ref{ex:staffel}
satisfies $\cl^{\rm rI}(\mathcal E)\subsetneq\overline{\mathcal E}$. This
exponential family is depicted in Figure~\ref{fig:Pythagoras}:
The norm closure $\overline{\mathcal E}$ is the union of $\mathcal E$
with the bold circle around $\mathcal E$ and the dashed upright segment
from $\rho(0)$ to $0_2\oplus 1$. The upright segment is missing in the 
rI-closure $\cl^{\rm rI}(\mathcal E)$ except for its top endpoint $c$. 
See \S{}IV.B in \cite{Weis_Knauf} for this analysis.
\end{Exa}
\par
The Staffelberg is a Gibbsian family (with 
$\tfrac{1}{{\rm tr}\id}\id\in\cE$). The rI-closure ${\rm cl}^{\rm rI}(\cE)$ 
is a set of maximizers of the von Neumann entropy under linear constraints.
The norm closure $\overline{\cE}$ is too large for this aim.
%
%
%
%
%
%
%
\subsection{Why poonems are essential}
\label{sec:why-poonems}
\par
Poonems are essential for the Complete projection theorem. Strictly speaking, 
the bijection in Lemma~\ref{lem:ext-m}
\[\textstyle
\pi_U|_{{\rm ext}(\cE)}:\;{\rm ext}(\cE)\;\to\;\bM(U)
\]
as well as the Complete Pythagorean theorem can be proved without any need 
for poonems---using the equivalent notion of face.
\par
The first reason to use poonems is that the algebraic algorithm in 
Remark~\ref{rem:pu} to compute the projection lattice $\cP^U$ proceeds along 
chains of faces of $\bM(U)$ 
\begin{equation}
\label{eq:chain}\textstyle
\bM(U)
\;=\;
F_0
\;\supsetneq\;
F_1
\;\supsetneq\;
\cdots
\;\supsetneq\;
F_m
\end{equation}
such that $F_{i+1}$ is an exposed face of $F_i$ for $i=0,\ldots,m-1$.
Elements of such chains are poonems by definition. Though the projection 
lattice $\cP^U$ is essential for defining the extension ${\rm ext}(\cE)$ 
in (\ref{def:ext}) we are not forced to use the algorithm, since 
$\cP^U$ is defined in (\ref{eq:main_iso}) in terms of lattice isomorphisms.
\par
Chains (\ref{eq:chain}) and the algorithm are essential in the 
Complete projection theorem. Certainly, if $\rho$ is a state, $p\in\cP^U$ 
and $\pi_U(\rho)\in\pi_U(\cE_p)$ then we can compute the minimum
\[\textstyle
{\rm d}_{\cE_p}(\rho)
\;=\;
\inf\{S_\rho(\sigma)\mid\sigma\in\cE_p\}\,.
\]
This will be done in Proposition~\ref{pro:intriguing} in the paragraph 
following (\ref{eq:rho_exposed}). But it is not clear how the value
${\rm d}_{\cE_p}(\rho)$ is related to ${\rm d}_{\cE_q}(\rho)$ for
$q\succeq p$, $q\in\cP^U$. The equality 
${\rm d}_{\cE_p}(\rho)={\rm d}_{\cE_q}(\rho)$ will be proved by approximation 
of $\cE_p$ from within $\cE_q$, using e-geodesics. E-geodesics guarantee a 
controlled limit of the (discontinuous) relative entropy 
(Lemma~\ref{lemma:infimum_at_infty}). It is not possible to exhaust the 
extension ${\rm ext}(\cE)$ in one step of such an approximation.
In fact, we show in Proposition~\ref{pro:geodesics} that the union $X$ 
of an exponential family $\cE$ with the two limit points of all in $\cE$ 
included e-geodesics covers only part of the mean value set $\bM(U)$ 
under the projection $\pi_U$. Relative interiors of non-exposed faces 
of $\bM(U)$ are missing. 
\par
The Swallow family $\cE=R_{\cA}(V)$ in 
Example~\ref{exa:swallow} makes this clear. Two non-exposed faces of 
the mean value set $\bM(V)$ are depicted in Figure~\ref{fig:faces}. 
As depicted in Figure~\ref{fig:two_families}, each of these non-exposed 
faces has a unique inverse projection under $\pi_U$, which is 
$\rho(\alpha)$ for $\alpha\in\{0,\tfrac{\pi}{2}\}$. So there exists no 
e-geodesic in $\cE$ with limit $\rho(\alpha)$ although $\rho(\alpha)$
belongs to the extension ${\rm ext}(\cE)$ and {\it a fortiori} to the 
norm closure $\overline{\cE}$. On the other hand, the open segment 
between $0_2\oplus 1$ and $\rho(\alpha)$ is included in $X$ and it is 
an exponential family $\cE_p$, corresponding to the rank-two projection 
$p=\rho(\alpha)+0_2\oplus 1\in\cP^V$. Now $\rho(\alpha)$ can be approximated
in a second step by the e-geodesic $\cE_p$. The corresponding
faces of the mean value set $\bM(V)$ are
\[\textstyle
\pi_V(\cS)=\bM(V)
\quad\supsetneq\quad
\pi_V(\text{segment between }0_2\oplus 1 \text{ and } \rho(\alpha))
\quad\supsetneq\quad
\pi_V(\rho(\alpha))\,.
\]
They are depicted schematically in Figure~\ref{fig:poonems} and they form a 
chain (\ref{eq:chain}).
\par
For details about missing e-geodesic limits at non-exposed faces we refer 
to the proof of Proposition~\ref{pro:geodesics}. For explicit calculations 
on the Swallow family we refer to Remark 29~a) in \cite{Weis_Knauf}.
%
%
%
%
%
\section{Analysis on the state space}
\label{sec:matrix_analysis}
\par
This a preparatory section summarizing techniques from convex geometry, 
geometry of state spaces and mean value sets including their algebraic 
formulation. Detailed perturbation theoretical proofs are provided in 
\S\ref{sec:perturb}, where the standard theory is not sufficient.
\par
Let us introduce important notations. The use of spectral values may seem 
unnecessary. We have decided to work with the extension 
${\rm ext}(\cE)=\bigcup_{p\in\cP^U\setminus\{0\}}\cE_p$ of an exponential 
family (\ref{def:ext}) as a single object in the Euclidean space $\sa$. 
Thus we need to distinguish between invertibility in several compressed 
algebras $p\cA p$.
\begin{Def}[Spectral values, norms and Euclidean vector spaces]~
\label{def:represent}
\begin{enumerate}[1.]
\item
Let $a\in\cA$. The {\it spectrum} of $a$ is  
\[\textstyle
{\rm spec}_\cA(a):=\{\lambda\in\bC\mid a-\lambda\id
\text{ is not invertible in } \cA\},
\]
its elements are the {\it spectral values} of $a$ in $\cA$. The matrix
$a$ is {\it positive semi-definite} if $a\in\sa$ and if $a$ has no 
negative spectral values, we then write $a\succeq 0$. If $a\succeq 0$, 
then there exists $b\in\cA$, $b\succeq 0$ with $a=b^2$, see e.g.\
\S2.2 in \cite{Murphy}. The matrix $b$ is unique and one defines 
$\sqrt{a}:=b$. We have $a^*a\succeq 0$ and put $|a|:=\sqrt{a^*a}$.
\item
In addition to the two-norm $\|\cdot\|_2$ introduced in \S\ref{intro:hist} 
we consider the {\it spectral norm} $\|a\|$, which is the square root of 
the largest eigenvalue of $a^*a$ and the {\it trace norm} $\|a\|_1:=\tr|a|$. 
The topology of any norm is the {\it norm topology} and convergence of a 
sequence $(a_i)_{i\in\bN}\subset\cA$ to $a\in\cA$ in any norm will be 
denoted by $\lim_{i\to\infty}a_i=a$. The three norms restrict to the 
real vector space of self-adjoint matrices $\sa$.
\item
In any Euclidean vector space $(\bE,\langle\cdot,\cdot\rangle)$ we denote 
the {\it two-norm} by $\|x\|_2:=\sqrt{\langle x,x\rangle}$ and write
$x\perp y:\iff\langle x,y\rangle=0$ for $x,y\in\bE$. For subsets 
$X,Y\subset\bE$ we write
$X\perp Y:\iff\langle x,y\rangle=0\,\forall x\in X,y\in Y$ and 
$X^\perp:=\{y\in\bE\mid x\perp y\,\forall x\in X\}$ (if $z\in\bE$ then 
$z\perp Y:\iff \{z\}\perp Y$ and $z^\perp:=\{z\}^\perp$). If $\bA$ is 
a non-empty affine subspace of $\bE$, then we denote the translation vector 
space of $\bA$ by
\[\textstyle
\lin(\bA)
\;:=\;
\bA-\bA
\;=\;
\{a-b\mid a,b\in\bA\}\,.
\]
We denote the orthogonal projection from $\bE$ onto $\bA$ by 
$\pi_\bA:\bE\to\bA$. This affine mapping is characterized for each
$x\in\bE$ by the equation $x-\pi_\bA(x)\perp\lin(\bA)$. 
\end{enumerate}
\end{Def}
%
%
%
%
%
%
%
\subsection{Lattices of faces and projections}
\label{ingredients:convex}
\par
In this section we settle the convex geometry and the algebraic description 
of the mean value set $\bM(U)=\pi_U(\cS)$, referring to 
\cite{Weis_touch,Weis_supp}. The mean value set (\ref{def:m_set}) is the 
orthogonal projection of the state space $\cS$ onto a linear subspace 
$U\subset\sa$. Notice that \cite{Weis_supp} erroneously uses eigenvalues 
and not spectral values. The corrected statements are cited below from the 
copy on the arXiv.
\par
A map $f:X\to Y$ between two partially ordered sets 
$(X,\leq)$ and $(Y,\leq)$ is {\it isotone} if for all $x,y\in X$ such that 
$x\leq y$ we have $f(x)\leq f(y)$. A lattice is a partially ordered set 
$(\cL,\leq)$ where the {\it infimum} $x\wedge y$ and {\it supremum}
$x\vee y$ of each two elements $x,y\in\cL$ exist. A {\it lattice 
isomorphism} is a bijection between two lattices that preserves the 
lattice structure. A lattice $\cL$ is {\it complete} if for an arbitrary 
subset $S\subset\cL$ the infimum $\bigwedge S$ and the supremum
$\bigvee S$ exist. The {\it least element} $\bigwedge\cL$ and the
{\it greatest element} $\bigvee\cL$ in a complete lattice $\cL$ are
{\it improper} elements of $\cL$, all other elements of $\cL$ are
{\it proper} elements. An {\it atom} of a complete lattice $\cL$ is an 
element $x\in\cL$, $x\neq\bigwedge\cL$, such that $y\leq x$ and $y\neq x$ 
implies $y=\bigwedge\cL$ for all $y\in\cL$.

\par
The projection lattice $\cP=\{p\in\cA\mid p^2=p^*=p\}$, defined in 
(\ref{def:proj}), with the partial ordering $\preceq$ is a 
complete lattice. For this and the following two statements see e.g.\
\cite{Alfsen} or Remark 2.6 in \cite{Weis_supp}. For a self-adjoint 
(also normal) matrix $a\in\sa$ we have
\begin{equation}
\label{eq:sa_dom}\textstyle
a\;=\;pap
\quad\iff\quad
pa\;=\;a
\quad\iff\quad
s(a)\;\preceq\;p\,.
\end{equation}
Hence the partial ordering for projections $p,q\in\cP$ simplifies to 
$p\preceq q\iff pq=p$.

\par
We need two distinct notions of ``\,face\,'' of a convex set, each
defining a lattice of subsets ordered by inclusion. We begin with a
general convex set.
\begin{Def}
\label{def:face-lattices}
Let $(\bE,\langle\cdot,\cdot\rangle)$ be a finite-dimensional Euclidean 
vector space.
\begin{enumerate}[1.]
\item
The {\it closed segment} between $x,y\in\bE$ is 
$[x,y]:=\{(1-\lambda)x+\lambda y\mid\lambda\in[0,1]\}$, the 
{\it open segment} is 
$]x,y[\,:=\{(1-\lambda)x+\lambda y\mid\lambda\in(0,1)\}$. A subset 
$C\subset\bE$ is {\it convex} if $x,y\in C$ $\implies$ $[x,y]\subset C$.  
\item 
Let $C$ be a convex subset of $\bE$. A {\it face} of $C$ is a convex 
subset $F$ of $C$, such that whenever for $x,y\in C$ the open segment 
$]x,y[$ intersects $F$, then the closed segment $[x,y]$ is included in 
$F$. If $x\in C$ and $\{x\}$ is a face, then $x$ is called an
{\it extreme point}. The set of faces of $C$ will be denoted by 
$\cF(C)$, called the {\it face lattice} of $C$.
\item
The {\it support function} of a convex subset $C\subset\bE$ is defined
by $\bE\to\bR\cup\{\pm\infty\}$,
$u\mapsto h(C,u):=\sup_{x\in C}\langle u,x\rangle$. For non-zero
$u\in\bE$ the set
\[
H(C,u)\;:=\;\{x\in\bE:\langle u,x\rangle=h(C,u)\}
\]
is an affine hyperplane unless it is empty, which can happen if 
$C=\emptyset$ or if $C$ is unbounded in $u$-direction. If
$C\cap H(C,u)\neq\emptyset$, then we call $H(C,u)$ a
{\it supporting hyperplane} of $C$. The {\it exposed face} of $C$ by $u$ 
is 
\[
F_{\perp}(C,u)\;:=\;C\cap H(C,u)
\]
and we put $F_\perp(C,0):=C$. The faces $\emptyset$ and $C$ are exposed 
faces of $C$ by definition. The set of exposed faces of $C$ will be 
denoted by $\cF_{\perp}(C)$, called the {\it exposed face lattice} of $C$. 
A face of $C$, which is not an exposed face is a {\it non-exposed face}
and we then say the face $F$ is not {\it exposed}. 
\item
Some topology is needed.
Let $X\subset\bE$ be an arbitrary subset. The {\it affine hull} of $X$, 
denoted by $\aff(X)$, is the smallest affine subspace of $\bE$ that 
contains $X$. The interior of $X$ with respect to the relative topology of 
$\aff(X)$ is the {\it relative interior} $\ri(X)$ of $X$. The complement 
$\rb(X):=X\setminus\ri(X)$ is the {\it relative boundary} of $X$. If 
$C\subset\bE$ is a non-empty convex subset then we consider the vector 
space $\lin(C)=\{x-y\mid x,y\in\aff(C)\}$. We define the {\it dimension} 
$\dim(C):=\dim(\lin(C))$ and $\dim(\emptyset):=-1$.
\end{enumerate}
\end{Def}
\begin{Rem}
\begin{enumerate}[1.]
\item
\par
As observed e.g.\ in \cite{Weis_Knauf,Weis_touch}, the mean value set $\bM(U)$ 
can have non-exposed faces even though all faces of $\cS$ are exposed.
An example is shown in Figure~\ref{fig:faces}.
\item
Let $C\subset\bE$ be a convex subset.
Different to Rockafellar or Schneider \cite{Rockafellar,Schneider} we
always include $\emptyset$ and $C$ to $\cF_\perp(C)$ so that this set is
a lattice. The inclusion $\cF_\perp(C)\subset\cF(C)$ is easy to show and
there are various ways to see that $\cF_\perp(C)$ and $\cF(C)$ are complete 
lattices ordered by inclusion where the infimum is the intersection, see e.g.\ 
\S1.1 in \cite{Weis_touch} or \S2.1 in \cite{Weis_supp}. The convex set $C$
admits by Theorem~18.2 in \cite{Rockafellar} a partition into relative
interiors of its faces
\begin{equation}\textstyle
\label{eq:stratification}
C\;=\;\bigcup\limits^{\bullet}{}_{F\in\cF(C)}\ri(F)\,.
\end{equation}
In particular, every proper face of $C$ 
is included in the relative boundary of $C$ and its dimension is
strictly smaller than the dimension of $C$.
\end{enumerate}
\end{Rem}
\par
We recall the algebraic description of the face lattice
$\cF(\cS_\cA)=\cF_\perp(\cS_\cA)$ of the state space $\cS_\cA$.
\begin{Def}
\label{def:sfaces}
Extreme points of $\cS$ are called {\it pure states}. For every
orthogonal projection $p\in\cP_\cA$ we set
\[\textstyle
\bF(p)\;=\;\bF_\cA(p)\;:=\;\cS_{p\cA p}
\]
and we denote the face lattice of the state space by
$\cF=\cF_\cA:=\cF(\cS_\cA)$. We use notation $\az=\{a\in\sa\mid{\rm tr}(a)=0\}$
resp.\ $\ao=\{a\in\sa\mid{\rm tr}(a)=1\}$ for the spaces of trace-less resp.\
trace one self-adjoint matrices.
\end{Def} 
\begin{Pro}[Proposition 2.9 in \cite{Weis_supp}]
\label{ss:st}
The state space $\cS$ is a convex body of dimension $\dim(\sa)-1$, the 
affine hull is $\aff(\cS)=\ao$, the translation vector space is 
$\lin(\cS)=\az$ and the relative interior consists of all invertible 
states. The support function at $a\in\sa$ is the maximal spectral value 
$h(\cS,a)=\lambda^+(a)$ of $a$. If $a\in\sa$ is non-zero, then the exposed 
face $F_\perp(\cS,a)=\bF(p)$ by $a$ is the state space of the compressed 
algebra $p\cA p$, where $p=p^+(a)$ is the maximal projection of $a$.
\end{Pro}
\begin{Cor}[Corollary 2.10 in \cite{Weis_supp}] 
\label{ss:iso}
All faces of the state space $\cS$ are exposed. The mapping 
$\bF:\cP\to\cF$, $p\mapsto\bF(p)$ is an isomorphism of complete lattices.
\end{Cor}
\begin{Rem}
\label{rem:state_faces}
It follows from Corollary~\ref{ss:iso}, Proposition~\ref{ss:st} and 
(\ref{eq:sa_dom}) that every face of $\cS$ can be written as
$\bF(p)=\{\rho\in\cS\mid s(\rho)\preceq p\}$ for some $p\in\cP$ and the 
relative interior is $\ri\bF(p)=\{\rho\in\cS\mid s(\rho)=p\}$.
\end{Rem}
\par
Let us turn to the mean value set $\bM(U)=\pi_U(\cS)$ defined 
in (\ref{def:m_set}), where $U\subset\sa$ is a linear subspace. A 
lifting construction connects to the isomorphism $\bF:\cP\to\cF$.
This leads to algebraic descriptions of the two face lattices 
$\cF_\perp(\bM(U))\subset\cF(\bM(U))$.
\begin{Def}
We define for subsets $C\subset\sa$ the (set-valued) {\it lift} by
\[\textstyle
L^U(C)\;=\;L_\cA^U(C)\;:=\;\cS_\cA\cap(C+U^{\perp})\,.
\]
We define the {\it lifted face lattice}
\[\textstyle
\cL^U\;=\;\cL^U_\cA\;:=\;\{\,L^U(F)\,\mid\,F\in\cF(\bM(U))\,\}
\]
and the {\it lifted exposed face lattice}
\[\textstyle
\cL^{U,\perp}\;=\;\cL^{U,\perp}_\cA
\;:=\;\{\,L^U(F)\,\mid\,F\in\cF_\perp(\bM(U))\,\}\,.
\]
\end{Def}
\begin{Lem}[\S5 in \cite{Weis_touch}]
\label{lem:lift_iso}
The lift $L$ restricts to the bijection
$\cF(\bM(U))\stackrel{L}{\longrightarrow}\cL^U$ and to the bijection
$\cF_\perp(\bM(U))\stackrel{L}{\longrightarrow}\cL^{U,\perp}$. These are
isomorphisms of complete lattices with inverse $\pi_U$. For $u\in U$ 
we have $\pi_U\,[F_{\perp}(\cS,u)]=F_{\perp}(\bM(U),u)$ and 
$L^U\,[F_{\perp}(\bM(U),u)]=F_{\perp}(\cS,u)$.
\end{Lem}
\par
The lifting construction defines useful lattice isomorphisms, if we use 
appropriate lattices of projections:
\begin{Def}
The {\it projection lattice} resp.\ {\it exposed projection lattice} 
of $U$ is
\begin{equation}\textstyle
\label{eq:projection_lattice_u}
\cP^U\;=\;\cP^U_\cA\;:=\;\bF^{-1}(\,\cL^U_\cA\,)\qquad\text{resp.}\qquad
\cP^{U,\perp}\;=\;\cP^{U,\perp}_\cA\;:=\;\bF^{-1}(\,\cL^{U,\perp}_\cA\,)\,.
\end{equation}
\end{Def}
\par
Corollary~\ref{ss:iso} and Lemma~\ref{lem:lift_iso} imply two lattice
isomorphisms defined for suitable projections $p$ by 
$p\mapsto\pi_U(\bF(p))$:
\begin{equation}
\label{eq:main_iso}
\cP^U\;\longrightarrow\;\cF(\bM(U))
\qquad\text{resp.}\qquad
\cP^{U,\perp}\;\longrightarrow\;\cF_\perp(\bM(U))
\end{equation}
between $\cP^U$ and the face lattice of the mean value set resp.\
between $\cP^{U,\perp}$ and the exposed face lattice.
Lemma~\ref{lem:lift_iso} characterizes the lifted exposed face 
lattice by
\[\textstyle
\cL^{U,\perp}\;=\;
\{F_\perp(\cS,u)\mid u\in U\}\cup\{\emptyset\}\,.
\]
The algebraic description in Proposition~\ref{ss:st} of faces
$F_\perp(\cS,u)$ of the state space $\cS$ translates therefore
to the exposed faces of the mean value set $\bM(U)$:
\begin{Cor}
\label{cor:exposed_projections}
The exposed projection lattice is
$\cP_\cA^{U,\perp}=\{p_\cA^+(u)\mid u\in U\}\cup\{0\}$.
\end{Cor}
\par
Sequences of faces allows an algebraical description of non-exposed 
faces of $\bM(U)$. 
\begin{Def}[Access sequence]~
\label{def:access}
\begin{enumerate}[1.]
\item
Let $C$ be a convex subset $C$ of the finite-dimensional 
Euclidean vector space $(\bE,\langle\cdot,\cdot\rangle)$.
We call a finite sequence $F_0,\ldots,F_m\subset C$ an
{\it access sequence} (of faces) for $C$ if $F_0=C$ and if
$F_{i+1}$ is a properly included exposed face of $F_i$ for 
$i=0,\ldots,m-1$, 
\begin{equation}
\label{def:access_sequence}\textstyle
F_0
\;\supsetneq\;
F_1
\;\supsetneq\;
\cdots
\;\supsetneq\;
F_m\,.
\end{equation}
\item
For $p\in\cP$ and 
$a\in\sa$ the orthogonal projection $\sa\to(p\cA p)_{\rm sa}$ is
\begin{equation}
\label{eq:projection_compression}\textstyle
c^p(a)
\;:=\;
\pi_{(p\cA p)_{\rm sa}}(a)
\;=\;
pap\,.
\end{equation}
\item
We call a finite sequence $p_0,\ldots,p_m\subset\cP^U$ an 
{\it access sequence} (of projections) for $U$ if $p_0=\id$ and if
$p_{i+1}$ belongs to the exposed projection lattice
$\cP^{c^{p_i}(U),\perp}_{p_i\cA p_i}$ for $i=0,\ldots,m-1$ and 
such that ($p_i\succ p_{i+1}$ $:\iff$ $p_i\succeq p_{i+1}$ and 
$p_i\neq p_{i+1}$)
\[\textstyle
p_0
\;\succ\;
p_1
\;\succ\;
\cdots
\;\succ
\;p_m\,.
\]
\end{enumerate}
\end{Def}
\par
Access sequences of faces are also used in \cite{Csiszar05}.
Gr\"unbaum \cite{Gruenbaum} defines a {\it poonem} as an element of an 
access sequence of faces. An example is depicted in Figure~\ref{fig:poonems}. 
In finite dimensions the notion of poonem is equivalent to the notion of 
face, see e.g.\ \S1.2.1 in \cite{Weis_touch}. 
\begin{Thm}[\S3.2 in \cite{Weis_supp}]
The lattice isomorphism $\cP^U\to\cF(\bM(U))$ in (\ref{eq:main_iso})
extends to a bijection from the set of access sequences of projections
for $U$ to the set of access sequences of faces for $\bM(U)$ by assigning
\[\textstyle
(p_0,\ldots,p_m)
\;\mapsto\;
(\pi_U(\bF(p_0)),\ldots,\pi_U(\bF(p_m)))\,.
\]
\end{Thm}
\par
We will also use the following results from \S3.2 in \cite{Weis_supp}.
\begin{Lem}
\label{lem:double_projection}
If $p\in\cP$ is a projection, then 
$c^p(U)\stackrel{\pi_U}{\longrightarrow}\pi_U((p\cA p)_{\rm sa})$ is a 
real linear isomorphism and the following diagrams commute.
\[
\xymatrix@!0@C=1.6cm@R=1.0cm{%
(p\cA p)_{\rm sa}\ar@{->>}[rr]^(0.45){\pi_U}\ar@{->>}[ddr]_{\pi_{c^p(U)}} 
&& \pi_U((p\cA p)_{\rm sa})\\\\ 
& c^p(U)\ar@{^{(}->>}[uur]_{\pi_U}}
\hspace*{0.2cm}
\xymatrix@!0@C=1.6cm@R=1.0cm{%
\bF(p)\ar@{->>}[rr]^(0.45){\pi_U}\ar@{->>}[ddr]_{\pi_{c^p(U)}} 
&& \pi_U(\bF(p))\\\\ 
& \bM_{p\cA p}(c^p(U))\ar@{^{(}->>}[uur]<-0.5ex>_{\pi_U}}
\hspace*{0.2cm}
\xymatrix@!0@C=1.6cm@R=1.0cm{%
\ri(\bF(p))\ar@{->>}[rr]^(0.45){\pi_U}\ar@{->>}[ddr]_{\pi_{c^p(U)}} 
&& \ri(\pi_U(\bF(p)))\\\\ 
& \ri(\bM_{p\cA p}(c^p(U)))\ar@{^{(}->>}[uur]<-0.5ex>_{\pi_U}}
\]
\end{Lem}
\begin{Cor}
\label{cor:char_pu}
A projection $p\in\cP$ belongs to the projection lattice $\cP^U$ if and 
only if $p$ belongs to an access sequence of projections for $U$.
\end{Cor}
\begin{Cor}
\label{cor:two_for_access}
For each two projections $p,q\in\cP^U$ such that $p\preceq q$ there exists 
an access sequence for $U$ including $p$ and $q$. 
\end{Cor}
\begin{Rem}
\label{rem:pu}
Corollary~\ref{cor:char_pu} implies a computation method for $\cP^U$,
which is an algebraic reformulation of the concept of poonem for the 
special case of a mean value set. One has to compute the maximal 
projection (see Definition~\ref{def:functional_calc}.2) of all elements 
of $U$, then the maximal projections of elements of $c^p(U)$ for each 
previously calculated projection $p$ and so on (see Remark 3.10 and 
\S3.3 in \cite{Weis_supp}).
\end{Rem}
\begin{Lem}[\S3.2 in \cite{Weis_supp}]
\label{lem:stratum}
If $\rho\in\cS$, then $\rho\in\ri(\bF(p))+U^\perp$ holds for a unique 
projection $p\in\cP^U$. We have
$p=\bigwedge\{q\in\cP^U\mid s(\rho)\preceq q\}$.
\end{Lem}
\par
Although we will not need it in the following, let us point out
an advantage that the coordinates of the convex support have over 
mean value sets. The algebraic decomposition of mean value sets in 
Lemma~\ref{lem:double_projection} into its faces becomes a simple 
inclusion ${\rm cs}_{p\cA p}\subset{\rm cs}_\cA$
of convex support sets:
\begin{Lem}
\label{lem:coords_on_cs_faces}
Let $u_1,\ldots,u_k\in\sa$ and put $U:={\rm span}_\bR(u_1,\ldots,u_k)$. 
Then for all $p\in\cP^U$ the following diagram commutes.
\[
\xymatrix@!0@C=1.6cm@R=1.0cm{%
 & & \pi_U(\bF(p)) \ar@{^{(}->>}[rrr]^{m_{u_1,\ldots,u_k}}
 & & & \parbox{1.6cm}{$m_{u_1,\ldots,u_k}(\bF(p))$}
 \ar@{}[rrr]|\subset & & & {\rm cs}_\cA(u_1,\ldots,u_k)\\
 \bF(p) \ar[rru]^{\pi_U} \ar[rrd]_{\pi_{c^p(U)}}
 \ar@(r,dl)[rrrrru]|(0.43){\hole\hole\hole\hole}_(0.72)%
 {m_{u_1,\ldots,u_k}} 
 \ar@(r,ul)[rrrrrd]|(0.43){\hole\hole\hole\hole}^(0.72)%
 {m_{c^p(u_1),\ldots,c^p(u_k)}}  \\
 & & \bM_{p\cA p}(c^p(U)) \ar@{^{(}->>}[rrr]_{m_{c^p(u_1),\ldots,c^p(u_k)}} 
 \ar@{^{(}->>}[uu]_{\pi_U}
 & & & \parbox{1.6cm}{${\rm cs}_{p\cA p}(c^p(u_1),\ldots,c^p(u_k))$}
 \ar@{^{(}->>}[uu]_{{\rm Id}|_{\bR^k}}
}
\]
\end{Lem}
{\em Proof:\/}
We extend the second diagram in Lemma~\ref{lem:double_projection}. We can use the 
bijection $m_{u_1,\ldots,u_k}|_{\bM(U)}:\bM(U)\to{\rm cs}_\cA(u_1,\ldots,u_k)$ in 
(\ref{eq:mean_iso}), it satisfies $m_{u_1,\ldots,u_k}\circ\pi_U=m_{u_1,\ldots,u_k}$. 
In the algebra $p\cA p$ this means 
$m_{c^p(u_1),\ldots,c^p(u_k)}\circ\pi_{c^p(U)}=m_{c^p(u_1),\ldots,c^p(u_k)}$ and 
\[\textstyle
m_{c^p(u_1),\ldots,c^p(u_k)}|_{\bM(c^p(U))}:\;
\bM(c^p(U))\;\to\;{\rm cs}_{p\cA p}(c^p(u_1),\ldots,c^p(u_k))
\]
is a bijection. For all $a\in p\cA p$ and $i=1,\ldots,k$ we have
\[\textstyle
\langle u_i,a\rangle
\;=\;\langle u_i,pap\rangle
\;=\;
\langle c^p(u_i),a\rangle
\]
hence $m_{u_1,\ldots,u_k}(a)=m_{c^p(u_1),\ldots,c^p(u_k)}(a)$ holds
and completes the proof.
\hspace*{\fill}$\Box$\\
%
%
%
%
%
%
\subsection{Projections and functional calculus}
\label{sec:p-lattice}
\par
We recall functional calculus for normal matrices. Definitions are
somewhat technical because we want to work in subalgebras of 
${\rm Mat}(n,\bC)$ not containing $\id_n$. The partial ordering on
$\sa$ induced by the positive semi-definite cone is central in the
following. We will consider this ordering in its restriction to the 
lattice of projections. 
\begin{Def}[The projection lattice and spectral decomposition]~
\begin{enumerate}[1.]
\item
Let $a\in\cA$ be a normal matrix, i.e.\ $a^*a=aa^*$. Let $N\in\bN$,
$\{c_i\}_{i=1}^N\subset\bC$ be mutually distinct numbers and let
$\{p_i\}_{i=1}^N\subset\cP_\cA$ be a family of non-zero projections
such that for $i,j=1,\ldots,N$ we have $p_ip_j=p_i\delta_{ij}$, where 
$\delta_{ij}=0$ unless $i=j$ with $\delta_{ii}=1$. If
$\sum_{i=1}^Np_i=\id$ and
\begin{equation}
\label{def:spectral_deco}\textstyle
a\;=\;\sum_{i=1}^Nc_i p_i\,,
\end{equation}
then the sum (\ref{def:spectral_deco}) is called {\it spectral decomposition} 
of $a$ in $\cA$, $\{p_i\}_{i=1}^N$ is a {\it spectral family} for $a$ in 
$\cA$ and its members are {\it spectral projections} of $a$ in $\cA$. 
\end{enumerate}
\end{Def}
\begin{Rem}[Spectral decomposition]~
\label{rem:spectral_form}
\begin{enumerate}[1.]
\item
It is a classical result of linear algebra, see e.g.\ \S\S79--80 in
\cite{Halmos}, that a normal matrix $a\in{\rm Mat}(n,\bC)$ has a unique 
spectral decomposition 
$a=\sum_{\lambda\in{\rm spec}_{{\rm Mat}(n,\bC)}(a)}\lambda p_\lambda$ in
${\rm Mat}(n,\bC)$. Moreover, for every 
$\lambda\in{\rm spec}_{{\rm Mat}(n,\bC)}(a)$ there exists a polynomial 
$f_\lambda$ in one variable and with complex coefficients, such that 
$p_{\lambda}=f_\lambda(a)$ . 
\item
Let $\cA\subset{\rm Mat}(n,\bC)$ be a C*-subalgebra of ${\rm Mat}(n,\bC)$
with identity $\id$ and $a\in\cA$ a normal matrix. If
$a=\sum_{\lambda\in{\rm spec}(a)}\lambda p_\lambda$ is the 
spectral decomposition of $a$ in ${\rm Mat}(n,\bC)$ then it is easy to 
show that
$a=\sum_{\lambda\in{\rm spec}_\cA(a)}\lambda(\id p_{\lambda})$
is the unique spectral composition of $a$ in $\cA$. Either
${\rm spec}_\cA(a)={\rm spec}_{{\rm Mat}(n,\bC)}(a)$ or
${\rm spec}_\cA(a)\subsetneq{\rm spec}_\cA(a)\cup\{0\}
={\rm spec}_{{\rm Mat}(n,\bC)}(a)$.
For all non-zero $\lambda\in{\rm spec}_\cA(a)$ we have
$\id p_{\lambda}=p_{\lambda}$. But $\id p_{0}\neq p_{0}$ is possible.
\end{enumerate}
\end{Rem}
\begin{Def}[Special projections and functional calculus]~
\label{def:functional_calc}
\begin{enumerate}[1.]
\item
If $a\in\cA$ is a normal matrix then we denote the spectral projections of 
$a$ by $p^\lambda(a)=p^\lambda_\cA(a)$ for $\lambda\in{\rm spec}_\cA(a)$. 
The {\it support projection} of $a$, also called {\it support} of $a$,
is $s(a):=\sum_{\lambda\in{\rm spec}_\cA(a)\setminus\{0\}}p^\lambda(a)$.
The {\it kernel projection} of $a$ in $\cA$ is $k_\cA(a):=\id-s(a)$.
\item
If $a$ is self-adjoint, then the maximum of ${\rm spec}_\cA(a)$ is denoted 
by $\lambda^+(a)=\lambda^+_\cA(a)$ and the corresponding spectral 
projection in $\cA$ is denoted by $p^+(a)=p^+_\cA(a)$ and
$p^+_\cA(a)$ is called the {\it maximal projection} of $a$ in $\cA$. 
\item
If a complex valued function $f$ is defined on the spectrum of a normal 
matrix $a\in\cA$, then 
$f(a)=f_\cA(a):=\sum_{\lambda\in{\rm spec}_\cA(a)} 
f(\lambda)p^\lambda_\cA(a)$ is defined by
{\it functional calculus} in $\cA$. If  $p\in\cP$ and
$f$ is defined on the spectrum of a normal matrix $a\in p\cA p$,
we abbreviate functional calculus in $p\cA p$ by
$f^{[p]}(a)=f^{[p]}_\cA(a):=f_{p\cA p}(a)$.
\end{enumerate}
\end{Def}
\begin{Rem}[Projections and functional calculus]~
\label{rem:func_calc}
\begin{enumerate}[1.]
\item
By Remark~\ref{rem:spectral_form}.2, the support projection $s(a)$ 
of a normal matrix $a\in\cA$ does not depend on the algebra 
$\cA\subset{\rm Mat}(n,\bC)$. But $k_\cA(a)$ depends on $\cA$.
Similarly, if $a$ is self-adjoint then $p_\cA^+(a)$ depends on $\cA$.
\item
If $a\in\cA$ is a normal matrix and $f:\bC\to\bC$ is defined on
${\rm spec}_\cA(a)$ and on ${\rm spec}_{{\rm Mat}(n,\bC)}(a)$, then we 
have $f_\cA(a)=\id f_{{\rm Mat}(n,\bC)}(a)$. E.g.\ for $a,u\in\cA$
\begin{equation}
\label{eq:lieb_derivative}\textstyle
\frac{\partial}{\partial t}|_{t=0}\exp_\cA(a+tu)
\;=\;\int_0^1\exp_\cA((1-y)a)u\exp_\cA(ya){\rm d}y
\end{equation}
holds. This follows from the analogue equation in ${\rm Mat}(n,\bC)$,
see e.g.\ \cite{Lieb}, by multiplication with $\id\in\cA$.
This method has to be applied carefully, e.g.\ $\log(1,0)$ 
is undefined in $\cA=\bC^2$ and the term 
$\log^{[(1,0)]}(1,0)=(0,0)$ is an example of functional calculus in 
the compressed algebra $\bC\oplus\{0\}$.
\item
The von Neumann entropy (\ref{def:von_Neumann}) of $\rho\in\cS$ can be 
defined by $S(\rho):=\tr\eta(\rho)$ in terms of functional calculus in 
the algebra ${\rm Mat}(n,\bC)$. Since the function 
$\eta:[0,1]\to\bR$, $\eta(x)=-x\log(x)$, is continuous on $[0,1]$, 
it follows that the von Neumann entropy is a continuous function, see e.g.\ 
Theorem VIII.20 in \cite{Reed}. The detailed definition of the relative 
entropy (\ref{eq:relative_entropy}) is for $\rho,\sigma\in\cS$ 
\[\textstyle
S(\rho,\sigma)
\;=\;
\tr\rho(\log^{[s(\rho)]}(\rho)-\log^{[s(\sigma)]}(\sigma))
\]
if $s(\rho)\preceq s(\sigma)$ and otherwise $S(\rho,\sigma):=\infty$.
By part 1 this definition restricts from ${\rm Mat}(n,\bC)$ to any 
C*-subalgebra of ${\rm Mat}(n,\bC)$. The relative entropy is not continuous.
\end{enumerate}
\end{Rem}
%
%
%
%
%
%
%
%
\subsection{Two perturbative statements}
\label{sec:perturb}
\par
The following perturbation analysis is essential since the 
relative entropy can not be defined directly in terms of functional 
calculus with respect to a continuous function, like e.g.\ the von 
Neumann entropy, see Remark~\ref{rem:func_calc}.3. The analysis 
will allow us to consider logarithmic functions depending on density 
matrices with some eigenvalues converging to zero. In spite of 
considering a C*-subalgebra $\cA\subset{\rm Mat}(n,\bC)$ we can argue 
mainly within the algebra ${\rm Mat}(n,\bC)$. 
\begin{Def}~
\begin{enumerate}[1.]
\item
In this section we denote the set of eigenvalues of $a\in{\rm Mat}(n,\bC)$ 
by ${\rm spec}(a)={\rm spec}_{{\rm Mat}(n,\bC)}(a)$ and we shall write
$\zeta$ in place of $\zeta\id_n$ for scalars $\zeta\in\bC$.
\item
The {\it resolvent set} of a matrix $a\in{\rm Mat}(n,\bC)$ is the 
complement of the spectrum ${\rm res}(a):=\bC\setminus{\rm spec}(a)$.
\item 
The {\it resolvent} of $a\in{\rm Mat}(n,\bC)$ is defined for 
$\zeta\in{\rm res}(a)$ by $(a-\zeta)^{-1}$. 
\item
The {\it second resolvent equation} for
$a,b\in{\rm Mat}(n,\bC)$  and $\zeta\in{\rm res}(a)\cap{\rm res}(b)$ is
\begin{equation}\textstyle
\label{eq:2nd_res}
(a-\zeta)^{-1}-(b-\zeta)^{-1}
\;=\;
(a-\zeta)^{-1}(b-a)(b-\zeta)^{-1}\,.
\end{equation}
\end{enumerate}
\end{Def}
\begin{Rem}
\begin{enumerate}[1.]
\item
If $a,b\in{\rm Mat}(n,\bC)$ are self-adjoint matrices, let 
$\lambda^{\downarrow}_1(a),\ldots,\lambda^{\downarrow}_n(a)$ denote the 
eigenvalues of $a$ arranged in decreasing order and counting 
multiplicities. {\it Weyl's perturbation theorem}, proved e.g.\ in
\S{}III.2 in \cite{Bhatia}, states that
\begin{equation}
\label{eq:weyl}\textstyle
\max_{k=1}^n|\lambda^{\downarrow}_k(a)-\lambda^{\downarrow}_k(b)|
\;\leq\;
\|a-b\|\,.
\end{equation}
Here the spectral norm from Definition~\ref{def:represent}.2 is used.
\item
According to Problem 5.7 on page 40 in \cite{Kato}, if $\zeta$ belongs
to ${\rm res}(a)$ for a normal matrix $a\in{\rm Mat}(n,\bC)$, then the 
resolvent of $a$ is bounded by
\begin{equation}
\label{eq:res_bound}
\|(a-\zeta)^{-1}\|\;\leq\;{\rm dist}(\zeta,{\rm spec}(a))^{-1}
\end{equation}
where ${\rm dist}(z,M):=\inf\{|z-m|\mid m\in M\}$ for $z\in\mathbb C$ and
$M\subset\mathbb C$. 
\item
Given a normal matrix $a\in{\rm Mat}(n,\bC)$, let
$\Gamma\subset{\rm res}(a)$ be a positively oriented circular curve of 
radius $r>0$. It is well-known, see e.g.\ Chapter~2 \S1.4 in \cite{Kato},
that
\begin{equation}\textstyle
\label{eq:res_projection}
P_\Gamma(a)\;:=\;
-\frac{1}{2\pi i}\int_\Gamma(a-\zeta)^{-1}{\rm d}\zeta
\end{equation}
is the sum of all spectral projections $p^\lambda(a)$ of $a$ in
${\rm Mat}(n,\bC)$, such that $\lambda$ lies inside $\Gamma$. 
\item
Let $a,b\in{\rm Mat}(n,\bC)$ be self-adjoint matrices and let
$\Gamma_\lambda$ be disjoint circular curves  of radius $r>0$ centered at 
$\lambda\in{\rm spec}(b)$. If $\|b-a\|<r$, then by Weyl's perturbation 
theorem (\ref{eq:weyl}) every eigenvalue of $a$ lies in exactly
one of the circles $\{\Gamma_\lambda\}_{\lambda\in{\rm spec}(b)}$. The
projections $Q^\lambda(a):=P_{\Gamma_\lambda}(a)$ in
(\ref{eq:res_projection}) are defined and $\id_n=\sum_\lambda Q^\lambda(a)$
holds (with summation over the eigenvalues $\lambda\in{\rm spec}(b)$
of $b$). The second resolvent equation (\ref{eq:2nd_res}) and the 
inequality (\ref{eq:res_bound}) imply for $\lambda\in{\rm spec}(b)$
\begin{equation}
\label{eq:total_pro}\textstyle
\|Q^\lambda(a)-p^\lambda(b)\|
\;\leq\;
\frac{1}{2\pi}\int_{\Gamma_\lambda}
\|(b-\zeta)^{-1}(b-a)(a-\zeta)^{-1}\|{\rm d}\zeta
\;\leq\;\frac{\|b-a\|}{r(r-\|b-a\|)}\,.
\end{equation}
Hence for fixed $b$, if $\|b-a\|\to 0$ then $Q^\lambda(a)$ converges in 
spectral norm to $p^\lambda(b)$.
\end{enumerate}
\end{Rem}
\par
The next proposition will characterize the rI-convergence in 
Proposition~\ref{pro:continuous_1st_2nd}. By Remark~\ref{rem:func_calc}.1
the support projection $s(a)$ of a self-adjoint matrix $a\in\cA$ does not 
depend on $\cA$, so we assume $\cA={\rm Mat}(n,\bC)$ in the proof.
\begin{Lem}
\label{lem:continuous_log}
Let $\rho,\sigma\in\cS$ and $(\tau_i)_{i\in\mathbb N}\subset\cS$ such that
$s(\rho)\preceq s(\sigma)\preceq s(\tau_i)$ holds for all $i\in\mathbb N$. 
Then $\lim_{i\to\infty}S(\sigma,\tau_i)=0$ implies
$\lim_{i\to\infty}S(\rho,\tau_i)=S(\rho,\sigma)$.
\end{Lem}
{\em Proof:\/}
By the Pinsker-Csisz\'ar inequality (\ref{eq:pinsker}) the sequence
$(\tau_i)_{i\in\bN}$ converges to $\sigma$ in norm. We view $\tau_i$
as a perturbation of $\sigma$ and take a sufficiently small circle 
$\Gamma$ of radius $r>0$ about $0\in\mathbb C$. Then, for large 
$i\in\mathbb N$ the projection $P_\Gamma(\tau_i)$ in 
(\ref{eq:res_projection}) is defined and satisfies
$k(\tau_i)\preceq P_\Gamma(\tau_i)$ where 
$k(\tau_i)=k_{{\rm Mat}(n,\bC)}(\tau_i)$ is the kernel projection.
Then two projections $p_i,q_i\in\cA$ are defined by
$p_i:=P_\Gamma(\tau_i)-k(\tau_i)$ and $q_i:=\id_n-P_\Gamma(\tau_i)$,
they satisfy $q_i+p_i=s(\tau_i)$. We think of $p_i$ as the negligible 
contribution to $s(\tau_i)$.
\par
According to Definition~\ref{def:functional_calc}.3 we split the
functional calculus into two compressed algebras $p_i\cA p_i$ and
$q_i\cA q_i$,
\[\textstyle
S(\sigma,\tau_i)\;=\;
-S(\sigma)
-\tr\sigma\log^{[p_i]}(p_i\tau_i)
-\tr\sigma\log^{[q_i]}(q_i\tau_i)
\,.
\]
We have $\tau_j\stackrel{j\to\infty}{\longrightarrow}\sigma$,
by (\ref{eq:total_pro}) we have 
$q_j\stackrel{j\to\infty}{\longrightarrow}s(\sigma)$ and the spectral
values of $q_j\tau_jq_j$ in $q_j\cA q_j$ are strictly larger than $r>0$,
hence the term 
$\log^{[q_i]}(q_i\tau_i)\stackrel{i\to\infty}{\longrightarrow}
\log^{[s(\sigma)]}(\sigma)$ converges. Using the assumption
$S(\sigma,\tau_i)\stackrel{i\to\infty}{\longrightarrow}0$ gives
$\lim_{i\to\infty}\tr\sigma\log^{[p_i]}(p_i\tau_i)=0$.
\par
Now we use a monotonicity argument. It is clear that 
$\rho/\lambda^+(\rho)\preceq s(\rho)$ holds and by assumption we have 
$s(\rho)\preceq s(\sigma)$. If $\lambda>0$ is the smallest non-zero 
eigenvalue of $\sigma$ then $\lambda s(\sigma)\preceq\sigma$. 
Hence $0\preceq\tfrac{\lambda}{\lambda^+(\rho)}\rho\preceq\sigma$. For all
$i\in\bN$ we have $\log^{[p_i]}(p_i\tau_i)\preceq 0$ hence
\[\textstyle
0\;=\;
\lim_{i\to\infty}\tr\sigma\log^{[p_i]}(p_i\tau_i)
\;\leq\; 
\tfrac{\lambda}{\lambda^+(\rho)}
\lim_{i\to\infty}\tr\rho\log^{[p_i]}(p_i\tau_i)
\;\leq\;
0
\]
proves $\lim_{i\to\infty}\tr\rho\log^{[p_i]}(p_i\tau_i)=0$. Now
\begin{align*}
S(\rho,\tau_i)&\;=\;
-S(\rho)-\tr\rho\log^{[p_i]}(p_i\tau_i)-\tr\rho\log^{[q_i]}(q_i\tau_i)\\
&\stackrel{i\to\infty}{\longrightarrow}\;
-S(\rho)-0-\tr\rho\log^{[s(\sigma)]}(\sigma)
\;=\;S(\rho,\sigma)
\end{align*}
completes the proof.
\hspace*{\fill}$\Box$\\
\par
The following statement is used in Proposition~\ref{pro:pert_R} to set
up the mean value chart of an exponential family and in 
Lemma~\ref{lem:e-limits} to study rI-closures of exponential families.
Part~1 is used implicitly in Lemma~7 in \cite{Wichmann}. 
\begin{Lem}
\label{lem:pert_exp}~
\begin{enumerate}[1.]
\item
Let $(x_j)_{j\in\bN}\subset\sa\setminus\{0\}$ such that 
$\lim_{j\to\infty}\|x_j\|=\infty$. We assume there exist $u,a\in\sa$
such that $\lim_{j\to\infty}\frac{x_j}{\|x_j\|}=u$ and 
$\lim_{j\to\infty}\exp_\cA(x_j)=a$. Then
${\rm spec}_\cA(u)\subset(-\infty,0]$ and $s(a)\preceq k_\cA(u)$. 
\item
Let $\theta,u\in\sa$ such that ${\rm spec}_\cA(u)\subset(-\infty,0]$.
Then 
\[\textstyle
\lim_{t\to+\infty}\exp_\cA(\theta+t u)
\;=\;
\exp_\cA^{[k_\cA(u)]}(k_\cA(u)\theta k_\cA(u))
\,.
\]
\end{enumerate}
\end{Lem}
{\em Proof:\/}
Using a C*-algebra embedding we can assume $\id=\id_n$ in the proof.
Then eigenvalues can be used in $\cA$.
The strategy in the first part is to consider $y_j:=\frac{x_j}{\|x_j\|}$
as perturbations of $u$ and to estimate spectral values of $e^{x_j}$ in
suitable compressed subalgebras. We choose disjoint circular curves
$\Gamma_\lambda$ in the complex plane about the eigenvalues
$\lambda\in{\rm spec}(u)$. Using Weyl's perturbation theorem
(\ref{eq:weyl}), the projections in (\ref{eq:res_projection}) 
\[\textstyle
Q^\lambda(x_j)
\;:=\;
P_{\Gamma_\lambda}(x_j)
\;=\;
P_{\Gamma_\lambda}(y_j)
\]
are defined for large $j$. Let $\lambda\in{\rm spec}(u)$ and
$\lambda\neq 0$. The projection $Q^\lambda(x_j)$ is a sum of spectral 
projections of $x_j$ in ${\rm Mat}(n,\bC)$ for non-zero eigenvalues of
$x_j$, so $Q^\lambda(x_j)\in\cA$ by Remark~\ref{rem:spectral_form}.2.
We consider functional calculus in the compressed algebra
$Q^\lambda(x_j)\cA Q^\lambda(x_j)$, 
\[\textstyle
h^\lambda(x_j)
\;:=\;
\exp_\cA^{[Q^\lambda(x_j)]}(Q^\lambda(x_j)x_j)
\;=\;
Q^\lambda(x_j)\exp(x_j)
\,.
\]
The spectral values of the self-adjoint matrix $Q^\lambda(x_j)y_j$
in $Q^\lambda(x_j)\cA Q^\lambda(x_j)$ converge for $j\to\infty$ to
$\lambda\neq 0$ because there is only one eigenvalue of $u$ in the circle
$\Gamma_\lambda$. Since $x_j=y_j\|x_j\|$ we have for $\lambda<0$
and for large $j$ the bound
$\|h^\lambda(x_j)\|\leq e^{\tfrac{\lambda}{2}\|x_j\|}$. Then 
\begin{equation}
\label{eq:pert_conv}\textstyle
\|x_j\|\stackrel{j\to\infty}{\to}\infty
\quad\text{implies}\quad 
h^\lambda(x_j)\stackrel{j\to\infty}{\to}0
\,. 
\end{equation}
If $\lambda>0$ then the 
analogous arguments show that the spectral norm
$\|h^\lambda(x_j)\|\geq e^{\tfrac{\lambda}{2}\|x_j\|}$ diverges to 
$+\infty$.
\par
For $\lambda\neq 0$ the projection $Q^\lambda(x_j)$ converges to
$p^\lambda(u)$ by (\ref{eq:total_pro}). Hence with summation over 
$\lambda\in{\rm spec}(u)\setminus\{0\}$ we have
$s(u)=\lim_{j\to\infty}\sum_{\lambda\neq0}Q^\lambda(x_j)$. Now the
assumed convergence of $\exp_\cA(x_j)\stackrel{j\to\infty}{\to}a$ 
gives 
\[\textstyle
s(u)a
\;=\;
\lim_{j\to\infty}\sum_{\lambda\neq0}
Q^\lambda(x_j)\exp(x_j)
\;=\;
\lim_{j\to\infty}\sum_{\lambda\neq0}
h^\lambda(x_j)
\;=\;
0
\]
and ${\rm spec}(u)\subset(-\infty,0]$. Then (\ref{eq:sa_dom}) and
the equation
\[\textstyle
k_\cA(u)a
\;=\;
(\id-s(u))a
\;=\;
a
\]
show $s(a)\preceq k_\cA(u)$.
\par
We prove convergence and calculate the limit in the second statement.
For small real parameter $c>0$ let $x_c:=u+c\theta$, then
$x_c\stackrel{c\to 0}{\to}u$. For $\lambda\in{\rm spec}(u)\cup\{0\}$
we choose disjoint circular curves $\Gamma_\lambda$ in the complex
plane about each such $\lambda$ and we define
\[\textstyle
Q^\lambda(x_c)
\;:=\;
P_{\Gamma_\lambda}(x_c)\,.
\]
For all $\lambda<0$ the argument in (\ref{eq:pert_conv}) shows
$Q^\lambda(x_c)\exp(x_c)\stackrel{c\to 0}{\to}0$. Since
$\id_n=\sum_{\lambda\in{\rm spec}(u)\cup\{0\}}Q^\lambda(x_c)$ holds
for small $c$, we have
\[\textstyle
\lim_{t\to+\infty}\exp(\theta+t u)
\;=\;
\lim_{c\to 0}\exp(\tfrac 1{c}x_c)
\;=\;
\lim_{c\to 0}Q^0(x_c)\exp(Q^0(x_c)\tfrac 1{c}x_c)
\,.
\]
By (\ref{eq:total_pro}) we have 
$Q^0(x_c)\stackrel{c\to 0}{\to}k(u)\in{\rm Mat}(n,\bC)$. The first order 
expansion is calculated in Chapter II \S1 equation (1.17) in \cite{Kato}:
With $\widetilde{Q}:=\frac{1}{2\pi i}
\int_{\Gamma_0}(u-\zeta)^{-1}\theta(u-\zeta)^{-1}{\rm d}\zeta$ 
we have%
\footnote{If $g$ is a positive real valued function and $f$ is any function
(here with values in ${\rm Mat}(n,\bC)$), then $f=o(g)$ means
$\tfrac f{g}\to 0$ and $o(g)$ is called {\it Landau symbol}.}  
\[\textstyle
Q^0(x_c)
\;=\;
k(u)+c\widetilde{Q}+o(c)
\,.
\]
We compute 
\[\textstyle
Q^0(x_c)\tfrac 1{c}x_c
\;=\;
\tfrac 1{c}Q^0(x_c)x_cQ^0(x_c)
\;=\;
k(u)\theta k(u) + o(1)
\]
and the continuity of the exponential gives
\[\textstyle
\lim_{t\to+\infty}\exp(\theta+t u)
\;=\;
\lim_{c\to0}Q^0(x_c)\exp(Q^0(x_c)\tfrac 1{c}x_c)
\;=\;
k(u)\exp(k(u)\theta k(u))\,.
\]
Multiplication of this formula with the identity $\id_n$ of $\cA$ completes 
the proof.
\hspace*{\fill}$\Box$\\
%
%
\section{Information topologies}
\label{sec:topo}
\par
We study in \S\ref{sec:topo-qm} the I-/rI-topology on the state space 
$\cS$ of a C*-subalgebra $\cA$ of ${\rm Mat}(n,\bC)$. The analysis  
is based on a socalled divergence function and its L*-convergence,
that we customize in \S\ref{sec:topo_seq}. 
%
%
%
%
\subsection{The sequential topology of a divergence function}
\label{sec:topo_seq}
\par
We consider a divergence function defined for pairs of elements in some set. 
A topology is defined by a natural convergence of countable sequences in 
terms a divergence function. Finally, we explore divergence functions 
having two properties, which are available for the relative entropy on 
${\rm Mat}(n,\bC)$. 
\begin{Def}[L*-convergence\footnote{Our definition
of an L*-space is taken from \cite{Engelking,Berge}.
An L*-space in the sense of \cite{Dudley64} has also the 
unique limit property d).}]
\label{def:l_star}
Let $X$ be any set.
A relation $C\subset X^{\bN}\times X$ between sequences and members of 
$X$ is a {\it convergence} on $X$. If $((x_n)_{n\in\mathbb N},x)\in C$ 
then we write $x_n\stackrel{C}{\longrightarrow}x$ and we say
$(x_n)_{n\in\mathbb N}$ $C$-{\it converges} to $x$ and $x$ is the
$C$-{\it limit} of $(x_n)_{n\in\mathbb N}$. A convergence $C$ on $X$ 
is an {\it L*-convergence} and $(X,C)$ is an {\it L*-space} if
\begin{enumerate}
\item[a)]
$x_n=x$ for all $n$ implies $x_n\stackrel{C}{\longrightarrow}x$,
\item[b)]
$x_n\stackrel{C}{\longrightarrow}x$ and $(y_n)_{n\in\mathbb N}$ is a 
subsequence of $(x_n)_{n\in\mathbb N}$ then 
$y_n\stackrel{C}{\longrightarrow}x$.
\item[c)]
$x_n\not\stackrel{C}{\longrightarrow}x$ (i.e.\ it is false that 
$x_n\stackrel{C}{\longrightarrow}x$) implies the existence
of a subsequence $(y_n)_{n\in\mathbb N}$ of $(x_n)_{n\in\mathbb N}$,
such that for any subsequence $(z_n)_{n\in\mathbb N}$ of
$(y_n)_{n\in\mathbb N}$ we have 
$z_n\not\stackrel{C}{\longrightarrow}x$.
\end{enumerate}
A convergence $C$ on $X$ is said to have {\it unique limits} if
\begin{enumerate}
\item[d)]
$x_n\stackrel{C}{\longrightarrow}x$ and $x_n\stackrel{C}{\longrightarrow}y$ 
implies $x=y$.
\end{enumerate}
We consider the family $\cT(C)$ of subsets $U\subset X$ such that
$x\in U$ and $x_n\stackrel{C}{\longrightarrow}x$ implies $x_n\in U$
for large $n$.
\end{Def}
\begin{Rem}[Sequential topologies and closures]
\label{rem:seq_closure}
It is well-known \cite{Dudley64} that $\cT(C)$ is a topology on $X$ if
$C$ is a convergence on $X$. Moreover, if  $Y\subset X$ is $\cT(C)$
closed then $(y_n)_{n\in\mathbb N}\subset Y$ and 
$y_n\stackrel{C}{\longrightarrow}y$ imply $y\in Y$. Important for our 
purpose is: If b) above holds, then the converse is also true,
$Y\subset X$ is $\cT(C)$ closed if and only if 
$(y_n)_{n\in\mathbb N}\subset Y$ and $y_n\stackrel{C}{\longrightarrow}y$ 
imply $y\in Y$. If $(X,C)$ is an L*-space then $\cT(C)$ is called the
{\it sequential topology} induced by $C$.
\end{Rem}
\par
We consider closures in a sequential space. 
\begin{Def}[Sequential closures]
\label{def:l_star_e}
Let $C$ be a convergence on $X$. The {\it sequential closure} of 
$Y\subset X$ is
\begin{equation}
\label{def:cl}\textstyle
{\rm cl}_C(Y)\;:=\;\{x\in X\mid (x_n)_{n\in\mathbb N}
\stackrel{C}{\longrightarrow} x
\text{ for a sequence } (x_n)_{n\in\mathbb N}\subset Y\}\,.
\end{equation}
The following property \cite{Dudley64}, will be proved in the context of  
the relative entropy:
\begin{enumerate}
\item[e)]
if $x_n\stackrel{C}{\longrightarrow}x$ and 
$(x^{(m)})_n\stackrel{C}{\longrightarrow}x_m$ for all $m\in\bN$, then there 
exists a function $n:\mathbb N\to\mathbb N$, such that 
$(x^{(m)})_{n(m)}\stackrel{C}{\longrightarrow}x$.
\end{enumerate}
A weaker property is defined in Problem 1.7.18 in \cite{Engelking}:
\begin{enumerate}
\item[e')]
if $x_n\stackrel{C}{\longrightarrow}x$ and 
$(x^{(m)})_n\stackrel{C}{\longrightarrow}x_m$ for all $m\in\bN$, then there 
exists sequences of natural numbers $m_1,m_2,\ldots$ and $n_1,n_2,\ldots$,
such that $(x^{(m_k)})_{n_k}\stackrel{C}{\longrightarrow}x$.
\end{enumerate}
\end{Def}
\par
Sequential closures in L*-spaces need not be topological closures.
\begin{Exa}[Information closures]
\label{exa:clcl}
The I-/rI-convergence of probability measures in (\ref{intro:info_conv})
is an L*-con\-ver\-gence \cite{Harremoes,Dudley98}. Harremo\"es
discusses a triangle $D$ in $\bP(\bN)$, the probability
simplex (\ref{def:prob_simplex}), where 
${\rm cl}_C(D)\subsetneq{\rm cl}_C({\rm cl}_C(D))$
holds for the I-convergence $C$. Csisz\'ar and Mat\'u\v s 
\cite{Csiszar04} discuss an exponential family $\mathcal E$ of Borel 
probability measures in $\mathbb R^3$ where 
${\rm cl}_C(\mathcal E)\subsetneq{\rm cl}_C({\rm cl}_C(\mathcal E))$
holds for the rI-convergence $C$. 
\end{Exa}
\par
Sequential closures and topological closures in L*-spaces are related 
as follows.
\begin{Rem}[Idempotent sequential closure]
\label{ref:it_lim2}
Let $C$ be a convergence on $X$ satisfying b). Then for each $Y\subset X$
the $\cT(C)$ closure of $Y$ equals the sequential closure 
${\rm cl}_C(Y)$ if and only if e') holds for $C$. 
Indeed, by Remark~\ref{rem:seq_closure}, since b) holds, a subset 
$Y\subset X$ is $\cT(C)$ closed if and only if ${\rm cl}_C(Y)=Y$. Hence
${\rm cl}_C(Y)$ is the $\cT(C)$ closure of $Y$ if and only if 
${\rm cl}_C({\rm cl}_C(Y))={\rm cl}_C(Y)$. The equation 
${\rm cl}_C({\rm cl}_C(Y))={\rm cl}_C(Y)$ is easily seen to be 
equivalent to e') for a arbitrary convergence $C$.
\end{Rem}
\par
Every L*-convergence $C$ can be computed from the topology $\cT(C)$.
\begin{Def}[The convergence of a topology]
If $(X,\cT)$ is a topological space then the convergence $C(\cT)$ is 
defined for sequences $(x_i)_{i\in\bN}\subset X$ and $x\in X$ by
\[\textstyle
(x_i)_{i\in\bN}\stackrel{C(\cT)}{\longrightarrow}x
\quad:\iff\quad
\text{if } x\in U\in\cT 
\text{ then } x_i\in U
\text{ for large }i.
\]
\end{Def}
\par
For any topological space $(X,\cT)$ it is easy to show $\cT\subset\cT(C(\cT))$. 
Similarly, if $C$ is a convergence on $X$, then $C\subset C(\cT(C))$ holds. An 
equality condition was proved by Kisy\'nski, see Problem 1.7.19 in 
\cite{Engelking}:
\begin{Thm}
\label{thm:kisynski}
If $(X,C)$ is an L*-space, then $C(\mathcal T(C))=C$.
\end{Thm}
\par
Divergence functions in Definition~\ref{def:divergence_function} will
generalize metric spaces.
\begin{Exa}[Metric spaces]
\label{exa:metric}
Let $(X,d)$ be a metric space for $d:X\times X\to\bR$. Then 
$x_n\stackrel{C_d}{\longrightarrow}x$ $:\iff$ 
$\lim_{i\to\infty}d(x,x_i)=0$ defines an L*-convergence $C_d$ on $X$,
such that e) holds. Moreover, a base of the {\it metric topology}
$\cT(C_d)$ at $x\in X$ is given by the open disks 
$B(x,\epsilon):=\{y\in X\mid d(x,y)<\epsilon\}$ for $\epsilon>0$.
Since rational values of $\epsilon$ suffice, a metric topology is first 
countable.
\end{Exa}
\par
Continuity will allow to generalize the idea that open disks define a base.
\begin{Def}[Continuity]
Let $f:X\to X'$ be a function and $C$ resp.\ $C'$ be a convergence on 
$X$ resp.\ $X'$. Then $f$ is {\it continuous} for $C$ 
and $C'$ at $x\in X$ if $f(x_n)\stackrel{C'}{\longrightarrow}f(x)$ 
whenever $x_n\stackrel{C}{\longrightarrow}x$. 
The function $f$ is {\it continuous} for $C$ and $C'$ if $f$ is continuous 
for $C$ and $C'$ at every $x\in X$. If $\mathcal T$ resp.\ $\mathcal T'$ 
is a topology on $X$ resp.\ $X'$, then $f$ is continuous for $\mathcal T$ 
and $\mathcal T'$ if $f^{-1}(U')$ is $\mathcal T$ open for every $\mathcal T'$ 
open set $U'\subset X'$.
\end{Def}
\par
The following statement is an excerpt of Theorem 2.2 in \cite{Dudley64}.
Dudley restricts to convergences satisfying a) and b) in 
Definition~\ref{def:l_star}. But the proof works for arbitrary convergences
as well.
\begin{Thm}\label{thm:dudley}
Let $f:X\to X'$ be a function and $C$ resp.\ $C'$ be a convergence on 
$X$ resp.\ $X'$.
\begin{enumerate}[1.]
\item
If $f$ is continuous for $C$ and $C'$, then $f$ is continuous for 
$\mathcal T(C)$ and $\mathcal T(C')$.
\item
If $(X',C')$ is an L*-space then $f$ is continuous for $C$ and $C'$ if and
only if $f$ is continuous for $\mathcal T(C)$ and $\mathcal T(C')$.
\end{enumerate}
\end{Thm}
\par
Subspaces will allow us to relate several topologies to each other.
\begin{Def}[Subspaces]
Let $B\subset X$. If $\cT$ is a topology on $X$, then the
{\it subspace topology}
\[\textstyle
\cT|_B
\;:=\;
\{ B\cap U\mid U\in\cT\}
\]
is defined. If $C$ is a convergence on $X$, we have the
{\it subspace convergence}
\[\textstyle
C|_B
\;:=\;
C\cap(B^\bN\times B)\,.
\]
\end{Def}
\par
We want to allow infinite ``distances'' appearing in the relative entropy.
\begin{Exa}[One point compactification]
\label{exa:alexandroff}
Let $I=\bR$ or $I=[a,\infty)$ for $a\in\bR$. The
{\it Alexandroff compactification} of $I$ is a topology
$\cT^{\rm c}$ on $I\cup\{\infty\}$, where $\cT^{\rm c}$ open sets are
norm open subsets of $I$ or they are of the form $I\cup\{\infty\}\setminus F$, 
where $F\subset I$ is norm compact. Theorem 3.5.11 in \cite{Engelking} 
shows that $(I\cup\{\infty\},\cT^{\rm c})$ is a compact Hausdorff topological
space. The convergence $C^{\rm c}:=C(\cT^{\rm c})$ is
\begin{align*}
& \{((x_i),x)\in[0,\infty]^\bN\times[0,\infty)\mid
x_i<\infty\text{ for large }i
\text{ and }\lim_{i\to\infty}x_i=x\}\\
\cup \;
& \{((x_i),\infty)\mid (x_i)\subset[0,\infty] \text{ such that }
\forall R\in[0,\infty)\text{ we have } x_i\geq R 
\text{ for large }i\}\,.
\end{align*}
\par
It is easy to show $\cT(C(\cT^{\rm c}))\subset\cT^{\rm c}$ (every 
$U\in\cT(C(\cT^{\rm c}))$ including $\infty$ has a bounded complement 
and with each real $x\in U$ there is a disk $B(x,\epsilon)$ in $U$). The 
converse inclusion holds for arbitrary topologies so we have 
\[\textstyle
\cT(C^{\rm c})
\;=\;
\cT^{\rm c}\,.
\]
It is easy to show that $C^{\rm c}$ is an L*-convergence, hence 
Theorem~\ref{thm:dudley}.2 show for every convergence $C$ on a set $X$
and any function $f:X\to I\cup\{\infty\}$ that $f$ is continuous for 
$C$ and $C^{\rm c}$ if and only if $f$ is continuous for $\cT(C)$ and 
$\cT^{\rm c}$.
\end{Exa}
\par
We are ready to study the I-/rI-topology abstractly. In the sequel we 
will use the compactification $[0,\infty]$ of the non-negative half-axis 
$[0,\infty)$. We shall frequently write $\lim_{i\to\infty}x_i=x$ in place 
of $x_i\stackrel{C^{\rm c}}{\longrightarrow}x$ for
$(x_i)_{i\in\bN}\subset[0,\infty]$ and $x\in[0,\infty]$. 
\begin{Def}[Divergence functions]
\label{def:divergence_function}~
\begin{enumerate}[1.]
\item
A {\it divergence function} on a set $X$ is a function 
$f:X\times X\to[0,\infty]$, such that for all $x\in X$ we have 
$f(x,x)=0$. Let $C_f$ be the convergence on $X$ defined by
\[\textstyle
x_n\stackrel{C_f}{\longrightarrow}x
\;:\iff\;
\lim_{n\to\infty}f(x,x_n)=0
\,. 
\]
Two extra assumptions on a divergence function $f$ on $X$ suffice for 
our purpose to analyze the I-/rI-convergence: 
\begin{enumerate}
\item[A)]
An abstract {\it Pinsker-Csisz\'ar inequality} holds, i.e.\ $(X,d)$ is a 
metric space and there is a function $g:[0,\infty]\to[0,\infty]$, continuous 
for $C^{\rm c}$ and $C^{\rm c}$ at $0$, such that $g(0)=0$ and such that for 
all $x,y\in X$ we have $d(x,y)\leq g(f(x,y))$.
\item[B)]
The divergence function $f$ is continuous in the second argument, i.e.\
for all $x\in X$ the function $X\to[0,\infty]$, $y\mapsto f(x,y)$ 
is continuous for $C_f$ and $C^{\rm c}$.
\end{enumerate}
\item
For $x\in X$ and $\epsilon\in(0,\infty]$ we define the {\it open}
resp.\ {\it closed $f$-disk}
\[\textstyle
V^f(x,\epsilon)
\;:=\;\{y\in X\mid f(x,y)<\epsilon\}
\quad\text{resp.}\quad
W^f(x,\epsilon)
\;:=\;\{y\in X\mid f(x,y)\leq\epsilon\}\,.
\]
\end{enumerate}
\end{Def}
\par
If $f$ is the relative entropy between probability measures on $(\bN,2^\bN)$,
then property B) fails and property A) holds by the Pinsker-Csisz\'ar inequality, 
see \S\ref{intro:hist}.
\begin{Lem}[Divergence functions]\label{lem:special_diverge}
Let $f$ be a divergence function on a set $X$. Then the convergence $C_f$ 
is an L*-convergence. In particular $C(\cT(C_f))=C_f$. The sequential closure 
of $Y\subset X$ is
\begin{align*}
{\rm cl}_{C_f}(Y)
&\;=\;
\{x\in X\mid\lim_{n\to\infty}f(x,y_n)=0
\text{ for a sequence }(y_n)_{n\in\bN}\subset Y\}\\
&\;=\;
\{x\in X\mid\inf_{n\in\bN}f(x,y_n)=0
\text{ for a sequence }(y_n)_{n\in\bN}\subset Y\}\\
&\;=\;
\{x\in X\mid\inf_{y\in Y}f(x,y)=0\}\,.
\end{align*}
\begin{enumerate}[1.]
\item
Let $f$ satisfy property A) in Definition~\ref{def:divergence_function}.1
for a metric $d:X\times X\to\bR$. Then $C_f$ has unique limits. We have 
$C_f\subset C_d$ and $\cT(C_f)\supset\cT(C_d)$. In particular $\cT(C_f)$ 
is a Hausdorff topology. 
\item
The property B) in Definition~\ref{def:divergence_function}.1 is equivalent 
with the property that for all $x\in X$ the function $X\to[0,\infty]$, 
$y\mapsto f(x,y)$ is continuous for $\cT(C_f)$ and $\cT^{\rm c}$. 
\par
Property B) implies that for each $x\in X$ and $\epsilon\in(0,\infty]$ the 
open $f$-disk $V^f(x,\epsilon)$ is $\cT(C_f)$ open and the closed $f$-disk
$W^f(x,\epsilon)$ is $\cT(C_f)$ closed. It follows that the open $f$-disks 
$\{V^f(x,\epsilon)\mid\epsilon>0\}$ are a base for $(X,\cT(C_f))$ at $x$. 
This shows that $\cT(C_f)$ is first countable and for any subset 
$Y\subset X$ we have $\cT(C_f)|_Y=\cT(C_f|_Y)$. 
\par
Property B) implies that the L*-convergence $C_f$ has property e) in 
Definition~\ref{def:l_star_e}. This shows for any subset $Y\subset X$ that
the sequential closure ${\rm cl}_{C_f}(Y)$ is the $\cT(C_f)$ closure of $Y$.
\end{enumerate}
\end{Lem}
{\em Proof:\/}
Clearly $C_f$ is an L*-convergences and then $C(\cT(C_f))=C_f$ follows from
Theorem~\ref{thm:kisynski}. The statements about the sequential closure 
are clear.
\par
{\it Property A).} Let us prove unique limits, i.e.\ d) in 
Definition~\ref{def:l_star}. Let $x\in X$ and 
$(x_i)_{i\in\mathbb N}\subset X$. Assuming 
$x_n\stackrel{C_f}{\longrightarrow}x$, i.e.\ 
$\lim_{i\to\infty}f(x,x_i)=0$, the continuity of $g$ at zero 
(for $C^{\rm c}$) gives
\[\textstyle
\lim_{i\to\infty}g\circ f(x,x_i)
\;=\;
0\,.
\]
For all $i\in\bN$ we have by assumption $d(x,x_i)\leq g\circ f(x,x_i)$, 
so $\lim_{i\to\infty}d(x,x_i)=0$. Limits are unique in a metric space 
so this translates to the convergence $C_f$. We have thereby proved 
$C_f\subset C_d$. If follows that $\cT(C_f)\supset\cT(C_d)$. Since 
$\cT(C_d)$ is Hausdorff, so is $\cT(C_f)$. 
\par
{\it Property B).} For all $x\in X$ the continuity of 
the function $X\to[0,\infty]$, $y\mapsto f(x,y)$ for $C_f$ and 
$C^{\rm c}$ is equivalent to the continuity for $\cT(C_f)$ and 
$\cT^{\rm c}$ according to the discussion in the last paragraph of 
Example~\ref{exa:alexandroff}.
\par
Hence, if property B) holds, then the preimage of every $\cT^{\rm c}$ open 
resp.\ closed subset of $[0,\infty]$ is $\cT(C_f)$ open resp.\ closed. In 
particular, every open resp.\ closed $f$-disk is $\cT(C_f)$ open resp.\ 
closed. The open $f$-disks $\{V^f(x,\epsilon)\mid\epsilon>0\}$ define a 
base for $(X,\cT(C_f))$ at $x\in X$: By contradiction, let $U$ be $\cT(C_f)$ 
open, $x\in U$ and let us assume that $U$ contains no open $f$-disk about 
$x$. Then there exists a sequence $(x_i)_{i\in\bN}\subset X\setminus U$ with 
\[\textstyle
(x_i)_{i\in\bN}\stackrel{C_f}{\longrightarrow}x\,.
\]
But $X\setminus U$ is $\cT(C_f)$ closed and so by
Remark~\ref{ref:it_lim2} it contains all $C_f$-limits 
of sequences in $X\setminus U$. So $x\in X\setminus U$ contradicts the 
assumption $x\in U$. The space $(X,\cT(C_f))$ is first countable, e.g.\ 
$\{V^f(x,1/n)\mid n\in\bN\}$ is a base at $x\in X$.
\par
Let us consider a subspace $Y\subset X$. Then $\cT(C)|_Y\subset\cT(C|_Y)$ 
is easy to show. Conversely, for all $y\in Y$ and $\epsilon>0$ 
we have
\[\textstyle
V^{f|_{Y\times Y}}(y,\epsilon)
\;=\;
V^f(y,\epsilon)\cap Y\,.
\]
The divergence function $f|_{Y\times Y}$ on $Y$ satisfies B), hence a set 
$U\in\cT(C|_Y)$ equals
\[\textstyle
U
\;=\;
\bigcup_{\alpha\in I}
V^{f|_{Y\times Y}}(y_\alpha,\epsilon_\alpha)
\;=\;
\left(\bigcup_{\alpha\in I}
V^f(y_\alpha,\epsilon_\alpha)\right)\cap Y
\]
for some $y_\alpha\in Y$ and $\epsilon_\alpha>0$,
$\alpha\in I$. We have proved $U\in\cT(C_f)|_Y$.
\par
To prove property e) we use for each $x\in X$ the continuity of the function 
$X\to[0,\infty]$, $y\mapsto f(x,y)$ for $C_f$ and $C^{\rm c}$ in an open 
$f$-disk $V^f(x,\epsilon)$ for some $\epsilon>0$.
If $(x_i)_{i\in\bN}\stackrel{C_f}{\longrightarrow}x$
then there exists a sequence of positive numbers
$(\epsilon_i)_{i\in\bN}\stackrel{i\to\infty}{\longrightarrow}0$, such that
$f(x,x_i)<\epsilon_i$ for all $i$. For every $i\in\bN$ we
choose a sequence $(x^i_j)_{j\in\bN}\subset X$ such that 
$(x^i_j)_{j\in\bN}\stackrel{C_f}{\longrightarrow}x_i$. By continuity of
$f(x,\cdot)$ for $C_f$ and $C^{\rm c}$ there exists $m_i\in\bN$ 
for all $i$ such that $f(x,x^i_j)<\epsilon_i$ for all $j\geq m_i$. Then 
$f(x,x^i_{m_i})\leq\epsilon_i$ for all $i$ implies 
\[\textstyle
\lim_{i\to\infty}f(x,x^i_{m_i})
\;\leq\;
\lim_{i\to\infty}\epsilon_i
\;=\;
0
\,.
\]
This proves property e) for $C_f$. A consequence for any $Y\subset X$
is that ${\rm cl}_{C_f}(Y)$ is the $\cT(C_f)$ closure of $Y$ (see 
Remark~\ref{ref:it_lim2}).
\hspace*{\fill}$\Box$\\
%
%
%
%
%
%
%
%
%
%
%
%
%
\subsection{The I-topology and the rI-topology}
\label{sec:topo-qm}
\par
The relative entropy $S:\cS\times\cS\to\mathbb[0,\infty]$ defines two
divergence functions. Some results are formulated in terms of the 
convex geometry of the state space. Corollary~\ref{cor:top} collects 
topological conditions for a commutative algebra. 
Several definitions appear already in \S\ref{sec:news_top}, e.g.\ for 
$\rho,\sigma\in\cS_\cA$ the functions $S^{\rm I}(\rho,\sigma)=S(\sigma,\rho)$ 
and $S^{\rm rI}(\rho,\sigma)=S(\rho,\sigma)$ are defined.
In the sequel let $\omega\in\{{\rm I},{\rm rI}\}$. 
\begin{Def}
\label{def:IrI}
Let $(\rho_i)_{i\in\mathbb N}\subset\cS$ be a sequence and let 
$\rho\in\cS$. We define the {\it $\omega$-convergence} $C^\omega$ on
$\cS$ by
\[\textstyle
\rho_i\stackrel{C^\omega}{\longrightarrow}\rho
\;:\iff\; 
\lim_{i\to\infty}S^\omega(\rho,\rho_i)=0\,.
\]
The topology $\cT^\omega=\mathcal T(C^\omega)$ on $\cS$
is called $\omega$-{\it topology}.
We denote the norm convergence on $\cS$ by $C^{\|\cdot\|}$ and the norm
topology on $\cS$ by $\cT^{\|\cdot\|}:=\cT(C^{\|\cdot\|})$.
\end{Def}
\par
We begin with continuity of the relative entropy, using the L*-convergence 
$C^{\rm c}$ on $[0,\infty]$ corresponding to the Alexandroff compactification,
see Example~\ref{exa:alexandroff}.
\begin{Pro}
\label{pro:continuous_1st_2nd}
For every state $\rho\in\cS$ the mapping $\cS\to[0,\infty]$, 
$\sigma\mapsto S^\omega(\rho,\sigma)$ is continuous for $C^\omega$ and 
$C^{\rm c}$.
\end{Pro}
{\em Proof:\/}
Concerning the I-convergence, we have to show for $\rho,\sigma\in\cS$
and $(\tau_i)_{i\in\mathbb N}\subset\cS$ that 
$\lim_{i\to\infty}S(\tau_i,\sigma)=0$ implies
$\lim_{i\to\infty}S(\tau_i,\rho)=S(\sigma,\rho)$. Let us first assume
that $s(\rho)\succeq s(\sigma)$ holds, i.e.\ $S(\sigma,\rho)<\infty$.
Since $\lim_{i\to\infty}S(\tau_i,\sigma)=0$ we have
$s(\sigma)\succeq s(\tau_i)$
for large $i$ and hence $s(\rho)\succeq s(\tau_i)$ holds for large $i$. By 
the Pinsker-Csisz\'ar inequality (\ref{eq:pinsker}) the sequence  
$(\tau_i)_{i\in\bN}$ converges to $\sigma$ in norm. Hence the continuity
of the von Neumann entropy, see e.g.\ \S{}II.A in \cite{Wehrl}, proves
\[\textstyle
S(\tau_i,\rho)
\;=\;
-S(\tau_i)-\tr\tau_i\log(\rho)
\;\stackrel{i\to\infty}{\longrightarrow}\;
-S(\sigma)-\tr\sigma\log(\rho)
\;=\;
S(\sigma,\rho)\,.
\]
Second, we consider $s(\rho)\not\succeq s(\sigma)$, i.e.\ 
$S(\sigma,\rho)=\infty$. By Remark~\ref{rem:conv_entropy}.1 the relative 
entropy is lower semi-continuous. We obtain
$\liminf_{i\to\infty}S(\tau_i,\rho)\geq S(\sigma,\rho)=\infty$ and this 
implies $\lim_{i\to\infty}S(\tau_i,\rho)=\infty$.
\par
Concerning the rI-convergence, we have to show that 
$\lim_{i\to\infty}S(\sigma,\tau_i)=0$ implies
$\lim_{i\to\infty}S(\rho,\tau_i)=S(\rho,\sigma)$. If
$s(\rho)\not\preceq s(\sigma)$ then $S(\rho,\sigma)=\infty$ and the 
lower semi-continuity of the relative entropy proves 
$\lim_{i\to\infty}S(\rho,\tau_i)=\infty$ as in the previous paragraph. 
Finally we consider $s(\rho)\preceq s(\sigma)$ with 
$S(\rho,\sigma)<\infty$. Since
$S(\sigma,\tau_i)\stackrel{i\to\infty}{\longrightarrow}0$  we have 
$s(\sigma)\preceq s(\tau_i)$ for large $i$. 
Lemma~\ref{lem:continuous_log} completes the proof.
\hspace*{\fill}$\Box$\\
\par
The norm topology is too coarse for a similar continuity result, see e.g.\ 
Example~\ref{ex:norm_coarseness}. We now prove that $\omega$-closures do 
not decrease the relative entropy. For $X\subset\cS_\cA$ we use 
$S^\omega(\rho,X)=\inf_{\sigma\in X}S^\omega(\rho,\sigma)$ and the 
$\omega$-closure from (\ref{def:rI-closure}).
\begin{Cor}
\label{cor:closures}
Let $\rho\in\cS$ and $X\subset\cS$. Then 
$S^\omega(\rho,X)=S^\omega(\rho,{\rm cl}^\omega(X))$ holds.
\end{Cor}
{\em Proof:\/}
For every state $\sigma\in{\rm cl}^\omega(X)$ there exists by
(\ref{def:rI-closure}) a sequence $(\sigma_i)_{i\in\bN}\subset X$, 
such that $\sigma_i\stackrel{C^\omega}{\longrightarrow}\sigma$.
Proposition~\ref{pro:continuous_1st_2nd} shows that the relative
entropies converge, 
$\lim_{i\to\infty}S^\omega(\rho,\sigma_i)=S^\omega(\rho,\sigma)$. 
Hence
\[\textstyle
S^\omega(\rho,X)
\;=\;
\inf_{\tau\in X}S^\omega(\rho,\tau)
\;\leq\;
\inf_{i\in\bN}S^\omega(\rho,\sigma_i)
\;\leq\;
\lim_{i\to\infty}S^\omega(\rho,\sigma_i)
\;=\;
S^\omega(\rho,\sigma)\,.
\]
Taking the infimum over all $\sigma\in{\rm cl}^\omega(X)$, we get
$S(\rho,X)\leq S(\rho,{\rm cl}^\omega(X))$. The converse inequality
is trivial.
\hspace*{\fill}$\Box$\\
\par
We now investigate the $\omega$-topology of the state space $\cS$. 
For $\rho\in\cS$ and $\epsilon\in(0,\infty]$ we use the open 
$\omega$-disk resp.\ closed $\omega$-disk
defined in (\ref{eq:open_disks}) resp.\ (\ref{eq:closed_disks})
and the face lattice $\cF$ of the state space $\cS$, introduced in 
\S\ref{ingredients:convex}. 
\begin{Thm}[Information topology and reverse information topology]~
\label{thm:info_top}
The convergence $C^\omega$ is an L*-convergence. In particular 
$C(\cT^\omega)=C^\omega$. The sequential closure (\ref{def:cl}) of 
$X\subset\cS$ equals the $\omega$-closure ${\rm cl}^\omega(X)$ from 
(\ref{def:rI-closure}),
\begin{align}
\label{eq:sq-closure}
{\rm cl}^\omega(X)
&\;=\;
\{\rho\in\cS\mid\lim_{i\to\infty}S^\omega(\rho,\rho_i)=0
\text{ for a sequence }(\rho_i)_{i\in\bN}\subset X\}\\\nonumber
&\;=\;
\{\rho\in\cS\mid\inf_{i\in\bN}S^\omega(\rho,\rho_i)=0
\text{ for a sequence }(\rho_i)_{i\in\bN}\subset X\}\\\nonumber
&\;=\;
\{\rho\in\cS\mid S^\omega(\rho,X)=0\}\,.
\end{align}
\begin{enumerate}[1.]
\item
The L*-convergence $C^\omega$ has unique limits. We have 
$C^\omega\subset C^{\|\cdot\|}$ and $\cT^\omega\supset\cT^{\|\cdot\|}$. 
In particular $\cT^\omega$ is a Hausdorff topology.
\item
For every $\rho\in\cS$ the mapping $\cS\to[0,\infty]$, 
$\sigma\mapsto S^\omega(\rho,\sigma)$ is continuous for $C^\omega$ and
$C^{\rm c}$ and continuous for $\cT^\omega$ and $\cT^{\rm c}$. 
\par
For each $\rho\in\cS$ and $\epsilon\in(0,\infty]$ 
the open $\omega$-disk $V^\omega(\rho,\epsilon)$ is $\cT^\omega$ open
and the closed $\omega$-disk $W^\omega(\rho,\epsilon)$ is
$\mathcal T^\omega$ closed. The open $\omega$-disks
$\{V^\omega(\rho,\epsilon)\mid\epsilon\in(0,\infty]\}$ are a base for 
$(\cS,\mathcal T^\omega)$ at $\rho$.
In particular $\cT^\omega$ is first countable and for any subset
$X\subset\cS$ we have $\cT^\omega|_X=\cT(C^\omega|_X)$.
\par
For any subset $X\subset\cS$ the sequential closure ${\rm cl}^\omega(X)$ 
is the $\cT^\omega$ closure of $X$.  
\item
Every term in the partition $\cS=\bigcup_{F\in\cF}{\ri}F$ is a 
$\cT^{\rm I}$ connected component of $\cS$. For all faces
$F\in\cF$ we have $C^{\rm I}|_{{\ri}F}=C^{\|\cdot\|}|_{{\ri}F}$
and $\cT^{\rm I}|_{{\ri}F}=\cT^{\|\cdot\|}|_{{\ri}F}$. 
\item
We have $\cT^{\|\cdot\|}\subset\cT^{\rm rI}\subset\cT^{\rm I}$ 
and $C^{\|\cdot\|}\supset C^{\rm rI}\supset C^{\rm I}$.
\item
The $\cT^{\rm rI}$ closure of ${\ri}\cS$ is $\cS$ and the topological space 
$(\cS,\cT^{\rm rI})$ is connected.
\end{enumerate}
\end{Thm}
{\em Proof:\/}
Both divergence functions $S^{\rm I}(\rho,\sigma)=S(\sigma,\rho)$ and 
$S^{\rm rI}(\rho,\sigma)=S(\rho,\sigma)$ defined for $\rho,\sigma\in\cS$ are 
divergence functions in the sense of Definition~\ref{def:divergence_function}.1. 
They satisfy condition A) and B) according to the Pinsker-Csisz\'ar inequality 
(\ref{eq:pinsker}) and Proposition~\ref{pro:continuous_1st_2nd}. So
Lemma~\ref{lem:special_diverge} proves the theorem up to part 2 inclusive.
\par
{\it We show part 3.} According to part 2, for 
every $\rho\in\cS$ the open I-disk of infinite radius is $\cT^{\rm I}$ 
open and has by Remark~\ref{rem:state_faces} the form
\begin{equation}
\label{eq:I-disk_infty}\textstyle
V^{\rm I}(\rho,\infty)
\;=\;
\{\sigma\in\cS\mid S(\sigma,\rho)<\infty\}
\;=\;
\{\sigma\in\cS\mid s(\sigma)\preceq s(\rho)\}
\;=\;
\bF(s(\rho))\,.
\end{equation}
By the lattice isomorphism $\bF:\cP\to\cF$ in Corollary~\ref{ss:iso} we 
obtain that every face $F$ of $\cS$ is $\cT^{\rm I}$ open. Let us show
that ${\ri}F$ is $\cT^{\rm I}$ open. The complement $\cS\setminus F$ of 
$F$ is $\cT^{\rm I}$ 
closed and the relative boundary ${\rm rb}F$ of $F$ is norm closed. By 
part 2 we have $\cT^{\|\cdot\|}\subset\cT^{\rm I}$ hence ${\rm rb}F$ is 
$\cT^{\rm I}$ closed as well. So
\[\textstyle
{\ri}F
\;=\;
\cS\setminus({\rm rb}F\cup(\cS\setminus F))
\]
is $\cT^{\rm I}$ open. Finally, the relative interior ${\ri}F$ is also 
$\cT^{\rm I}$ closed because by the stratification (\ref{eq:stratification}) 
we have ${\ri}F=\cS\setminus(\bigcup_{G\neq F}{\ri}G)$,
the union extending over faces $G\in\cF$.
\par
Let $F\in\cF$ be an arbitrary face. Since the relative interior of $F$
consists of states of constant support (see Remark~\ref{rem:state_faces}),
the relative entropy is norm continuous on ${\ri}F\times{\ri}F$.
Hence we have 
$C^{\|\cdot\|}|_{{\ri}F}\subset C^{\rm I}|_{{\ri}F}$. This shows 
$C^{\|\cdot\|}|_{{\ri}F}=C^{\rm I}|_{{\ri}F}$ as the converse inclusion 
follows from the Pinsker-Csisz\'ar inequality. With part 2 we have
\[\textstyle
\cT^{\rm I}|_{{\ri}F}
\;=\;
\cT(C^{\rm I}|_{{\ri}F})
\;=\;
\cT(C^{\|\cdot\|}|_{{\ri}F})
\;=\;
\cT^{\|\cdot\|}|_{{\ri}F}\,.
\]
\par
{\it We show part 4.} We begin with a proof of $\cT^{\rm rI}\subset\cT^{\rm I}$.
We first notice $C^{\rm rI}|_{{\ri}F}=C^{\rm I}|_{{\ri}F}$
for every face $F$ of $\cS$. This follows from 
$C^{\|\cdot\|}|_{{\ri}F}=C^{\rm I}|_{{\ri}F}$ proved in part 3 and from
$C^{\|\cdot\|}|_{{\ri}F}=C^{\rm rI}|_{{\ri}F}$, which can be proved 
analogously. Also
\[\textstyle
\cT(C^{\rm rI})|_{{\ri}F}
\;=\;
\cT(C^{\rm I})|_{{\ri}F}\,.
\]
Let $U\in\cT^{\rm rI}$. Then $U\cap{\ri}F\in\cT^{\rm rI}|_{{\ri}F}=
\cT^{\rm I}|_{{\ri}F}$ and since ${\ri}F$ is $\cT^{\rm I}$ open by
part~3, this shows $U\cap{\ri}F\in\cT^{\rm I}$. Now
$U=\bigcup_{F\in\cF}(U\cap{\ri}F)\in\cT^{\rm I}$ and we have proved
$\cT^{\rm rI}\subset\cT^{\rm I}$. Part 1 adds the inequality
$\cT^{\|\cdot\|}\subset\cT^{\rm rI}\subset\cT^{\rm I}$. Since these
three topologies arise from L*-convergences we get
$C^{\|\cdot\|}\supset C^{\rm rI}\supset C^{\rm I}$ from
Theorem~\ref{thm:kisynski}.
\par
{\it We show part 5.} We first show that any non-empty $\cT^{\rm rI}$ open 
set $U\subset\cS$ intersects $\ri\cS$. By part 2 the set $U$ contains
an open ${\rm rI}$-disk 
$V^{\rm rI}(\rho,\epsilon)=\{\sigma\in\cS\mid S(\rho,\sigma)<\epsilon\}$
for some density matrix $\rho$ and $\epsilon>0$. We show that 
$V^{\rm rI}(\rho,\epsilon)$ intersects $\ri\cS$. The relative interior
$\ri\cS$ consists of all invertible density matrices by 
Proposition~\ref{ss:st}. So for any fixed $\tau\in\ri\cS$ we have 
$S(\rho,\tau)<\infty$ and then (\ref{eq:continuous_straight}) implies
\[\textstyle
0\;=\;S(\rho,\rho)
\;=\;\lim_{\lambda\nearrow 1}S(\rho,(1-\lambda)\tau + \lambda\rho)\,,
\]
whence $(1-\lambda)\tau + \lambda\rho\in V^{\rm rI}(\rho,\epsilon)$
for $\lambda\nearrow 1$. Since $(1-\lambda)\tau + \lambda\rho\in\ri\cS$
is invertible for $\lambda<1$, this shows that $U$ intersects $\ri\cS$. 
\par
As shown in the previous paragraph, the relative boundary $\rb\cS$ does
not contain a $\cT^{\rm rI}$ open set so the $\cT^{\rm rI}$ closure of
$\ri\cS$ equals $\cS$. 
We show that $\cS$ is $\cT^{\rm rI}$ connected. By part 3 and 4 we have 
$\cT^{\rm rI}|_{\ri\cS}=\cT^{\|\cdot\|}|_{\ri\cS}$. The convex set $\ri\cS$ 
is connected in the norm topology hence in the $\cT^{\rm rI}$ topology. 
The claim follows since the closure of a connected set is connected, 
see e.g.\ \S{}IV.7 in \cite{Berge}.
\hspace*{\fill}$\Box$\\
\par
The following conditions have applications to exponential families in 
\S\ref{sec:non-commutative}.
\begin{Cor}
\label{cor:top}
If ${\rm dim}_\bC(\cA)>1$, then $C^{\rm I}\subsetneq C^{\rm rI}$,
$\cT^{\rm I}\supsetneq\cT^{\rm rI}$ and $\cS$ is not 
$\cT^{\rm I}$ compact. The following assertions are equivalent.\\[1mm]
\begin{tabular}{lcl}
\parbox{5mm}{1.}
$\cA$ is commutative, & \hspace{5mm} &
\parbox{5mm}{4.}
$C^{\rm rI}=C^{\|\cdot\|}$,\\
\parbox{5mm}{2.}
$\cT^{\rm I}$ is second countable, & &
\parbox{5mm}{5.}
$\cT^{\rm rI}=\cT^{\|\cdot\|}$,\\
\parbox{5mm}{3.}
$\cT^{\rm rI}$ is second countable, & &
\parbox{5mm}{6.}
$\cS$ is $\cT^{\rm rI}$ compact.
\end{tabular}
\end{Cor}
{\em Proof:\/}
{\it Item 1 implies 4.} If $\cA$ is commutative, then by (\ref{eq:direct_rep})
it is isomorphic to $\bC^n$. We can argue by convergence in components
of $\bC^n$ and find $C^{\|\cdot\|}=C^{\rm rI}$.
\par
{\it Statements in the headline.} If ${\rm dim}_\bC(\cA)>1$, then
by (\ref{eq:direct_rep}) $\cA$ contains a C*-subalgebra $\cB\cong\bC^2$ 
and by Example~\ref{ex:norm_coarseness} we have 
$C^{\rm I}|_{\cS_\cB}\subsetneq C^{\|\cdot\|}|_{\cS_\cB}$ while 
$C^{\|\cdot\|}|_{\cS_\cB}=C^{\rm rI}|_{\cS_\cB}$ was shown in the previous 
paragraph. Now the inclusion $C^{\rm I}\subset C^{\rm rI}$
in Theorem~\ref{thm:info_top}.4 shows
$C^{\rm I}\subsetneq C^{\rm rI}$. In terms of topology, since
$C^\omega=C(\cT^\omega)$ holds for $\omega\in\{{\rm I},{\rm rI}\}$
by Theorem~\ref{thm:info_top}.1,
we have also $\cT^{\rm rI}\subsetneq\cT^{\rm I}$.
\par
We show that $\cS$ is not $\cT^{\rm I}$ compact if ${\rm dim}_\bC(\cA)>1$. 
Theorem~\ref{thm:info_top}.3 shows
$({\ri}\cS,\cT^{\rm I}|_{{\ri}\cS})
=({\ri}\cS,\cT^{\|\cdot\|}|_{{\ri}\cS})$.
But $({\ri}\cS,\cT^{\|\cdot\|}|_{{\ri}\cS})$ is not a compact topological
space since ${\ri}\cS$ is the relative interior of a convex set of
dimension $>0$. Then $\cS$ is not $\cT^{\rm I}$ compact 
because ${\ri}\cS$ is its $\cT^{\rm I}$ connected component. 
\par
{\it Item 4 implies 5.} By definition $\cT^{\rm rI}=\cT(C^{\rm rI})$ and
$\cT^{\|\cdot\|}=\cT(C^{\|\cdot\|})$.
\par
{\it Item 5 implies 3 and 6.} By Proposition~\ref{ss:st} the state
space $\cS$ is a convex body, hence is a (norm) compact metric space.
On the other hand, a compact metric space is second countable, 
see e.g.\ \S{}V.4--5 in \cite{Berge}. Since
$\cT^{\rm rI}=\cT^{\|\cdot\|}$ is assumed, the state space is
$\cT^{\rm rI}$ compact and $\cT^{\rm rI}$ second countable.
\par
{\it Item 1 implies 2.} For every face $F\in\cF$ we have 
$\cT^{\rm I}|_{{\ri}F}=\cT^{\|\cdot\|}|_{{\ri}F}$ by 
Theorem~\ref{thm:info_top}.3. As shown in the previous paragraph,
$\cT^{\|\cdot\|}|_F$ is second countable. Since $\ri F$ is an
$\cT^{\|\cdot\|}|_F$ open subset of $F$, the topology 
$\cT^{\rm I}|_{{\ri}F}=\cT^{\|\cdot\|}|_{{\ri}F}$ is second
countable. The simplex $\cS$ is partitioned into finitely many relative 
interiors ${\ri}F$ of faces $F$ by (\ref{eq:stratification}). These sets 
are $\cT^{\rm I}$ connected components of $\cS$, so the proof 
is complete.
\par
{\it  Auxiliary calculation.} 
To show that each of items 2, 3 or 6 implies 1, we show that $\cS$
has an open cover, indexed by pure states, without a proper subcover. 
If the algebra $\cA$ is non-commutative, then this cover is uncountable.
By Remark~\ref{rem:state_faces} we can write for any state $\rho\in\cS$ the
open rI-disk of infinite radius in the form
\begin{equation}
\label{eq:open_rI}\textstyle
V^{\rm rI}(\rho,\infty)
\;=\;
\{\sigma\in\cS\mid s(\rho)\preceq s(\sigma)\}
\;=\;
\bigcup_{\substack{p\in\cP\\p\succeq s(\rho)}}{\ri}\bF(p)\,.
\end{equation}
Here $\cP$ denotes the projection lattice of $\cA$. The open rI-disks
are $\cT^{\rm rI}$ open by Theorem~\ref{thm:info_top}.2.
For pure state $p,q\in\cP\cap\cS$ we have
\[\textstyle
p
\;\not\in\;
V^{\rm rI}(q,\infty)\quad\text{ if }p\neq q\,.
\]
If $\cA$ is non-commutative then $\cA$ contains a C*-subalgebra isomorphic 
to ${\rm Mat}(2,\bC)$, see (\ref{eq:direct_rep}), hence 
$\cP\cap\cS$ is uncountable.
\par
{\it Item 6 implies 1.} The open cover  
$\bigcup_{p\in\cP\cap\cS}V^{\rm rI}(p,\infty)$ of $\cS$ has no 
finite subcover.
\par
{\it Item 3 implies 1.} If $\cB$ is a base of $\cT^{\rm rI}$, then for all
$p\in\cP\cap\cS$ there is a $\cT^{\rm rI}$ open set $U_p\in\cB$
such that $p\in U_p\subset V^{\rm rI}(p,\infty)$. The map
$\cP\cap\cS\to\cB$, $p\mapsto U_p$ is injective. This prove that $\cB$
is not countable.
\par
{\it Item 2 implies 1.} Theorem~\ref{thm:info_top}.4 shows
$\cT^{\rm rI}\subset\cT^{\rm I}$ so the arguments in the previous 
paragraph apply unmodified.
\hspace*{\fill}$\Box$\\
%
%
%
%
\section{Exponential families}
\label{sec:exp}
\par
We study an exponential family $\cE$ in a C*-subalgebra $\cA$ of ${\rm Mat}(n,\bC)$. 
The analysis is based on the mean value parametrization of $\cE$, developed in 
\S\ref{sec:charts}. The family $\cE$ of states is defined by the real analytic 
function (\ref{def:R})
\[\textstyle
R_\cA:\;
\sa\;\to\;\sa\,,
\quad
R(\theta)
\;=\;
R_\cA(\theta)
\;=\;
\exp_\cA(a)/\tr(\exp_\cA(a))\,.
\]
We consider a non-empty affine subspace $\Theta\subset\sa$ of self-adjoint matrices, 
its translation vector space
\[\textstyle
U
\;:=\;
\lin(\Theta)
\;=\;
\Theta-\Theta
\]
and we define the exponential family $\cE:=R_\cA(\Theta)$. In \S\ref{sec:exted} we 
define an extension ${\rm ext}(\cE)$ of $\cE$. And we prove the bijection
$\pi_U|_{{\rm ext}(\cE)}:{\rm ext}(\cE)\to\bM(U)$ to the mean value set 
$\bM(U)=\pi_U(\cS)$, which is a projection of the state space (\ref{def:m_set}). Then 
we prove the Complete Pythagorean theorem in \S\ref{sec:pytha}. The Complete projection 
theorem is proved in \S\ref{sec:project}. Application to quantum correlations are 
described in \S\ref{sec:max}. In \S\ref{sec:non-commutative} we discuss 
necessary conditions for commutativity of the algebra $\cA$.
%
%
%
%
%
%
%
%
\subsection{The mean value chart}
\label{sec:charts}
\par
We settle the mean value chart of an exponential family. Its inverse is the real 
analytic mean value parametrization. The mean value chart was established for 
linear spaces $\Theta=U$ in \cite{Wichmann}. Examples are shown in 
Figure~\ref{fig:two_families}.
\par
Restrictions to affine subspaces of $\sa$ are the rule in subsequent 
arguments, hence we accept relatively open convex subsets of $\sa$
(in place of open subset of $\mathbb R^d$) as domains of
differentiable maps and as ranges of diffeomorphisms and charts.
We recall that the relative interior of the state space consists 
of all invertible states,
\begin{equation}
\label{eq:ris}\textstyle
\ri\cS
\;=\;
\{\rho\in\cS\mid\rho^{-1}\text{ exists in }\cA\}\,,
\end{equation}
and is (norm) open in $\ao=\{a\in\sa\mid\tr(a)=1\}$, see e.g\ \S2.3 in 
\cite{Weis_supp}.
\begin{Pro}\label{pro:mean_value_chart}
Let $\id\not\in U$ for the multiplicative identity $\id$ of $\cA$.
\begin{enumerate}[1.]
\item
The set $\pi_U(\cE)$ is open relative to $U$ and 
$\pi_U\circ R_\cA|_{\Theta}:\Theta\to\pi_U(\cE)$ is a real 
analytic diffeomorphism. 
\item
If $\Theta$ has codimension one in $\sa$, then
$R_\cA|_\Theta:\Theta\to\ri\cS$ is a real analytic diffeomorphism.
\item
The bijections
$(R_\cA|_\Theta)^{-1}:\cE\to\Theta$ and
$\pi_U|_\cE:\cE\to\pi_U(\cE)$ are global charts for $\cE$ and 
$(\pi_U|_\cE)^{-1}:\pi_U(\cE)\to\cE$ is real analytic.
\end{enumerate}
\end{Pro}
{\em Proof:\/}
{\it Part 1.} The derivative of $\sa\to\bR$, $a\mapsto\tr\exp_\cA(a)$ can
be computed from (\ref{eq:lieb_derivative}) using cyclic reordering
under the trace. For $a,u\in\sa$ we have
\[\textstyle
\frac{\partial}{\partial t}|_{t=0}\tr\exp_\cA(a+tu)
\;=\;
\langle u,\exp_\cA(a)\rangle\,.
\]
Hence the free energy (\ref{def:free_energy}) has the
derivative at $\theta\in\sa$ in the direction $u\in\sa$ 
\begin{equation}
\label{eq:dF}\textstyle
\frac{\partial}{\partial t}|_{t=0}F_\cA(\theta+tu)
\;=\;
\langle u,R_\cA(\theta)\rangle\,.
\end{equation}
From the product rule and (\ref{eq:lieb_derivative}) we get
\begin{equation}
\label{eq:dR}\textstyle
\frac{\partial}{\partial t}|_{t=0}R_\cA(\theta+tu)
\;=\;
\int_0^1 R_\cA(\theta)^{1-y} u R_\cA(\theta)^y{\rm d}y
-\langle u,R_\cA(\theta)\rangle R_\cA(\theta)\,.
\end{equation}
For $\theta,u,v\in\sa$ we consider the real symmetric bilinear form
\begin{equation}
\label{def:ddF}\textstyle
\langle\!\langle u,v\rangle\!\rangle_\theta
\;:=\;
\frac{\partial^2}{\partial s\partial t}|_{s=t=0}F_\cA(\theta+su+tv)\,.
\end{equation}
If restricted to $\theta\in\Theta$ and $u,v\in U$ this bilinear form
is called BKM-metric, see Remark~\ref{rem:bkm}. We recall the
well-known fact that it defines a Riemannian matric. We obtain from
(\ref{eq:dF}) and 
(\ref{eq:dR})
\begin{equation}
\label{eq:ddF}\textstyle
\langle\!\langle u,v\rangle\!\rangle_\theta
\;=\;
\langle u,\frac{\partial}{\partial t}|_{t=0}R_\cA(\theta+tv)\rangle
\;=\;
\int_0^1\big\langle
\xi(u,y),\xi(v,y)\big\rangle{\rm d}y
\end{equation}
with the not necessarily self-adjoint matrix
\[\textstyle
\xi(u,y)
\;:=\;
R_\cA(\theta)^{\tfrac y{2}}
\left[\,u-\langle u,R_\cA(\theta)\rangle\,\id\right]
R_\cA(\theta)^{\tfrac{1-y}{2}}\,.
\]
We have $\langle\!\langle u,u\rangle\!\rangle_\theta>0$ unless $u\in\sa$ 
is a (real) scalar multiple of $\id$. Hence
$\langle\!\langle\cdot,\cdot\rangle\!\rangle_\theta$ is a non-degenerate 
bilinear form on $U$.
\par
Since $R|_{\Theta}$ is real analytic, the composition 
$\pi_U\circ R|_{\Theta}$ with the orthogonal projection to $U$ is also
real analytic. If $\{u_i\}_{i=1}^k$ is an orthonormal 
basis of $U$ then the directional derivative at $\theta\in\Theta$  along 
$u\in U$ is by (\ref{eq:ddF})
\begin{equation}
\label{eq:dChart_Change}\textstyle
\frac{\partial}{\partial t}|_{t=0}\pi_U\circ R(\theta+tu)
\;=\;
\pi_U\left(\frac{\partial}{\partial t}|_{t=0} R(\theta+tu)\right)
\;=\;
\sum_{i=1}^k \langle\!\langle u,u_i\rangle\!\rangle_\theta u_i\,.
\end{equation}
Since $\langle\!\langle\cdot,\cdot\rangle\!\rangle_\theta$ is 
non-degenerate on $U$, the Jacobian of $\pi_U\circ R|_{\Theta}$ is 
invertible everywhere. Then the inverse function theorem implies that 
$\pi_U\circ R|_{\Theta}$ is locally invertible and its local inverses are 
real analytic functions, see e.g.\ \S2.5 in \cite{Krantz}. This implies that
the image $\pi_U\circ R(\Theta)$ is an open subset relative to $U$.
\par
The global injectivity of $\pi_U\circ R|_{\Theta}$ follows from the projection 
theorem (\ref{thm:projection_E}): If there are $\theta,\theta'\in\Theta$ such 
that $\pi_U\circ R(\theta)=\pi_U\circ R(\theta')$, then $R(\theta)=R(\theta')$.
Taking the logarithm on both sides one has
$\theta-F(\theta)\id=\theta'-F(\theta')\id$ so the difference 
$\theta-\theta'$ is proportional to $\id$. Hence $\theta=\theta'$ by
the assumption $\id\not\in U$. 
\par
{\it Part 2.} If $\Theta=\az$ is the space of traceless matrices, then
\begin{equation}
\label{def:L}\textstyle
\log_0:\;\ri\cS\;\to\;\az\,,\quad 
\rho\;\mapsto\;\log(\rho)-\tfrac{\tr\log(\rho)}{\tr\id}\id
\end{equation}
is inverse to $R|_{\az}$ and this shows $R(\Theta)=\ri\cS$. Since
$R(\theta+\id)=R(\theta)$ for all $\theta\in\sa$, we have
$R(\Theta)=\ri\cS$ for every affine subspace $\Theta\subset\sa$ of
codimension one and with $\id\not\in\lin(\Theta)$. 
\par
{\it Part 3.} By virtue of the real analytic diffeomorphism in 1 it is 
sufficient to prove that $R_\cA|_{\Theta}:\Theta\to\cE$ is a real
analytic bijection. The function $R_\cA$ is real analytic by definition
and $R_\cA|_{\Theta}$ is invertible on $\cE$ by 2.
\hspace*{\fill}$\Box$\\
\begin{Rem}[The BKM-metric]
\label{rem:bkm}
If $\Theta$ has codimension one in $\sa$ and if $\id\not\in U$ then 
$\Theta\cong\ri\cS$. On this manifold the family (\ref{eq:ddF}) of scalar 
products, parametrized by $\theta\in\Theta$ and defined for $u,v\in U$ by
\[\textstyle
\langle\!\langle u,v\rangle\!\rangle_\theta
\;=\;
\langle u,\frac{\partial}{\partial t}|_{t=0}R_\cA(\theta+tv)\rangle\,,
\]
is a Riemannian metric, called {\it BKM-metric} (an acronym for 
Bogoliubov, Kubo and Mori, see e.g.\ \cite{Amari_Nagaoka,Petz08}). 
\par
Indeed, by Proposition~\ref{pro:mean_value_chart}.2 the map
$R_\cA|_\Theta:\Theta\to\ri\cS$ is a diffeomorphism. For $\theta\in\Theta$
and $u\in U$ let us use the curve $\gamma_{\theta,u}:\bR\to\Theta$,
$t\mapsto \theta+tu$ to represent a tangent vector at the footpoint 
$\theta$. The $(-1)$-representation $u^{(-1)}$ of $u$ is by definition 
taken in the identity chart ${\rm id}:\ri\cS\to\ri\cS$, so
\[\textstyle
u^{(-1)}
\;=\;
\frac{\partial}{\partial t}|_{t=0}R_\cA(\theta+tu)\,.
\]
The $(+1)$-representation $u^{(+1)}$ of $u$ is by definition taken in 
the logarithmic representation $\log:\ri\cS\to\sa$, so
\[\textstyle
u^{(+1)}
\;=\;
D\log_\cA(u^{(-1)})
\;=\;
\frac{\partial}{\partial t}|_{t=0}\log_\cA\circ R_\cA(\theta+tu)
\;=\;
u+\lambda\id
\] 
for some $\lambda\in\bR$. Since $v^{(-1)}$ has trace zero, we arrive
at the {\it mixed representation}
$\langle\!\langle u,v\rangle\!\rangle_\theta=\tr(u^{(+1)}v^{(-1)})$
of the BKM-metric, see e.g.\ \cite{Grasselli}.
\end{Rem}
\par
Let us calculate the range of the chart $\pi_U|_\cE$. The following
statement gives us an upper bound on the norm closure $\overline{\cE}$. 
It is used implicitly in Lemma~7 in \cite{Wichmann}.
\begin{Pro}
\label{pro:pert_R}
Let $(x_i)_{n\in\bN}\subset\Theta$ and assume that the states 
$R_\cA(x_i)$, $i\in\bN$, converge in norm to the state $\rho$. If 
$\rho\not\in\cE$ then $s(\rho)\preceq p_\cA^+(u)$ holds for every 
accumulation point of $(\frac{x_i}{\|x_i\|})_{i\in\bN}$.
\end{Pro}
{\em Proof:\/}
If $x_i$ has a bounded subsequence, then by continuity of $R_\cA$
we have $\rho=\lim_{i\to\infty}R_\cA(x_i)\in\cE$. So we can assume
$\|x_i\|\stackrel{i\to\infty}{\rightarrow}\infty$. By selecting a 
subsequence let us choose an accumulation point
\[\textstyle
u
\;:=\;
\lim_{i\to\infty}\frac{x_i}{\|x_i\|}
\;\in\;
U\,.
\]
To apply Lemma~\ref{lem:pert_exp}.1 we need to bound the free energy 
(\ref{def:free_energy}). Let $\lambda^-_\cA(a)$ resp.\ $\lambda^+_\cA(a)$
denote the smallest resp.\ largest spectral value of a self-adjoint matrix
$a\in\sa$. Then 
\[\textstyle
\log(\tr\id)+\lambda^-_\cA(a)
\;\leq\;
F(a)
\;\leq\;
\log(\tr\id)+\lambda^+_\cA(a)
\]
and it implies $|F(a)|\leq \log(\tr\id) + \|a\|$ for the spectra norm
$\|\cdot\|$. 
\par
From the bounded sequence $(\frac{F(x_i)}{\|x_i\|})_{i\in\bN}$
we select another subsequence, such that
\[\textstyle
\lambda
\;:=\;
\lim_{i\to\infty}\frac{F(x_i)}{\|x_i\|}
\;\in\;
\bR
\]
converges. Defining $y_i:=x_i-F(x_i)\id$ gives 
$e^{y_i}=R(x_i)\stackrel{i\to\infty}{\rightarrow}\rho$ and
\[\textstyle
\frac{y_i}{\|y_i\|}
\;=\;
(\tfrac{x_i}{\|x_i\|}-\tfrac{F(x_i)}{\|x_i\|}\id)/
\|\tfrac{x_i}{\|x_i\|}-\tfrac{F(x_i)}{\|x_i\|}\id\|
\;\stackrel{i\to\infty}{\rightarrow}\;
\frac{u-\lambda\id}{\|u-\lambda\id\|}\,.
\]
Since $(u-\lambda)/\|u-\lambda\|$ has the same maximal projection as
$u$, the claim follows from Lemma~\ref{lem:pert_exp}.1.
\hspace*{\fill}$\Box$\\
\par
The following statement is an idea from Theorem 2~b) in
\cite{Wichmann}. 
\begin{Lem}
\label{lem:topology_mc}
Let $f:V\to W$ be a continuous map between two finite-dimensional
real vector spaces. Let $K\subset V$
be non-empty and bounded, $L\subset W$ be connected and
$f(K)\subset L$. If $f(K)$ is open and
$f(\overline{K}\setminus K)\cap L=\emptyset$, then $f(K)=L$.
\end{Lem}
{\em Proof:\/}
Since $f(\overline{K}\setminus K)\cap L=\emptyset$ we have
$L\setminus f(K)=L\setminus f(\overline{K})
=(W\setminus f(\overline{K}))\cap L$ and 
$f(\overline{K})\cap L=f(K)\cap L$, hence
\begin{equation}
\label{eq:disconnect}\textstyle
L
\;=\;
\left(f(\overline{K})\cap L\right)\cup
\left(L\setminus f(\overline{K})\right)
\;=\;
\left(f(K)\cap L\right)\cup
\left((W\setminus f(\overline{K}))\cap L\right)\,.
\end{equation}
The set $f(K)$ is open in $W$ by assumption and since $f(\overline{K})$
is compact $W\setminus f(\overline{K})$ is open in $W$. Since 
$f(K)\cap L\neq\emptyset$ by assumption, (\ref{eq:disconnect})
is a disconnection of $L$ unless $L\setminus f(K)=\emptyset$. 
Since $L$ is connected by assumption, $f(K)\supset L$ follows.
\hspace*{\fill}$\Box$\\
\par
We have collected all arguments needed to compute $\pi_U(\cE)$.
The mean value set $\bM(U)$ plays a crucial role (\ref{def:m_set}).
\begin{Thm}
\label{thm:mean_value_chart}
Let $\id\not\in U$. Then $\ri\bM_\cA(U)$ is open in the norm topology
of $U$ and the chart change
$\pi_U\circ R_\cA|_{\Theta}:\Theta\to\ri\bM_\cA(U)$ is a real analytic 
diffeomorphism. We have 
$\pi_U(\overline{\cE}\setminus\cE)=\rb\bM_\cA(U)$.
\end{Thm}
{\em Proof:\/}
The map $\pi_U\circ R|_{\Theta}:\Theta\to\pi_U(\cE)$ is a
real analytic diffeomorphism by Proposition~\ref{pro:mean_value_chart}.1
and $\pi_U(\cE)$ is open relative to $U$. We shall first show
\begin{equation}
\label{eq:boundary_incl}\textstyle
\pi_U(\overline{\cE}\setminus\cE)
\;\subset\;
\rb\bM(U)\,.
\end{equation}
\par
Let $\rho\in\overline{\cE}\setminus\cE$. Proposition~\ref{pro:pert_R}
shows that the support projection of $\rho$ satisfies
$s(\rho)\preceq p^+(u)$ for a non-zero $u\in U$ and 
Proposition~\ref{ss:st} shows that $\rho$ lies in the exposed face 
$\bF(p^+(u))=F_\perp(\cS,u)$ of the state space. Then
Lemma~\ref{lem:lift_iso} shows that $\pi_U(\rho)$ lies in the exposed
face $F_\perp(\bM(U),u)$ of the mean value set. The mean value set 
$\bM(U)$ has non-empty interior because it contains $\pi_U(\cE)$
and then Theorem 13.1 in \cite{Rockafellar} proves that the exposed
face $F_\perp(\bM(U),u)$ is included in the boundary of $\bM(U)$.
This proves $\pi_U(\rho)\in\rb\bM(U)$.
\par
In order to prove that 
$\pi_U\circ R|_{\Theta}:\Theta\to\ri\bM(U)$ is a real analytic 
diffeomorphism it suffices to prove $\pi_U(\cE)=\ri\bM(U)$. The convex 
body $\bM(U)$ is the projection of the whole state space, so 
$\pi_U(\cE)\subset\bM(U)$. But $\cE\subset\ri(\cS)$ holds by
(\ref{eq:ris}) and thanks to the equality
$\pi_U\circ\ri(\cS)=\ri\circ\pi_U(\cS)$ (see e.g.\ Theorem 6.6
in \cite{Rockafellar}) we have $\pi_U(\cE)\subset\ri\bM(U)$. 
We meet the conditions of Lemma~\ref{lem:topology_mc} with
\[\textstyle
V=\sa,\quad
W=U,\quad
f=\pi_U,\quad
K=\cE\quad 
\text{and}\quad
L=\ri\bM(U)\,.
\]
Indeed, $K=\cE\subset\cS$ is non-empty and bounded, the convex set 
$L=\ri\bM(U)$ is connected. We have proved above that 
$f(K)=\pi_U(\cE)$ is included in $L=\ri\bM(U)$ and $f(K)=\pi_U(\cE)$
is open relative to $W=U$. Moreover 
$\pi_U(\overline{\cE}\setminus\cE)\subset\rb\bM(U)$ in
(\ref{eq:boundary_incl}) implies
\[\textstyle
f(\overline{K}\setminus K)\cap L
\;=\;
\pi_U(\overline{\cE}\setminus\cE)\cap\ri\bM(U)
\;\subset\;
\rb\bM(U)\cap\ri\bM(U)
\;=\;\emptyset\,.
\]
Then $\pi_U(\cE)=\ri\bM(U)$ follows from Lemma~\ref{lem:topology_mc}.
\par
Finally we show $\pi_U(\overline{\cE}\setminus\cE)=\rb\bM(U)$.
Since $\overline{\cE}\subset\cS$ is compact and $\pi_U(\cE)=\ri\bM(U)$ 
we have
\[\textstyle
\bM(U)
\;=\;
\overline{\pi_U(\cE)}
\;\subset\;
\pi_U(\overline{\cE})
\;\subset\;
\bM(U)
\]
hence $\pi_U(\overline{\cE})=\bM(U)$. Then $\pi_U(\cE)=\ri\bM(U)$ 
proves $\pi_U(\overline{\cE}\setminus\cE)\supset\rb\bM(U)$. The
opposite inclusion is (\ref{eq:boundary_incl}).
\hspace*{\fill}$\Box$\\
\par
In the sequel $\id\in U$ will naturally occur in our constructions.
Let us drop the condition $\id\not\in U$. We use the spaces $\az$ and
$\ao$ from Definition~\ref{def:sfaces}.
\begin{Cor}
\label{cor:mean_chart}
The map $\pi_U|_{\cE}:\cE\to\ri\bM(U)$ is a bijection and the inverse
$(\pi_U|_{\cE})^{-1}:\ri\bM(U)\to\cE$ is real analytic.
\end{Cor}
{\em Proof:\/}
If $\id\not\in U$ then the claim follows from
Theorem~\ref{thm:mean_value_chart} and 
Proposition~\ref{pro:mean_value_chart}.3.
We assume $\id\in U$ and we define $\Theta_0:=\pi_\az(\Theta)$
and $U_0:=\pi_\az(U)$. We have $U_0=\lin(\Theta_0)$ and since
$\az^\perp=\id\bR$ holds, we have $U=U_0+\id\bR$ and
$\Theta=\Theta_0+\id\bR$. Clearly
$\cE=R_\cA(\Theta)=R_\cA(\Theta_0)$ and 
$(\pi_{U_0}|_\cE)^{-1}:\ri\bM(U_0)\to\cE$ is a real analytic
bijection because $\id\not\in U_0$. We have
$\pi_U=\pi_{U_0}+\pi_{\id\bR}$ and
\[\textstyle
\pi_U|_{\ao}
\;=\;
\pi_{U_0}|_{\ao}+\tfrac 1{\tr\id}\id\,.
\]
So $\ri\bM(U)=\ri\bM(U_0)+\tfrac 1{\tr\id}\id$ holds and for 
$u\in\ri\bM(U)$ the equality
$(\pi_U|_\cE)^{-1}(u)=(\pi_{U_0}|_\cE)^{-1}(u-\tfrac 1{\tr\id}\id)$
completes the proof.
\hspace*{\fill}$\Box$\\
\begin{Def}
\label{def:mean_para}
We call the continuous bijection $\pi_U|_{\cE}:\cE\to\ri\bM_\cA(U)$
in Corollary~\ref{cor:mean_chart} the {\it mean value chart} of $\cE$.
The real analytic inverse 
\begin{equation}
\label{eq:mean:par}\textstyle
(\pi_U|_{\cE})^{-1}:\;\ri\,\bM_\cA(U)\;\to\;\cE
\end{equation}
is the {\it mean value parametrization} of $\cE$.
\end{Def}
%
%
%
%
%
%
%
%
%
%
%
%
%
%
%
%
\subsection{The extension of an exponential family}
\label{sec:exted}
\par
We define an extension ${\rm ext}(\cE)$ of $\cE$ composed of exponential families
in compressed algebras $p\cA p$ for orthogonal projections $p$, one for each face 
of the mean value set $\bM(U)$ (of the vector space $U$). We obtain a bijective
mean value parametrization
\[\textstyle
(\pi_U|_{{\rm ext}(\cE)}):\;\bM(U)\;\to\;{\rm ext}(\cE)\,.
\]
We obtain a projection with linear fibers
\[\textstyle
\pi_\cE:\;\cS_\cA\;\to\;{\rm ext}(\cE)
\]
from the state space $\cS_\cA$ to the extension. We finish by providing 
analogues of the {\it natural} and {\it canonical} parameters known in 
statistics.
\par
We recall from (\ref{eq:projection_compression}) for $p\in\cP$ and
$a\in\sa$ the projection to the compressed algebra $p\cA p$
\[\textstyle
c^p:\;\sa\;\to\;(p\cA p)_{\rm sa}\,,\quad
a\;\mapsto\;pap\,.
\]
Only the projections in the lattice $\cP^U$ are interesting. Through
(\ref{eq:main_iso}) they correspond to the faces the mean value set $\bM(U)$ 
and they can in principle be computed by spectral analysis, see 
Remark~\ref{rem:pu}. See also \S\ref{sec:poonem-algorithm} about the 
construction of $\cP^U$.
\par
Let us investigate the extension defined in (\ref{def:ext}).
\begin{Def}
\label{def:extension}
For orthogonal projections $p\in\cP$  we consider the exponential family 
\[\textstyle
\cE_p
\;:=\;
R_{p\cA p}(c^p(\Theta))\,.
\]
The {\it extension} of $\cE$ is defined in terms of the projection
lattice $\cP^U$ 
\[\textstyle
{\rm ext}(\cE)
\;:=\;
\bigcup_{p\in\cP^U\setminus\{0\}}\cE_p\,.
\]
\end{Def}
\par
We begin by parametrizing the extension ${\rm ext}(\cE)$ of the 
exponential family $\cE=R_\cA(\Theta)$ by mean values. Let 
$\theta_0,u_1,\ldots,u_k\in\sa$ such that
$\Theta:=\theta_0+{\rm span}_\bR(u_1,\ldots,u_k)$ and denote
${\bf u}=(u_1,\ldots,u_k)$. Then in the notation of (\ref{def:E}) the 
translation vector space of $\Theta$ is $U={\rm span}_\bR(u_1,\ldots,u_k)$.
The following parametrization by vectors in $\bM(U)$ has a formulation
in terms of the mean value map $m_{\bf u}$ and the convex support 
${\rm cs}({\bf u})$, defined respectively in (\ref{def:mean_value_map}) 
and (\ref{def:cs}).
\begin{Lem}
\label{lem:ext-m}
The projection
$\pi_U|_{{\rm ext}(\cE)}:{\rm ext}(\cE)\to\bM(U)$ is a bijection.
The mean value map
$m_{\bf u}|_{{\rm ext}(\cE)}:{\rm ext}(\cE)\to{\rm cs}({\bf u})$
is a bijection.
\end{Lem}
{\em Proof:\/}
For every non-zero projection $p\in\cP$, the mean value chart in
Corollary~\ref{cor:mean_chart} proves the bijection
\[\textstyle
\pi_{c^p(U)}|_{\cE_p}:\;\cE_p
\;\to\;
\ri\bM_{p\cA p}(c^p(U))\,.
\]
Using the third diagram in Lemma~\ref{lem:double_projection} we
have the bijection
\begin{equation}
\label{eq:bi_p}\textstyle
\pi_U|_{\cE_p}:\;\cE_p
\;\to\;
\ri\pi_U(\bF_\cA(p))\,.
\end{equation}
The map $\cP^U\to\cF(\bM(U))$, $p\mapsto\pi_U(\bF_\cA(p))$ is
a lattice isomorphism from the projection lattice $\cP^U$ to the face
lattice of $\bM(U)$, see (\ref{eq:main_iso}).
Then the stratification (\ref{eq:stratification}) of $\bM(U)$ into
the relative interiors of its faces shows that the bijections 
(\ref{eq:bi_p}) assemble to a bijection ${\rm ext}(\cE)\to\bM(U)$.
The second claim follows from (\ref{eq:mean_iso}).
\hspace*{\fill}$\Box$\\
\par
We renew and extend the definition of $\pi_\cE$ from (\ref{eq:pro_thm}).
\begin{Def}
The projection to the extension ${\rm ext}(\cE)$ is well-defined by 
Lemma~\ref{lem:ext-m} as the map
\begin{equation}
\label{def:pi_ce}\textstyle
\pi_\cE:\;\cS_\cA\;\to\;{\rm ext}(\cE)\,,\quad
\rho\;\mapsto\;(\pi_U|_{{\rm ext}(\cE)})^{-1}\circ\pi_U(\rho)\,.
\end{equation}
\end{Def}
\par
Coordinates can be put on the canonical parametrization as well. They 
depend on the projection lattice $\cP^U$. The free energy $F$, defined in 
(\ref{def:free_energy}), relates them to the mean value coordinates.
\begin{Cor}
\label{cor:rI-coords}
For every $\rho\in{\rm ext}(\cE)$ and $x:=m_{\bf u}(\rho)\in{\rm cs}(\bf u)$
there exists a unique projection $p\in\cP^U\setminus\{0\}$ and some 
(in general non-unique) $\lambda_1,\ldots,\lambda_k\in\bR$, such that
\begin{equation}
\label{eq:rI-coords}\textstyle
\left\{\tfrac{\partial}{\partial\lambda_j}\,
F_{p\cA p}(p(\theta_0+\sum_{i=1}^k\lambda_iu_i)p)\right\}_{j=1}^k
\;=\;
x\,.
\end{equation}
Conversely, if $x\in{\rm cs}({\bf u})$, $p\in\cP^U\setminus\{0\}$
and $(\lambda_1,\ldots,\lambda_k)\in\bR^k$ solve (\ref{eq:rI-coords}), then 
the state $\rho:=R_{p\cA p}(p(\theta_0+\sum_{i=1}^k\lambda_iu_i)p)$ is the
unique state in ${\rm ext}(\cE)$ such that $m_{\bf u}(\rho)=x$.
\end{Cor}
{\em Proof:\/}
If $\rho\in{\rm ext}(\cE)$ then by definition of ${\rm ext}(\cE)$ there exist 
$p\in\cP^U$ and $\lambda_1,\ldots,\lambda_k\in\bR$ such that for 
$\theta:=\theta_0+\sum_{i=1}^k\lambda_iu_i$ we have 
$\rho=R_{p\cA p}(c^p(\theta))$. 
Since $s(\rho)=p$, the projection $p$ is unique. The derivative (\ref{eq:dF}) 
of the free energy is
\[\textstyle
\tfrac{\partial}{\partial\lambda_j}F_{p\cA p}
(c^p(\theta))
\;=\;
\langle c^p(u_j),R_{p\cA p}(c^p(\theta))\rangle
\;=\;
\langle u_j,R_{p\cA p}(c^p(\theta))\rangle\,,
\qquad 
j=1,\ldots,k\,.
\]
The collection of these $k$ equations solves the existence part of
the claim:
\begin{equation}\textstyle
\label{eq:d_F_m}
\left\{\tfrac{\partial}{\partial\lambda_j}
\,F_{p\cA p}(c^p(\theta))\right\}_{j=1}^k
\;=\;
m_{\bf u}(R_{p\cA p}(c^p(\theta)))
\;=\;
x\,.
\end{equation}
Conversely, if $x\in{\rm cs}({\bf u})$, $p\in\cP^U\setminus\{0\}$
and $(\lambda_1,\ldots,\lambda_k)\in\bR^k$ solve (\ref{eq:rI-coords}), then
the state $\rho:=R_{p\cA p}(p(\theta_0+\sum_{i=1}^k\lambda_iu_i)p)$ has, by 
(\ref{eq:d_F_m}), the mean values $m_{\bf u}(\rho)=x$. By definition of 
the extension ${\rm ext}(\cE)$ the state $\rho$ belongs to ${\rm ext}(\cE)$ 
and Lemma~\ref{lem:ext-m} shows that $\rho$ is unique in ${\rm ext}(\cE)$ 
with the mean value $m_{\bf u}(\rho)=x$.
\hspace*{\fill}$\Box$\\
\par
The parameters $(\lambda_1,\ldots,\lambda_k)\in\bR^k$ resp.\
$m_{\bf u}(\rho)\in\bR^k$ are analogues of {\it canonical parameters} 
resp.\ {\it natural parameters} of ${\rm ext}(\cE)$ in statistics, see 
\S20 in \cite{Cencov}. In the restriction to $p=\id$, they put coordinates 
on the canonical parametrization (\ref{def:E}) resp.\ on the mean value 
parametrization (\ref{eq:mean:par}). Notice that the non-uniqueness of 
canonical parameters $\lambda_1,\ldots,\lambda_k$ can not be resolved by
choosing linearly independent observables $\id,u_1,\ldots,u_k$. Even if
$(\lambda_1,\ldots,\lambda_k)\mapsto
R_{p\cA p}(p(\theta_0+\sum_{i=1}^k\lambda_iu_i)p)$ is a bijection
for $p=\id$, this will no longer be true for smaller projections $p\neq\id$ 
where linear independence of $p,pu_1p,\ldots,pu_kp$ can be lost.
%
%
%
%
%
%
\subsection{The Complete Pythagorean theorem}
\label{sec:pytha}
\par
We prove the Complete Pythagorean theorem for an exponential family $\cE$. 
It is applied  to the maximization of the von Neumann entropy in 
\S\ref{sec:informal-pythagorean}.
\par
We use the projection $\pi_\cE:\cS_\cA\to{\rm ext}(\cE)$, defined in 
(\ref{def:pi_ce}), with linear fibers parallel to $U^\perp$.
\begin{Thm}[Complete Pythagorean theorem]
\label{thm:pyth_with_proof}
If $\rho\in\cS_\cA$ and if $\sigma$ lies in the extension ${\rm ext}(\cE)$, 
then $S(\rho,\pi_\cE(\rho))+S(\pi_\cE(\rho),\sigma)=S(\rho,\sigma)$ holds.
\end{Thm}
{\em Proof:\/}
By Definition~\ref{def:extension} of the extension ${\rm ext}(\cE)$
there exist projections $p,q\in\cP^U$, such that $\pi_\cE(\rho)\in\cE_p$
and $\sigma\in\cE_q$. If $p\not\preceq q$ then $s(\rho)\not\preceq q$
follows from Lemma~\ref{lem:stratum} since $q\in\cP^U$. We get
\[\textstyle
S(\pi_\cE(\rho),\sigma)
\;=\;
S(\rho,\sigma)
\;=\;
\infty
\]
and the non-negativity of the relative entropy proves the claim.
\par
Let $p\preceq q$. Then by Lemma~\ref{lem:stratum} we have
$s(\rho)\preceq p=s(\pi_\cE(\rho))\preceq q=s(\sigma)$ and only 
finite relative entropies appear in the claimed equation. We subtract
the trivial equation
\[\textstyle
S(\pi_\cE(\rho),\pi_\cE(\rho))+S(\pi_\cE(\rho),\sigma)
\;=\;
S(\pi_\cE(\rho),\sigma)
\]
and continue to show that the resulting difference (see Remark~\ref{rem:func_calc}.3
for notation of functional calculus)
\begin{align*}
x
&\;:=\;
S(\rho,\pi_\cE(\rho))-S(\rho,\sigma)
-[S(\pi_\cE(\rho),\pi_\cE(\rho))-S(\pi_\cE(\rho),\sigma)]\\
&\;=\;
\tr[\rho-\pi_\cE(\rho)][\log^{[q]}\sigma-\log^{[p]}\pi_\cE(\rho)]
\end{align*}
is zero. By definition of $\cE_p$ resp.\ $\cE_q$ there exist 
$\theta,\widetilde{\theta}\in\Theta$ and $y,\widetilde{y}\in\bR$, such that
$\log^{[p]}(\pi_\cE(\rho))=c^p(\theta)+yp$ resp.\
$\log^{[q]}(\sigma)=c^q(\widetilde{\theta})+\widetilde{y}q$. Since
$s(\rho)\preceq s(\pi_\cE(\rho))=p\preceq q$ this gives
\[\textstyle
x
\;=\;
\tr[\rho-\pi_\cE(\rho)][q\widetilde{\theta}q-p\theta p]
\;=\;
\tr[\rho-\pi_\cE(\rho)][\widetilde{\theta}-\theta]\,.
\]
Since the translation vector space $U$ of $\Theta$ is perpendicular to
the fibers of the projection $\pi_\cE$,  the difference $\widetilde{\theta}-\theta$
is perpendicular to $\rho-\pi_\cE(\rho)$, hence $x=0$. 
\hspace*{\fill}$\Box$\\
%
%
%
%
%
%
%
\subsection{The Complete projection theorem}
\label{sec:project}
\par
We prove the Complete projection theorem. In \S\ref{sec:select-com-pro}
we show the corollary that the extension ${\rm ext}(\cE)$ in 
Definition~\ref{def:extension} is the rI-closure (\ref{def:rI-closure})
and we consider the example of the Staffelberg family.
Application to quantum correlations are described in \S\ref{sec:max}.
\par
We are going to use the full scope of convex geometry introduced in
\S\ref{ingredients:convex}. A discussion why poonems are essential
is given in \S\ref{sec:why-poonems}. Let us start by citing 
Lemma~9 and 13 in \cite{Weis_Knauf}. The first lemma follows also from 
Lemma~\ref{lem:pert_exp}.2 in this article. The free energy
(\ref{def:free_energy}) is denoted by $F$.
\begin{Lem}
\label{lem:e-limits}
Suppose $\theta,u\in\sa$ and $p:=p_\cA^+(u)$ is the maximal projection of 
$u$. We have
\begin{equation}
\label{asy:exp1}\textstyle
\lim_{t\to\infty}R_\cA(\theta+t\,u)
\;=\;R_{p\cA p}(c^p(\theta))\,.
\end{equation}
and
\begin{equation}
\label{asy:free_energy}\textstyle
\lim_{t\to\infty}\big(F_\cA(\theta+t\,u)-t\,\lambda_\cA^+(u)\big)
\;=\;F_{p\cA p}(c^p(\theta))\,.
\end{equation}
\end{Lem}
\par
For technical reasons we denote the relative entropy with the first
argument $\rho\in\cS$ fixed by $S_\rho:\cS\to[0,\infty]$,
$S_\rho(\sigma):=S(\rho,\sigma)$.
\begin{Lem}
\label{lemma:infimum_at_infty}
Suppose $\theta,u\in\sa$ and $u$ is not proportional to the 
identity $\id$ in $\cA$. If the state $\rho$ belongs to 
the exposed face $F_\perp(\cS_\cA,u)$ of the state space, then 
$S_{\rho}(R_\cA(\theta+t\,u))$ is strictly monotone decreasing in
$t\in\bR$ and
\[\textstyle
\inf_{t\in\bR}S_{\rho}\big(R_\cA(\theta+t\,u)\big)%
\;=\;
\lim_{t\to\infty}S_{\rho}\big(R_\cA(\theta+t\,u)\big)%
\;=\;
S_{\rho}\big(\lim_{t\to\infty}R_\cA(\theta+t\,u)\big)\,.
\]
\end{Lem}
\par
The following approximation along e-geodesics is described in 
\S\ref{sec:why-poonems} with the example of the Swallow family. We use the entropy 
distance ${\rm d}_\cE$ defined in (\ref{def:entropy_dist}).
\begin{Pro}\label{pro:intriguing}
Let $\rho\in\cS$ and $\sigma\in{\rm ext}(\cE)$, such that
$\sigma\neq\pi_\cE(\rho)$ and $S_\rho(\sigma)<\infty$. There exists a 
norm continuous curve $\gamma:[0,1]\to{\rm ext}(\cE)$ from $\gamma(0)=\sigma$
to $\gamma(1)=\pi_\cE(\rho)$, such that
\begin{enumerate}[1.]
\item
$S_\rho(\gamma(t))$ is strictly monotone decreasing in $t$ and
\item
${\rm d}_\cE(\rho)\leq S_\rho(\gamma(1))$. 
\end{enumerate}
\end{Pro}
{\em Proof:\/}
We construct $\gamma$ by concatenation of several e-geodesics. Since
$S_\rho(\sigma)<\infty$ we have $s(\rho)\preceq s(\sigma)$. By 
Lemma~\ref{lem:stratum} there exists a unique projection $p\in\cP^U$, such that 
$\rho\in\ri\bF_\cA(p)+U^\perp$ and then it follows from Lemma~\ref{lem:ext-m} 
that $\pi_\cE(\rho)\in\cE_p$. We denote by $q\in\cP^U$ the projection such that 
$\sigma\in\cE_q$, i.e.\ $s(\sigma)=q$. By Lemma~\ref{lem:stratum} we have
$p\preceq s(\sigma)=q$. By Corollary~\ref{cor:two_for_access} there exists an 
access sequence of projections for $U$ including both $p$ and $q$, say
\[\textstyle
\id
\;=\;
p_0
\;\succ\;
p_1
\;\succ\;
\cdots
\;\succ\;
p_m
\;=\;
p\,,
\]
where $p=p_m$ for $m\geq 0$ and $q=p_l$ for $l\geq 0$ and $l\leq m$.
\par
We define a number of $(m-l)$ e-geodesic rays in 
$\cE_{p_l},\cE_{p_{l+1}},\ldots,\cE_{p_{m-1}}$. From the 
Definition~\ref{def:access}.3 of an access sequence and by 
Corollary~\ref{cor:exposed_projections}, for each $k=l,\ldots,m-1$
there exists $u_k\in c^{p_k}(U)$, such that
\begin{equation}
\label{eq:acc_direction}\textstyle
p_{k+1}
\;=\;
p_{p_k\cA p_k}^+(u_k)\,.
\end{equation}
Moreover, $u_k$ is not a multiple of the identity $p_k$ in $p_k\cA p_k$ 
(because $p_{k+1}\neq p_k$). Let $\theta\in\Theta$ such that
$\sigma=R_{q\cA q}(c^q(\theta))$. We define for $k=l,\ldots,m-1$ the 
e-geodesic
\begin{equation}
\label{def:e-rays}\textstyle
g_k:\;
\bR\;\to\;\cE_{p_k}\,,
\quad
t\;\mapsto\;R_{p_k\cA p_k}(c^{p_k}(\theta)+t u_k)\,.
\end{equation}
Using (\ref{asy:exp1}) we define
\begin{equation}
\label{eq:e-limit}\textstyle
\sigma_{k+1}
\;:=\;
\lim_{t\to\infty}g_k(t)
\;=\;
R_{p_{k+1}\cA p_{k+1}}(c^{p_{k+1}}(\theta))
\;\in\;
\cE_{p_{k+1}}\,.
\end{equation}
After reparametrization $t=\tfrac s{1-s}$, each e-geodesic ray
$g_l|_{[0,\infty)},g_{l+1}|_{[0,\infty)},\ldots,g_{m-1}|_{[0,\infty)}$
is defined on the segment $[0,1)$.
\par
We concatenate the reparametrized e-geodesic rays to a continuous curve 
$\widetilde{\gamma}:[0,m-l]\to{\rm ext}(\cE)$. The pieces fit together
by (\ref{eq:e-limit}). If $\sigma_m\neq\pi_\cE(\rho)$ then we add 
the e-geodesic segment in $\cE_p$ from $\sigma_m$ to $\pi_\cE(\rho)$.
This is parametrized under $R_{p\cA p}$ by a straight line segment in 
$c^p(\Theta)$, which we parametrize linearly by the unit interval
$[0,1]$. Since $\sigma\neq\pi_\cE(\rho)$, one of the inequalities
$m-l>0$ or $\sigma_m\neq\pi_\cE(\rho)$ must be true so we obtain a
curve $\gamma:[0,1]\to{\rm ext}(\cE)$ from $\widetilde{\gamma}$ by
linear reparametrization with the strictly positive factor $m-l$ or 
$m-l+1$.
\par
We argue that $S_\rho$ is strictly monotone decreasing along $\gamma$.
Let us begin with the rays in the canonical parametrization (\ref{def:e-rays}) 
for $k=l,\ldots,m-1$. Since $p\preceq p_{k+1}\preceq p_k$ we have by 
Proposition~\ref{ss:st} and (\ref{eq:acc_direction})
\begin{equation}
\label{eq:rho_exposed}\textstyle
\rho
\;\in\;
\bF_\cA(p)
\;\subset\;
\bF_\cA(p_{k+1})
\;=\;
\bF_{p_k\cA p_k}(p_{k+1})
\;=\;
F_\perp(\cS_{p_k\cA p_k},u_k)\,.
\end{equation}
Since $u_k$ is not a multiple of the identity $p_k$, 
Lemma~\ref{lemma:infimum_at_infty} can be invoked and it shows that 
$S_\rho$ is strictly monotone decreasing along $g_k$.
\par
The fact that $S_\rho$ is strictly monotone decreasing along the 
e-geodesic segment from $\sigma_m$ to $\pi_\cE(\rho)$ uses the strict
convexity of $S_\rho$ on $\cE_p$ in the canonical parametrization 
$c^p(\Theta)\to\cE_p$ under $R_{p\cA p}$. Let us recall the details.
In order to have an injective parametrization we project
$c^p(\Theta)$ onto $(p\cA p)_0=\{a\in(p\cA p)_{\rm sa}\mid\tr(a)=0\}$
so that $\Theta_0:=\pi_{(p\cA p)_0}(c^p(\Theta))$ satisfies
$\cE_p=R_{p\cA p}(c^p(\Theta))=R_{p\cA p}(\Theta_0)$. For all
$\theta_0\in\Theta_0$ an elementary calculation shows 
\[\textstyle
S_\rho(R_{p\cA p}(\theta_0))
\;=\;
-S(\rho)-\tr(\rho\theta_0)+F_{p\cA p}(\theta_0)
\]
where $S(\rho)$ is the von Neumann entropy and $F$ is the free energy
(\ref{def:free_energy}). Hence for
$u_0,v_0\in\pi_{(p\cA p)_0}(c^p(U))=\lin(\Theta_0)=\Theta_0-\Theta_0$
the second derivative
\[\textstyle
\tfrac{\partial^2}{\partial s\partial t}|_{s=t=0}
S_\rho(R_{p\cA p}(\theta_0+su_0+tv_0))
\;=\;
\tfrac{\partial^2}{\partial s\partial t}|_{s=t=0}
F_{p\cA p}(\theta_0+su_0+tv_0))
\;=\;
\langle\!\langle u_0,v_0\rangle\!\rangle_{\theta_0}
\]
equals the BKM-metric (\ref{def:ddF}). As discussed in the paragraph
following (\ref{eq:ddF}) this is a Riemannian metric, so
$S_\rho(R_{p\cA p}(\theta_0))$ has a positive definite Hessian 
throughout $\Theta_0$ and is strictly convex there. 
Since $\pi_\cE(\rho)\in\cE_p$, the function $S_\rho$
has on $\cE_p\cong\Theta_0$ a global minimum at $\pi_\cE(\rho)$, this
follows from the projection theorem (\ref{eq:pro_thm}). Hence $S_\rho$
is strictly monotone decreasing along the e-geodesic from
$\sigma_m\in\cE_p$ to $\pi_\cE(\rho)\in\cE_p$. 
\par
Second, we prove ${\rm d}_\cE(\rho)\leq S_\rho(\pi_\cE(\rho))$ by showing for 
$k=0,\ldots,m-1$ that ${\rm d}_{\cE_{p_k}}(\rho)\leq{\rm d}_{\cE_{p_{k+1}}}(\rho)$ 
holds. Then 
\[\textstyle
{\rm d}_\cE(\rho)
\;=\;
{\rm d}_{\cE_{p_0}}(\rho)
\;\leq\;
{\rm d}_{\cE_{p_1}}(\rho)
\;\leq\;
\cdots
\;\leq\;
{\rm d}_{\cE_{p_m}}(\rho)
\;=\;
{\rm d}_{\cE_p}(\rho)
\;\leq\;
S_\rho(\pi_\cE(\rho))
\]
will follow, the last inequality since $\pi_\cE(\rho)\in\cE_p$. Let 
$\tau\in\cE_{p_{k+1}}$ and let $\theta\in\Theta$ such that
$\tau=R_{p_{k+1}\cA p_{k+1}}(c^{p_{k+1}}(\theta))$. The construction 
from (\ref{eq:acc_direction}) to (\ref{eq:rho_exposed}) can be extended 
to $k=0,\ldots,m-1$. Now Lemma~\ref{lemma:infimum_at_infty} shows
\[\textstyle
{\rm d}_{\cE_{p_k}}(\rho)
\;\leq\;
\inf_{t\in\bR}S_\rho(R_{p_k\cA p_k}(c^{p_k}(\theta)+t u_k))
\;=\;
S_\rho(R_{p_{k+1}\cA p_{k+1}}(c^{p_{k+1}}(\theta)))
\;=\;
S_\rho(\tau)\,.
\]
Taking the infimum over all $\tau\in\cE_{p_{k+1}}$ the claim follows.
\hspace*{\fill}$\Box$\\
\par
Local minimizers in the following theorem are understood in the norm topology: 
If $(X,\cT)$ is a topological space, $Y\subset X$ and $f:Y\to\bR$, then 
$x_0\in Y$ is a {\it local minimizer} ({\it maximizer}) of $f$ on $Y$, if there 
is a $\cT$ open subset $V\subset X$ including $x_0$, such that for all 
$x\in V\cap Y$ we have $f(x_0)\leq f(x)$ ($f(x_0)\geq f(x)$). We now consider 
the entropy distance ${\rm d}_\cE$, the rI-closure ${\rm cl}^{\rm rI}(\cE)$ and 
the extension ${\rm ext}(\cE)$ defined respectively in (\ref{def:entropy_dist}), 
(\ref{def:rI-closure}) and (\ref{def:ext}).
\begin{Thm}[Complete projection theorem]
\label{thm:ext_rI}
For each $\rho\in\cS$ the relative entropy $S_\rho$ has a unique local minimizer 
on the extension ${\rm ext}(\cE)$ at $\pi_\cE(\rho)$. The entropy distance is 
${\rm d}_\cE(\rho)=\min_{\sigma\in{\rm ext}(\cE)}S_\rho(\sigma)=S_\rho(\pi_\cE(\rho))$.
\end{Thm}
{\em Proof:\/}
For each $\rho\in\cS$ we observe from Proposition~\ref{pro:intriguing}.1 and from 
the fact that $S_\rho$ is finite on $\cE$, that $\pi_\cE(\rho)$ is the unique 
global minimizer of $S_\rho$ on ${\rm ext}(\cE)$. We denote its value by 
${\rm d}_{{\rm ext}(\cE)}(\rho)$. With Proposition~\ref{pro:intriguing}.2 we have
for all $\sigma\in{\rm ext}(\cE)$
\[\textstyle
{\rm d}_\cE(\rho)
\;\leq\; 
S_\rho(\pi_\cE(\rho))
\;\leq\; 
S_\rho(\sigma)\,.
\]
Taking the infimum over $\sigma\in{\rm ext}(\cE)$ this shows 
${\rm d}_\cE(\rho)\leq{\rm d}_{{\rm ext}(\cE)}(\rho)$. Since the converse inequality 
is trivial, we have proved
\[\textstyle
{\rm d}_\cE(\rho)
\;=\;
{\rm d}_{{\rm ext}(\cE)}(\rho)
\,.
\]
Now ${\rm cl}^{\rm rI}(\cE)={\rm cl}^{\rm rI}({\rm ext}(\cE))$ follows from the 
definition of the rI-closure. We show 
${\rm cl}^{\rm rI}({\rm ext}(\cE))={\rm ext}(\cE)$ to conclude
${\rm ext}(\cE)={\rm cl}^{\rm rI}(\cE)$. The inclusion ``$\supset$'' 
is trivial and the converse ``$\subset$'' follows from the distance-like
properties of the relative entropy (\ref{eq:relative_entropy}) and since
the minimum ${\rm d}_{{\rm ext}(\cE)}(\rho)$ is attained on ${\rm ext}(\cE)$
for each state $\rho\in\cS$.
\par
It remains to discuss local minimizers $\sigma$ of $S_\rho$ on ${\rm ext}(\cE)$. 
If $S_\rho(\sigma)<\infty$, then Proposition~\ref{pro:intriguing}.1 shows that 
$\sigma$ is not a local minimizer unless $\sigma=\pi_\cE(\rho)$. If 
$S_\rho(\sigma)=\infty$ we observe that $\cE$ is norm dense in 
${\rm cl}^{\rm rI}(\cE)$ by the Pinsker-Csiz\'ar inequality (\ref{eq:pinsker}). 
Since $S_\rho$ has finite values on $\cE$, the state $\sigma$ is not a local 
minimizer.
\hspace*{\fill}$\Box$\\
%
%
%
%
%
%
\subsection{Maximizers of the entropy distance}
\label{sec:max}
\par
The {\it mutual information} is a measure of the total correlation in a 
bipartite quantum system \cite{Modi,Nielsen}. To see how it is related to an 
exponential family we consider two identical quantum systems, described by 
the algebra $\cA:={\rm Mat}(n,\bC)$. The algebra of the joint system is the 
tensor product $\cA\otimes\cA$. If $f:\cA\otimes\cA\to\bC$ is its 
state, then the state of the subsystems are $f_1(a):=f(a\otimes\id_n)$ and 
$f_2(a):=f(\id_n\otimes a)$ for $a\in\cA$. The density matrices $\rho$, 
$\rho_1$ and $\rho_2$ associated respectively to $f$, $f_1$ and $f_2$ can 
be used to define the mutual information 
\[\textstyle
I(\rho)
\;:=\;
S(\rho_1)+S(\rho_2)-S(\rho)\,.
\]
The mutual information is a continuous function on $\cS_{\cA\otimes\cA}$.
Using the vector space $L:=\{a\otimes\id_n+\id_n\otimes b\mid a,b\in\sa\}$ of 
{\it local observables}, the exponential family 
\[\textstyle
\cF
\;:=\;
R(L)
\;=\;
\{\rho_1\otimes\rho_2\mid\rho_1,\rho_2\in\cS(\cA)\text{ invertible }\}
\]
contains all invertible product states $\rho_1\otimes\rho_2$, having no
correlation. The mean value chart (\ref{eq:wichmanns-bijection}) and the 
projection theorem (\ref{thm:projection_E}) show for invertible 
$\rho\in\cS_{\cA\otimes\cA}$ that mutual information is the entropy distance 
(\ref{def:entropy_dist})
\[
I(\rho)
\;=\;
{\rm d}_\cF(\rho)
\;=\;
\inf\{S(\rho,\sigma)\mid\sigma\in\cF\}\,.
\]
\par
Maximization of correlation measures in terms of the entropy distance 
from an exponential family is proposed in \cite{Ay} as a structuring principle 
in natural systems, see also \cite{Knauf,Weis_Knauf} and the references therein.
We prove two necessary conditions for a local maximizer of the entropy
distance ${\rm d}_\cE$ from an exponential family $\cE$ in a C*-subalgebra
$\cA$ of ${\rm Mat}(n,\bC)$. See the beginning of \S\ref{sec:exp} for
notation.
\par
The first condition, an upper bound on the rank, enforces a certain degree of 
determinism on local maximizers. We shall use the fact that a unique face of 
$\cS_\cA$ exists, which contains a given state $\rho\in\cS_\cA$ in its relative 
interior (\ref{eq:stratification}).
\begin{Pro}
\label{pro:support_bound}
Let $\rho\in\cS_\cA$ be a local maximizer of the entropy distance 
${\rm d}_\cE$ from $\cE$ and assume that $F$ is the face of the
state space $\cS_\cA$ which contains $\rho$ in its relative interior.
Then ${\rm dim}(F)\leq{\rm dim}(\cE)$.
\end{Pro}
{\em Proof:\/}
We consider the convex body $K:=F\cap(\rho+U^\perp)$.
If two convex sets $X,Y\subset\bE^n$ in the finite-dimensional
Euclidean vector space $(\bE,\langle\cdot,\cdot\rangle)$ share a
relative interior point, then $\ri(X\cap Y)=\ri(X)\cap\ri(Y)$ 
follows, see e.g.\ Theorem 6.5 in \cite{Rockafellar}. Hence 
$\rho\in\ri(K)$ follows.
\par
If $\sigma\in K$, then by definition (\ref{def:pi_ce}) of the projection $\pi_\cE$ 
we have $\pi_\cE(\sigma)=\pi_\cE(\rho)$ and Theorem~\ref{thm:ext_rI} allows to 
rewrite the entropy distance
\[\textstyle
{\rm d}_\cE(\sigma)
\;=\;
S(\sigma,\pi_\cE(\sigma))
\;=\;
S(\sigma,\pi_\cE(\rho))\,.
\]
We have $p:=s(\pi_\cE(\rho))\succeq s(\sigma)$ by Lemma~\ref{lem:stratum}
and, using functional calculus in the algebra $p\cA p$
(Definition~\ref{def:functional_calc}.3), we get
\[\textstyle
{\rm d}_\cE(\sigma)
\;=\;
-S(\sigma)-\tr\sigma\log^{[p]}(\pi_\cE(\rho))\,.
\]
The von Neumann entropy $S(\sigma)$ is strictly concave, see e.g.\ 
\S{}II.B in \cite{Wehrl}. Hence ${\rm d}_\cE(\sigma)$ is a sum of 
a strictly convex function and a linear function, it is strictly convex
on $K$. Since $\rho$ is a local maximizer of ${\rm d}_\cE$ on $\cS_\cA$
it is a local maximizer on $K$. Since $\rho\in\ri(K)$ holds and
${\rm d}_\cE$ is strictly convex on $K$, we get $K=\{\rho\}$. 
Then
\[\textstyle
{\rm dim}(F)+{\rm dim}(U^\perp)
\;\leq\;
{\rm dim}(\sa)
\;=\;
{\rm dim}(U)+{\rm dim}(U^\perp)
\]
follows, hence ${\rm dim}(F)\leq {\rm dim}(U)$. If we choose a parametrization 
of $\cE$, such that $\id\not\in U$ (e.g.\ by replacing $\Theta$ by
$\pi_\az(\Theta)$) then Proposition~\ref{pro:mean_value_chart}.1
and~\ref{pro:mean_value_chart}.3 show ${\rm dim}(U)={\rm dim}(\cE)$,
completing the proof.
\hspace*{\fill}$\Box$\\
\par
Let us compare the physically relevant cases of $\cA=\bC^n$ and
$\cA={\rm Mat}(n,\bC)$. 
\begin{Rem}[Rank estimates]
In a C*-subalgebra $\cA$ of ${\rm Mat}(n,\bC)$ let $\rho\in\cS_\cA$ and let
$F$ be the face of $\cS_\cA$ containing $\rho$ in its relative interior. Let 
$p:=s(\rho)$ be the support projection of $\rho$ and let ${\rm rk}(\rho)$ be
the rank of $\rho$. Then Proposition~\ref{ss:st} shows $F=\cS_{p\cA p}$ and 
$\dim(\cS_{p\cA p})=\dim((p\cA p)_{\rm sa})-1$ hence
\[\textstyle
\dim(F)
\;=\;
\dim((p\cA p)_{\rm sa})-1
\;=\;
\dim_{\bC}(p\cA p)-1\,.
\]
If $\rho$ is a local maximizer of the entropy distance ${\rm d}_\cE$ from
an exponential family $\cE$ then Proposition~\ref{pro:support_bound} shows
\begin{equation}
\label{eq:inequality}\textstyle
\dim_{\bC}(p\cA p)
\;=\;
\dim(F)+1
\;\leq\;
\dim(\cE)+1\,.
\end{equation}
If $\cA\cong\bC^n$ is the algebra of diagonal matrices in ${\rm Mat}(n,\bC)$
we have $\dim_\bC(p\cA p)=\rk(\rho)$ hence (\ref{eq:inequality}) shows
\begin{equation}
\label{eq:inequality_ay}\textstyle
\rk(\rho)
\;\leq\;
\dim(\cE)+1\,.
\end{equation}
If $\cA={\rm Mat}(n,\bC)$ is the full matrix algebra of size $n$, then
$p\cA p$ is unitarily equivalent to the algebra of block diagonal matrices
${\rm Mat}(\rk(\rho),\bC)\oplus 0_{n-\rk(\rho)}$ and $p\cA p$ has
dimension $\dim_\bC(p\cA p)=\rk(\rho)^2$. Then (\ref{eq:inequality})
proves
\begin{equation}
\label{eq:inequality_weis}\textstyle
\rk(\rho)
\;\leq\;
\sqrt{\dim(\cE)+1}\,.
\end{equation}
\end{Rem}
\par
The second condition identifies local maximizers as the cutoff of their 
projection $\pi_\cE$ to the extension ${\rm ext}(\cE)$. The entropy distance of 
a local maximizer is a difference of free energies (\ref{def:free_energy}). We 
use functional calculus in compressed algebras 
(Definition~\ref{def:functional_calc}.3). 
\begin{Cor}
\label{cor:trunc}
Let $\rho\in\cS_\cA$ be any state. We denote $p:=s(\rho)$ the support 
projection of $\rho$, $q:=s(\pi_\cE(\rho))$ the support projection of the 
projection $\pi_\cE(\rho)$ and we fix a matrix $\theta\in\Theta$ such
that $\pi_\cE(\rho)=R_{q\cA q}(q\theta q)$. 
\begin{enumerate}[1.]
\item
If $u\in(p\cA p)_{\rm sa}$ is a traceless matrix, then
$\tfrac{\partial}{\partial t}{\rm d}_\cE(\rho+tu)|_{t=0}
=\langle u,\log^{[p]}(\rho)-p\theta p\rangle$.
\item
If $\rho$ is a local maximizer of the entropy distance ${\rm d}_\cE$ on
the state space $\cS_\cA$, then $\rho=R_{p\cA p}(p\theta p)$ and
${\rm d}_\cE(\rho)=F_{q\cA q}(q\theta q)-F_{p\cA p}(p\theta p)$.
\end{enumerate}
\end{Cor}
{\em Proof:\/}
By definition (\ref{def:pi_ce}) of the projection $\pi_\cE$ 
there exits a parameter $\theta\in\Theta$ and a projection $q\in\cP^U$ 
such that $\pi_\cE(\rho)=R_{q\cA q}(q\theta q)$. We notice $p\preceq q$ from 
Lemma~\ref{lem:stratum}. Corollary~\ref{cor:trunc} is proved in Theorem 31 in 
\cite{Weis_Knauf} for $q=\id$ (i.e.\ $\pi_\cE(\rho)$ invertible in $\cA$),  
$0\in\Theta$ (Gibbsian families) and for $\Theta$ consisting of traceless 
matrices. Using the mean value chart (\ref{def:mean_para}), the proof of
Theorem 31 in \cite{Weis_Knauf} is valid for affine parameter spaces $\Theta$
of trace-less matrices, including $0\not\in\Theta$: All 
assertions of the theorem are invariant under the substitution of 
$\theta\mapsto\theta+\lambda q$ for real $\lambda$, e.g.\ 
\begin{align*}\textstyle
F_{q\cA q}(q(\theta+\lambda\id)q)-F_{p\cA p}(p(\theta+\lambda\id)p)
&\;=\;
F_{q\cA q}(q\theta q)+\lambda-[F_{p\cA p}(p\theta p)+\lambda]\\
&\;=\;
F_{q\cA q}(q\theta q)-F_{p\cA p}(p\theta p)\,.
\end{align*}
This proves our claim for arbitrary non-empty affine subspaces
$\Theta\subset\sa$ if $q=\id$. 
\par
Otherwise, if $q\neq\id$, then
Theorem~\ref{thm:ext_rI} shows 
${\rm d}_\cE(\rho)={\rm d}_{\cE_q}(\rho)$. 
We argue analogously as before but with the algebra $q\cA q$ in place
of $\cA$. This is possible since $\pi_\cE(\rho)$ is invertible in 
$q\cA q$ and since $\rho\in q\cA q$. The latter is true since
$s(\rho)=p\preceq q$.
\hspace*{\fill}$\Box$\\
\begin{Rem}[Earlier results]
The idea to Proposition~\ref{pro:support_bound} goes back to Proposition~3.2
in \cite{Ay} where (\ref{eq:inequality_ay}) is proved on a subset of the 
probability simplex (\ref{def:prob_simplex}) on a finite set $\Omega$. This 
inequality was extended in Corollary~2 in \cite{Matus_Ay} to the whole 
probability simplex. Ay has also proved (\ref{eq:inequality_ay}) on a subset 
of the state space of ${\rm Mat}(n,\bC)$. The extension (\ref{eq:inequality})
to the whole state space of a C*-subalgebra of ${\rm Mat}(n,\bC)$ as well as 
the improvement from (\ref{eq:inequality_ay}) to (\ref{eq:inequality_weis})
are new.
\par
Corollary~\ref{cor:trunc} was first proved in Proposition 3.1
in \cite{Ay} for a subset of the probability simplex on a finite set
$\Omega$ and was proved in Theorem 5.1 in \cite{Matus} on the whole
probability simplex.
\end{Rem}
%
%
%
%
%
%
\subsection{Equality conditions for closures}
\label{sec:non-commutative}
\par
In addition to the rI-closure ${\rm cl}^{\rm rI}(\cE)$ and the norm closure 
$\overline{\cE}$ of an exponential family $\cE$ let us define the 
{\it geodesic closure}
\begin{equation}
\label{def:geo}\textstyle
{\rm cl}^{\rm geo}(\cE)
\;:=\;
\{\rho\in\cS_\cA\mid
\rho\text{ is the norm limit of an e-geodesic in }\cE\}
\end{equation}
where e-geodesics are one-dimensional exponential families (\ref{def:E}). 
The inclusions
\begin{equation}
\label{eq:inclusions}\textstyle
{\rm cl}^{\rm geo}(\cE)
\;\subset\;
{\rm cl}^{\rm rI}(\cE)
\;\subset\;
\overline{\cE}
\end{equation}
are already proved in Corollary~15 in \cite{Weis_Knauf}. The first inclusion 
follows from relative entropy estimates along e-geodesics. The second 
inclusion follows from the Pinsker-Csisz\'ar inequality. Below we argue that 
strict inclusions are only possible for a non-commutative algebra $\cA$. 
Examples in $\cA={\rm Mat}(2,\bC)\oplus\bC$ are the {\it Swallow family} 
with ${\rm cl}^{\rm geo}(\cE)\subsetneq{\rm cl}^{\rm rI}(\cE)$ and the 
{\it Staffelberg family} with ${\rm cl}^{\rm rI}(\cE)\subsetneq\overline{\cE}$,
see \S\ref{sec:select-com-pro}, \S\ref{sec:why-poonems} and also 
\S{}IV.B and \S{}IV.D in \cite{Weis_Knauf}. 
\par
We recall from Definition~\ref{def:face-lattices}.3 the lattice 
$\cF_\perp(\bM(U))$ of exposed faces of $\bM(U)$. 
\begin{Pro}
\label{pro:geodesics}
The projection $\pi_U({\rm cl}^{\rm geo}(\cE))$ of the geodesic closure 
can be a proper subset of the mean value set $\bM(U)$. In fact, 
\begin{equation}
\label{eq:geo-closure}\textstyle
\pi_U({\rm cl}^{\rm geo}(\cE))
\;=\;
\bigcup_{F\in\cF_\perp(\bM(U))}{\rm ri}(F)\,.
\end{equation}
In particular, ${\rm cl}^{\rm geo}(\cE)=\cl^{\rm rI}(\cE)$ holds if and only 
if all faces of the mean value set $\bM(U)$ are exposed faces.
\end{Pro}
{\em Proof:\/}
Using (\ref{asy:exp1}) and Corollary~\ref{cor:exposed_projections} we can
write\footnote{This is also proved in Proposition~10 in \cite{Weis_Knauf}.}
the geodesic closure of $\cE$ in the form 
${\rm cl}^{\rm geo}(\cE)=\bigcup_p\cE_p$, where the union extends over all
non-zero projections $p$ in the exposed projection lattice
$\cP^{U,\perp}_\cA$ defined in (\ref{eq:projection_lattice_u}). In
(\ref{eq:bi_p}) we have shown that each of the families $\cE_p$ projects 
to $\pi_U(\cE_p)=\ri\pi_U(\bF_\cA(p))$. So (\ref{eq:geo-closure}) follows 
from the lattice isomorphism (\ref{eq:main_iso}).
Theorem~\ref{thm:ext_rI} shows the disjoint union 
${\rm cl}^{\rm rI}(\cE)=\bigcup_p\cE_p$, where the union extends over all
non-zero projections $p$ in the projection lattice $\cP^U_\cA$.
So the equality ${\rm cl}^{\rm geo}(\cE)=\cl^{\rm rI}(\cE)$ is equivalent 
to $\cP^{U,\perp}_\cA=\cP^U_\cA$ and this, by the lattice isomorphism 
(\ref{eq:main_iso}), mean that all faces of $\bM_\cA(U)$ are exposed faces.
\hspace*{\fill}$\Box$\\
\par
We prove a condition for ${\rm cl}^{\rm rI}(\cE)=\overline{\cE}$ in terms of 
the entropy distance (\ref{def:entropy_dist}).
\begin{Pro}
\label{pro:continuity_with_proof}
We have ${\rm cl}^{\rm rI}(\cE)=\overline{\cE}$ if and only if the
entropy distance ${\rm d}_\cE$ is norm continuous on $\cS_\cA$.
\end{Pro}
{\em Proof:\/}
If the inclusion ${\rm cl}^{\rm rI}(\cE)\subset\overline{\cE}$ in
(\ref{eq:inclusions}) is strict, then there exists a norm convergent 
sequence $(\rho_i)_{i\in\bN}\subset{\rm cl}^{\rm rI}(\cE)$ with limit 
$\rho\in\cS_\cA\setminus{\rm cl}^{\rm rI}(\cE)$ in the compact
state space (see Proposition~\ref{ss:st}). By 
Theorem~\ref{thm:ext_rI} we have ${\rm d}_{\cE}(\rho)>0$ while 
${\rm d}_{\cE}(\rho_i)=0$ for $i\in\bN$ hence ${\rm d}_{\cE}$ is
discontinuous at $\rho\in\cS_\cA$.
\par
Conversely, let us prove that ${\rm d}_{\cE}$ is lower semi-continuous
if $\overline{\cE}={\rm cl}^{\rm rI}(\cE)$. Since $\overline{\cE}$ is a 
compact subset of $\sa$, lower semi-continuity of relative entropy 
(Remark~\ref{rem:conv_entropy}.1) implies lower semi-continuity 
of the mini\-mum
\[\textstyle
\cS_\cA\to\bR,\qquad
\rho\mapsto\min\{S(\rho,\sigma):\sigma\in\overline{\cE}\}\,.
\]
The proof given in Theorem 2, p.\ 116 in \cite{Berge}, uses a covering 
of $\overline{\cE}$ by open balls. This minimum function equals
${\rm d}_{\cE}$ by Theorem~\ref{thm:ext_rI}. 
Continuity of ${\rm d}_\cE$ follows from the lower semi-continuity 
of ${\rm d}_\cE$, see e.g.\ Lemma 4.2 in \cite{Ay}.
\hspace*{\fill}$\Box$\\
\par
The statements just proved can be formulated as necessary condition of 
commutativity of $\cA$. If $\cA\cong\bC^n$ then $\cS_\cA$ is a simplex, 
hence $\bM_\cA(U)$ is a polytope and all faces of $\bM_\cA(U)$ are exposed 
faces. Then Proposition~\ref{pro:geodesics} proves 
${\rm cl}^{\rm geo}(\cE)={\rm cl}^{\rm rI}(\cE)$. Moreover we have 
$\cT^{\rm rI}=\cT^{\|\cdot\|}$ by Corollary~\ref{cor:top} hence 
${\rm cl}^{\rm rI}(\cE)=\overline{\cE}$. In particular, 
Proposition~\ref{pro:continuity_with_proof} shows that the entropy distance 
${\rm d}_\cE$ is norm continuous in the setting of probability distributions
on a finite measurable space (\ref{def:commutative_inclusion}). This
was first proved in Lemma 4.2 in \cite{Ay}.
%
%
%
%
%
%
\section{Comments on the representation}
\label{sec:conc}
\par 
We show that our results about the I-/rI-topology hold for an
arbitrary finite-dimensional C*-algebra $\cA$.
In order to define exponential families in $\cA$ we chose a 
representation of $\cA$ as a C*-subalgebra of ${\rm Mat}(n,\bC)$.
We show that the Complete projection theorem and Pythagorean theorem 
are independent of this choice.
\par
The first object, the relative entropy, is monotone under 
C*-morphisms 
\[\textstyle
\Phi:\;\cB\;\to\;\cA
\]
between two unital C*-algebras \cite{Uhlmann}, i.e.\ if 
$f,g:\cA\to\bC$ are two states and if $\Phi^*(f):=f\circ\Phi$, then
$S(\Phi^*(f),\Phi^*(g))\leq S(f,g)$ holds. If $\Phi:\cB\to\cA$ is a
C*-isomorphism then $\Phi^{-1}$ provides the opposite inequality 
and $S(\Phi^*(f),\Phi^*(g))=S(f,g)$ follows. In particular, our results
about the I-/rI-topology in \S\ref{sec:topo-qm} are valid in any 
finite-dimensional C*-algebra independent of the representation.
\par
Our definition of exponential family needs normal states on a 
finite-dimensional von Neumann algebra $\cA$, represented by an algebra 
of linear operators on a Hilbert space. The normal states on $\cA$
are represented by positive and normalized trace class operators $\rho$
in $\cA$, with associated linear functional
\[\textstyle
a
\;\mapsto\;
\langle a,\rho\rangle
\;=\;\tr(a\rho)\,,\quad(a\in\cA)\,,
\]
see Theorem 2.4.21 in \cite{Bratteli} and \S~\ref{intro:hist}. 
However, not all representation are equally suitable. 
E.g.\ $\bC$ represented as $\{x=(x_i)_{i\in\bN}\in l^\infty\mid
\exists\lambda\in\bC,\forall i\in\bN:x_i=\lambda\}$ has no normal 
state. So we restrict to representation on finite-dimensional 
Hilbert spaces. We want to see if our results are independent 
this choice.
\par
Every finite-dimensional C*-subalgebra is, according to Theorem III.1.1
in \cite{Davidson}, C*-isomorphic to the direct sum 
\begin{equation}
\label{eq:direct_rep}\textstyle
\cB\;:=\;\bigoplus_{i=1}^N{\rm Mat}(k_i,\mathbb C)\,,
\end{equation}
where $N\in\bN_0$ and $k\in\bN^N$ is a multi-index. 
Any C*-algebra $\cA$ of linear operators on a finite-dimensional 
Hilbert space, which is C*-isomorphic to $\cB$, has the form of
\[\textstyle
\cA\;:=\;
\bigoplus_{i=1}^N\{\bigoplus_{j=1}^{m_i}a_i
\mid a_i\in{\rm Mat}(k_i,\mathbb C)\}\oplus 0_l
\]
up to unitary equivalence. Here $m_i\geq 1$ for $i=1,\ldots,N$ and 
$l\geq 0$ are integers see Corollary III.2.1 in \cite{Davidson}.
Moreover, there is a C*-isomorphism $\Phi:\cB\to\cA$
\begin{equation}
\label{eq:Phi}\textstyle
\Phi(\bigoplus_{i=1}^N b_i)
\;=\;
\bigoplus_{i=1}^N\bigoplus_{j=1}^{m_i}b_i\oplus 0_l\,,
\qquad (b_1,\ldots,b_N)\in\cB\,.
\end{equation}
\par
Let us begin to discuss mean values. While the mean value set 
(\ref{def:m_set}) pleases with a simple Euclidean geometry, the 
isomorphic convex support (\ref{def:cs}) has other advantages. 
Firstly, it is equivariant under the isomorphism (\ref{eq:Phi}):  
$f(b)=(\Phi^{-1})^*f(\Phi(b))$ holds for all states $f$ on $\cB$ and
$b\in\cB$. The mean value set is not equivariant. Another advantage of
the convex support is that its algebraic decomposition into faces
becomes a simple inclusion ${\rm cs}_{p\cA p}\subset{\rm cs}_\cA$,
see Lemma~\ref{lem:coords_on_cs_faces}.
\par
Let us discuss exponential families. The adjoint $\Phi^*:\cA^*\to\cB^*$
of (\ref{eq:Phi}) is given for $F_i\in{\rm Mat}(k_i,\mathbb C)$,
$i=1,\ldots,N$, by
\begin{equation}
\label{intro:adjoint}\textstyle
\Phi^*(\bigoplus_{i=1}^N\bigoplus_{j=1}^{m_i}F_i\oplus 0_l)\;=\;
\bigoplus_{i=1}^Nm_iF_i\,.
\end{equation}
\begin{Lem}
\label{lem:change_rep}
Let $\Theta\subset\cB_{\rm sa}$ be a non-empty affine subspace and 
$\cE:=R_\cB(\Theta)$.
Let $\theta_0:=\bigoplus_{i=1}^N\ln(m_i)\id_{k_i}\in\cB_{\rm sa}$.
Then the affine space 
$\widetilde{\Theta}:=\Phi\left(\Theta-\theta_0\right)
\subset\cA_{\rm sa}$ satisfies 
$(\Phi^*)^{-1}(\cE)=R_\cA(\widetilde{\Theta})$.
\end{Lem}
{\em Proof:\/}
By (\ref{intro:adjoint}) we have $\Phi^*\circ R\circ\Phi(\theta)
=R(\,\theta+\bigoplus_{i=1}^N\ln(m_i)\id_{k_i})$ for $\theta\in\Theta$.
\hspace*{\fill}$\Box$\\
\par
Lemma~\ref{lem:change_rep} shows that the class of exponential families
is preserved under the isomorphism (\ref{eq:Phi}). Clearly
$\Phi^*(\rho+U^\perp)=\Phi^*(\rho)+\Phi^{-1}(U)^\perp$ holds for all 
$\rho\in\cS_\cA$ and $U\subset\sa$. So the Complete
Pythagorean theorem and the Complete projection theorem
(Theorem~\ref{thm:pyth_with_proof} and Theorem~\ref{thm:ext_rI}) are
valid for C*-algebras represented on finite-dimensional Hilbert
spaces independent of the choice of representation. Gibbsian families
are not equivariant under the isomorphism (\ref{eq:Phi}) because 
$\widetilde{\Theta}$ in Lemma~\ref{lem:change_rep} is not necessarily 
a linear space even though $\Theta$ is a linear space.\\[1.0cm]
%
%
%
%
%
%
%
{\bf Acknowledgments:} 
I would like to thank Arleta Szko{\l}a for helpful and critical remarks about 
the final form of this article, Franti\v sek Mat\'u\v s for an introduction to 
sequential convergences and Nihat Ay for an introduction to maximizing relative 
entropy from exponential families. The Complete projection theorem was proved
2006--2009 in numerous discussions with my Ph.D.\ advisor Andreas Knauf, to 
whom I would like to express my sincere and deep gratitude. The thesis is 
available on the web \cite{Diss}. This work was supported by the DFG projects 
``Geometry and Complexity in Information Theory'' and
``Quantenstatistik: Entscheidungsprobleme und entropische Funktionale 
auf Zustandsr\"aumen''.
%
%
%
%
%
%
\bibliographystyle{amsalpha}

\begin{thebibliography}{10}
\renewcommand{\baselinestretch}{1}\normalsize
%
\bibitem[AS]{Alfsen} E.\,M.\ Alfsen and F.\,W.\ Shultz,
{\it State Spaces of Operator Algebras}, Birkh\"auser (2001).
%
\bibitem[Am1]{Amari_LNS} S.\ Amari, 
{\it Differential-geometrical methods in statistics},
Lecture Notes in Statistics {\bf 28},  Springer-Verlag New York
(1985).
%
\bibitem[Am2]{Amari} S.\ Amari, {\it Information Geometry on Hierarchy of 
Probability Distributions}, IEEE Trans.\ Inf.\ Theory
{\bf 47} 1701--1711 (2001).
%
\bibitem[AN]{Amari_Nagaoka} S.\ Amari and H.\ Nagaoka, {\it Methods of
Information Geometry}, AMS Translations of Mathematical Monographs
{\bf 191} (2000).
%
\bibitem[Ay]{Ay} N.\ Ay, {\it An Information-Geometric Approach to a 
Theory of Pragmatic Structuring}, Ann.\ Probab.\ {\bf 30} 416--436 (2002).
%
\bibitem[AK]{Knauf} N.\ Ay and A.\ Knauf, {\it Maximizing multi-information}, 
Kybernetika {\bf 42} 517--538 (2006).
%
\bibitem[Ba]{Barndorff} O.\ Barndorff-Nielsen, {\it Information and 
Exponential Families in Statistical Theory}, John Wiley \& Sons New York 
(1978).
%
\bibitem[Be]{Benatti} F.\ Benatti, {\it Dynamics, Information and 
Complexity in Quantum Systems}, Springer-Verlag (2009).
%
\bibitem[BW]{BWZ} I.\ Bengtsson, S.\ Weis and K.\ \.Zyczkowski ,
{\it Geometry of the Set of Mixed Quantum States: An Apophatic Approach},
Geometric Methods in Physics, Trends in Mathematics 175--197 (2013).
%
\bibitem[BZ]{Bengtsson} I.\ Bengtsson and K.\ \.Zyczkowski,
{\it Geometry of Quantum states. An Introduction to Quantum Entanglement},
Cambridge University Press (2006).
%
\bibitem[Br]{Berge} C.\ Berge, {\it Topological Spaces},
Oliver and Boyd Ltd (1963).
%
\bibitem[Bh]{Bhatia} R.\ Bhatia, {\it Matrix Analysis}, Springer-Verlag 
(1997).
%
\bibitem[Bi]{Birkhoff} G.\ Birkhoff, {\it Lattice Theory}, AMS 
Colloquium Publications 3rd.~ed.\ (1973).
%
\bibitem[BR]{Bratteli} O.\ Bratteli and D.\,W.\ Robinson, {\it Operator 
Algebras and Quantum Statistical Mechanics 1}, 2nd ed.\ Springer (2002).
%
\bibitem[BS]{Bjelakovic} I.\ Bjelakovi\'c, J.-D.\ Deuschel,
T.\ Kr\"uger, R.\ Seiler, R.\ Siegmund-Schultze and A.\ Szko{\l}a,
{\it A Quantum Version of Sanov's Theorem},
Comm.\ Math.\ Phys.\ {\bf 260} 659--671 (2005).
%
%
\bibitem[Cs2]{Csiszar67} I.\ Csisz\'ar, {\it On topological properties of 
f-divergences}, Studia Sci.\ Math.\ Hungar.\ {\bf 2} 329--339 (1967).
%
\bibitem[CM1]{Csiszar03} I.\ Csisz\'ar and F.\ Mat\'u\v s, 
{\it Information Projections Revisited}, IEEE Trans.\ Inf.\ Theory
{\bf 49} 1474--1490 (2003).
%
\bibitem[CM2]{Csiszar04} I.\ Csisz\'ar and F.\ Mat\'u\v s, {\it On 
Information Closures of Exponential Families: A Counterexample},
IEEE Trans.\ Inform.\ Theory {\bf 50} 922--924 (2004).
%
\bibitem[CM3]{Csiszar05} I.\ Csisz\'ar and F.\ Mat\'u\v s, 
{\it Closures of Exponential Families}, Ann.\ Probab.\ {\bf 33} 582--600 
(2005).
%
\bibitem[\v{C}e]{Cencov} N.\,N.\ \v{C}encov, {\it Statistical Decision 
Rules and Optimal Inference}, AMS Translations of Mathematical Monographs 
{\bf 53} (1982).
%
\bibitem[Da]{Davidson} K.\,R.\ Davidson, {\it C*-Algebras by Example}, 
Fields Institute Monographs {\bf 6} (1996).
%
\bibitem[Du1]{Dudley64} R.\,M.\ Dudley, {\it On Sequential Convergence},
Trans.\ Amer.\ Soc.\ {\bf 112} 483--507 (1964); correction ibid.\ 
{\bf 148} (1970).
%
\bibitem[Du2]{Dudley98} R.\,M.\ Dudley, {\it Consistency of M-Estimators 
and One-Sided Bracketing}, Progr.\ Probab.\ {\bf 43} 33--58, Birkh\"auser
(1998).
%
\bibitem[DS]{Dunford_Schwartz} N.\ Dunford and J.\,T.\ Schwartz, 
{\it Linear Operators}, I General Theory, Interscience Publishers
London (1958).
%
\bibitem[En]{Engelking} R.\ Engelking, {\it General Topology},
Sigma Series in Pure Mathematics {\bf 6} Heldermann (1989).
%
\bibitem[Gi]{Gilardoni} G.\,L.\ Gilardoni,
{\it On Pinsker's and Vajda's Type Inequalities for Csisz\'ar's f-Divergences},
IEEE Trans.\ Inf.\ Theory {\bf 56} (2010).
%
\bibitem[GS]{Grasselli} M.\,R.\ Grasselli and R.\,F.\ Streater, {\it On 
the Uniqueness of the Chentsov Metric in Quantum Information Geometry},
Infinite Dim.\ Anal.\ Quantum Info.\ and Related Topics {\bf 4} 173--182 
(2001).
%
\bibitem[Gr]{Gruenbaum} B.\ Gr\"unbaum, {\it Convex Polytopes},
Springer-Verlag 2nd.~ed.\ (2003).
%
\bibitem[Ha]{Halmos} P.\,R.\ Halmos, {\it Finite-Dimensional Vector 
Spaces}, Springer-Verlag (1987).
%
\bibitem[Hs]{Harremoes} P.\ Harremo\"es, {\it Information Topologies 
With Applications}, Entropy, Search, Complexity, Bolyai Society 
Mathematical Studies {\bf 16} (2007). 
%
\bibitem[Hi]{Hayashi} M.\ Hayashi, 
{\it Quantum information. An introduction},
Springer-Verlag Berlin (2006).
%
\bibitem[Ho]{Holevo} A.\,S.\ Holevo, {\it Probabilistic and statistical 
aspects of quantum theory}, 2nd ed. Edizioni della Normale (2011).
%
\bibitem[IO]{Ingarden} R.\,S.\ Ingarden, A.\ Kossakowski and M.\ Ohya, 
{\it Information Dynamics and Open Systems}, Kluwer Academic Publishers 
Group (1997).
%
\bibitem[Ja]{Jaynes} E.\,T.\ Jaynes,
{\it Information Theory and Statistical Mechanics I/II.} 
Phys.\ Rev.\ {\bf 106} 620--630 and {\bf 108} 171--190 (1957).
%
\bibitem[Je]{Jencova} A.\ Jen\v cov\'a, 
{\it Geometry of quantum states:\ dual connections and divergence functions},
Rep.\ Math.\ Phys.\ {\bf 47} 121--138 (2001).
%
\bibitem[Ka]{Kato} T.\ Kato, {\it Perturbation Theory for Linear 
Operators}, Springer-Verlag (1995).
%
%
\bibitem[KP]{Krantz} S.\,G.\ Krantz and H.\,R.\ Parks, 
{\it A Primer of Real Analytic Functions}, Birkh\"auser (2002).
%
\bibitem[KL]{Kullback} S.\ Kullback and R.\,A.\ Leibler,
{\it On Information and Sufficiency},
Ann.\ Math.\ Stat.\ {\bf 22} 79--86 (1951).
%
\bibitem[Li]{Lieb} E.\,H.\ Lieb, {\it Convex Trace Functions and
the Wigner-Yanase-Dyson Conjecture}, Adv.\ Math.\ {\bf 11} 267--188 (1973).
%
\bibitem[Ld]{Lindblad} G.\ Lindblad, 
{\it Entropy, Information and Quantum Measurement},
Comm.\ Math.\ Phys.\ {\bf 33} 305--322 (1973).
%
\bibitem[Ma]{Matus} F.\ Mat\'u\v s, 
{\it Optimality conditions for maximizers of the information divergence
from an exponential family}, Kybernetika {\bf 43} 731--746 (2007).
%
\bibitem[MA]{Matus_Ay} F.\ Mat\'u\v s and N.\ Ay,
{\it On maximization of the information divergence from an exponential
family}, Proceedings of WUPES'03, University of Economics Prague
199--204 (2003).
%
\bibitem[MW]{Modi} K.\ Modi, T.\ Paterek, W.\ Son, V.\ Vedral and 
M.\ Williamson,
{\it Unified View of Quantum and Classical Correlations},
Phys.\ Rev.\ Lett.\ {\bf 104} 080501 (2010). 
%
\bibitem[Mu]{Murphy} G.\,J.\ Murphy, 
{\it C*-Algebras and Operator Theory}, Academic Press (1990).
%
\bibitem[Na]{Nagaoka} H.\ Nagaoka, 
{\it Differential Geometrical Aspects of Quantum State Estimation
and Relative Entropy}, in: Quantum Communication, Computing and 
Measurement (eds.\ Hirota et al.) Plenum Press, New York (1994).
%
\bibitem[NC]{Nielsen} M.\,A.\ Nielsen and I.\,L.\ Chuang, {\it Quantum 
Computation and Quantum Information}, Cambridge University Press 
(2000).
%
\bibitem[OP]{Ohya} M.\ Ohya and D.\ Petz,
{\it Quantum Entropy and Its Use}, Springer-Verlag (1993).
%
\bibitem[Pe1]{Petz86} D.\ Petz, 
{\it Quasi-Entropies for Finite Quantum Systems},
Rep.\ Math.\ Phys.\ {\bf 23} 57--65 (1986).
%
\bibitem[Pe2]{Petz94} D.\ Petz,
{\it Geometry of canonical correlation on the state space of a quantum system}, 
J.\ Math.\ Phys.\ {\bf 35} 780--795 (1994).
%
\bibitem[Pe3]{Petz08} D.\ Petz, {\it Quantum Information Theory and 
Quantum Statistics}, Springer-Verlag (2008).
%
\bibitem[RS]{Reed} M.\ Reed and B.\ Simon, {\it Methods of modern 
mathematical physics I}, 2nd  ed., Academic Press, Inc.\ (1980).
%
\bibitem[Ro]{Rockafellar} R.\,T.\ Rockafellar, {\it Convex Analysis}, 
Princeton University Press (1972).
%
\bibitem[Ru]{Ruelle} D.\ Ruelle, {\it Statistical mechanics.
Rigorous results}, World Scientific Publishing (1999).
%
\bibitem[SS]{Sanyal} R.\ Sanyal, F.\ Sottile and B.\ Sturmfels,
{\it Orbitopes}, Mathematika {\bf 57} 275--314 (2011).
%
\bibitem[Sc]{Schneider} R.\ Schneider, {\it Convex Bodies: The 
Brunn-Minkowski Theory}, Cambridge University Press (1993).
%
\bibitem[Uh]{Uhlmann} A.\ Uhlmann, {\it Relative Entropy and the 
Wigner-Yanase-Dyson-Lieb Concavity in an Interpolation Theory}, 
Commum.\ Math.\ Phys.\ {\bf 54} 21--32 (1977).
%
\bibitem[We]{Wehrl} A.\ Wehrl, {\it General Properties of Entropy}, 
Rev.\ Mod.\ Phys.\ {\bf 50} (1978).
%
\bibitem[We1]{Diss} S.\ Weis, {\it Exponential Families with 
Incompatible Statistics and Their Entropy Distance}, PhD Dissertation,
University of Erlangen (2010).\\
\verb+http://www.opus.ub.uni-erlangen.de/opus/volltexte/2010/1580/+
%
\bibitem[We2]{Weis_touch} S.\ Weis,
{\it A Note on Touching Cones and Faces},\\
Journal of Convex Analysis {\bf 19} 323--353 (2012).
%
\bibitem[We3]{Weis_supp} S.\ Weis, {\it Quantum Convex Suppport},
Lin.\ Alg.\ Appl.\ {\bf 435} 3168--3188 (2011);
correction: ibid.\ {\bf 436} xvi (2012). 
\verb+http://arxiv.org/abs/1101.3098+
%
\bibitem[WK]{Weis_Knauf} S.\ Weis and A.\ Knauf, 
{\it Entropy Distance: New Quantum Phenomena},\\
J.~Math.\ Phys.\ {\bf 53} 102206 (2012).
%
\bibitem[Wr]{Werner} D.\ Werner, {\it Funktionalanalysis}, 
3rd ed.\ Springer-Verlag (2000).
%
\bibitem[Wi]{Wichmann} E.\,H.\ Wichmann, {\it Density matrices arising 
from incomplete measurements}, J.~Math.\ Phys.\ {\bf 4} 884--896 (1963).
%
\end{thebibliography}

\end{document}